\documentclass[useAMS, usenatbib]{mn2e}

\usepackage{aasmacros}
\usepackage{amssymb} 
\usepackage{subcaption} 
\usepackage{booktabs} 
\usepackage{multirow} 
\usepackage{amsfonts}
\usepackage{lmodern}
\usepackage{amsmath}
\usepackage{float}
\usepackage{graphicx}
\usepackage{xcolor}

\title[Quenching star formation with trapped IR radiation]{Quenching star formation with quasar outflows launched by trapped IR radiation} 

\author[Costa et al.]{Tiago Costa$^{1}$\footnotemark[1], Joakim Rosdahl$^{1, 2}$, Debora Sijacki$^{3}$ and Martin G. Haehnelt$^{3}$\\\
  $^{1}$Leiden Observatory, Leiden University, PO Box $9513$, NL-$2300$ RA Leiden, the Netherlands \\
  $^{2}$CRAL, Universit\'e de Lyon I, CNRS UMR 5574, ENS-Lyon, 9 Avenue Charles Andr\'e, 69561, Saint-Genis-Laval, France\\
  $^{3}$Institute of Astronomy and Kavli Institute for Cosmology,
  University of Cambridge, Madingley Road, Cambridge CB$3$ $0$HA, UK}

\begin{document}
 
\maketitle

\begin{abstract}
We present cosmological radiation-hydrodynamic simulations, performed with the code {\sc Ramses-RT}, of radiatively-driven outflows in a massive quasar host halo at $z \, = \, 6$. 
Our simulations include both single- and multi-scattered radiation pressure on dust from a quasar and are compared against simulations performed with thermal feedback.
For radiation pressure-driving, we show that there is a critical quasar luminosity above which a galactic outflow is launched, set by the equilibrium of gravitational  and radiation forces.
While this critical luminosity is unrealistically high in the single-scattering limit for plausible black hole masses, it is in line with a $\approx 3 \times 10^9 \, \rm M_\odot$ black hole accreting at its Eddington limit, if infrared (IR) multi-scattering radiation pressure is included.
The outflows are fast ($v \, \gtrsim \, 1000 \, \rm km \, s^{-1}$) and strongly mass-loaded with peak mass outflow rates $\approx 10^3 \-- 10^4 \, \rm M_\odot \, yr^{-1}$, but short-lived ($< 10 \, \rm Myr$).
Outflowing material is multi-phase, though predominantly composed of cool gas, forming via a thermal instability in the shocked swept-up component.
Radiation pressure- and thermally-driven outflows both affect their host galaxies significantly, but in different, complementary ways. 
Thermally-driven outflows couple more efficiently to diffuse halo gas, generating more powerful, hotter and more volume-filling outflows.
IR radiation, through its ability to penetrate dense gas via diffusion, is more efficient at ejecting gas from the bulge.
The combination of gas ejection through outflows with internal pressurisation by trapped IR radiation leads to a complete shut down of star formation in the bulge. 
We hence argue that radiation pressure-driven feedback may be an important ingredient in regulating star formation in compact starbursts, especially during the quasar's `obscured' phase.
\end{abstract}


\renewcommand{\thefootnote}{\fnsymbol{footnote}}
\footnotetext[1]{E-mail: costa@strw.leidenuniv.nl}

\section{Introduction}

In the present day Universe, most galaxies with stellar masses greater than $\sim 10^{11} \, \rm M_\odot$ are passive, with star formation rates per unit stellar mass well below $10^{-2} \, \rm Gyr^{-1}$ \citep[e.g.][]{Peng:10, Moustakas:13, Bauer:13}.
With stellar populations dominated by old stars, such passive galaxies typically exhibit red optical colours and are therefore said to be `red and dead' \citep[e.g.][]{Kauffmann:03, Bell:04, Brammer:09}.
Though rarer in number, inactive massive systems appear to exist out to redshifts as high as $z \,=\, 4$ \citep[see e.g.][]{Gobat:12, Muzzin:13, Straatman:14, Belli:15, Glazebrook:17}.
The mechanisms responsible for the suppression of star formation in massive galaxies thus appear to have been operating since the earliest stages of galaxy formation.

These observations put strain on galaxy formation models based on $\Lambda$CDM cosmologies.
The challenge for these models, most clearly illustrated by the discrepancy between the shapes of the observed stellar mass function of galaxies and the predicted dark matter halo mass function \citep{Read:05, Moster:10, Behroozi:10}, is to prevent excessive star formation in both low- and high-mass systems.
In order to reconcile theory and observations, virtually all models invoke a series of feedback processes that tap into the energy released by stars (through the stellar radiation field, supernovae, stellar winds and cosmic rays), and accreting black holes (through radiative heating, winds and jets).

While sufficiently energetic to regulate star formation in galaxies with stellar masses below $\sim 10^{11} \, M_\odot$, supernovae likely fall short of providing the required energy in the deepest gravitational potential wells \citep{Dekel:86, Benson:03}. 
In order to reproduce the bulk properties of massive galaxies, most successful models instead appeal to the energy output by active galactic nuclei (AGN) powered by rapidly accreting supermassive black holes \citep{Scannapieco:04, Churazov:05, DiMatteo:05, Croton:06, Bower:06}.
The underlying assumption is that energy and momentum liberated by AGN couples to the surrounding interstellar- and intergalactic media (ISM, IGM) efficiently, heating and pushing the surrounding gas and, possibly, expelling it via powerful large-scale outflows.
Accordingly, AGN feedback has been incorporated into practically all state-of-the-art cosmological simulations of galaxy formation \citep{Sijacki:07, McCarthy:11, Vogelsberger:14, Dubois:14, Schaye:15, Khandai:15, Weinberger:17}.
The limitations imposed by insufficient numerical resolution in even the most sophisticated cosmological simulations as well as the absence of key physical processes from the models (e.g. radiative transfer, magneto-hydrodynamics), however, prevent an \emph{ab-initio} treatment of AGN feedback.
Consequently, despite the many successes of cosmological simulations in reproducing the bulk properties of massive galaxies, the physical mechanism that couples AGN energy to its surroundings and its efficiency remain elusive.

One possible AGN feedback channel is direct radiation pressure on dusty gas at galactic scales \citep{Fabian:99, Murray:05, Thompson:15, Ishibashi:15}.
The importance of dust can be appreciated from the following argument.
As in the classical Eddington limit, the radiation force exerted by the radiation field of a central source with luminosity $L_{\rm cen}$ and flux $F \, = \, L_{\rm cen} / (4 \pi R^2)$, where $R$ is the radial distance to the source, can balance the gravitational weight of a galaxy with gas mass $M_{\rm gal} (R)$, here assumed to dominate over other matter components, if 
\begin{equation}
L_{\rm cen} \, = \, \frac{4 \pi G M_{\rm gal}(R) c}{\kappa}\, ,
\end{equation}
where $G$ is the gravitational constant, $c$ the speed of light in vacuum and $\kappa$ the scattering cross-section per unit mass. For instance, $\kappa \, = \, \kappa_{\rm T} \, = \,  0.346 \, \rm cm^2 \, g^{-1} $ if electron (Thomson) scattering is the dominant coupling mechanism.

If the central source is an AGN powered by a black hole with mass $M_{\rm BH}$, accreting at its Eddington limit, then it must emit at a luminosity
\begin{equation}
L_{\rm AGN} \, = \, \frac{4 \pi G M_{\rm BH} c}{\kappa_{\rm T}}\, .
\label{eq_eddlimit}
\end{equation}

The mass of the galaxy that can be supported by radiation pressure from its central AGN is obtained by setting $L_{\rm cen} \, = \, L_{\rm AGN}$, which gives
\begin{equation}
M_{\rm gal} (R) \, = \, M_{\rm BH} \frac{\kappa}{\kappa_{\rm T}} \, .
\end{equation}
We require $\kappa \gg \kappa_{\rm T}$ if radiation pressure is to have an appreciable impact on the gas of the host galaxy \citep{Fabian:06}.

Opacities greater than $\kappa_{\rm T}$ are easily achieved by continuum dust absorption and photon scattering \citep{Draine:84}.
At optical and UV frequencies, the dust opacity $\kappa_{\rm D}$ can reach values $\sim 10^3 \, \rm cm^2 \, g^{-1} \gg\kappa_{\rm T}$ for a Milky Way dust-to-gas ratio $f_{\rm D} \, \approx \, 0.01$ \citep[e.g.][]{Li:01}.
The dust grains, which are expected to be charged even at large ($\sim \, \rm kpc$) distances from the AGN due to its strong radiation field, are expected to be dynamically coupled to at least the cold phase of the interstellar medium \citep{Murray:05} and to therefore impart the full radiation force on the gas, though hydrodynamic decoupling may occur in shocks or in the presence of strong radiation fields \citep[e.g.][]{Hopkins:16}.
It is possible for the dust opacities to be important also at infrared (IR) frequencies.
Calculations for dusty media composed of grains of various sizes and chemical composition predict temperature-dependent Rosseland mean opacities that can reach $\kappa_{\rm IR} \approx (1 \-- 10) \, \rm cm^2 \, g^{-1}$ at dust temperatures $T_{\rm D} \, \sim \, 100 \, \rm K$ \citep[e.g.][]{Pollack:94, Semenov:03}.

At galactic scales, the highest optical depths are measured in systems hosting intense star formation activity, such as ULIRGs, sub-millimetre galaxies \citep[e.g.][]{Lutz:16, Simpson:17}, reddened quasars \citep{Banerji:12, Banerji:17, Hamann:17} and hot dust obscured galaxies \citep{Eisenhardt:12, Assef:15, Tsai:15}.
For Arp220, a late stage galaxy merger and a local ULIRG, the two $\sim 100 \, \rm pc$-scale dusty nuclei are optically thick in the continuum at far-infrared (FIR) wavelengths $\lambda\,=\, 600 \, \mathrm{\mu m} \-- 2.6 \, \rm mm$, and the estimated luminosity-to-mass ratio of $\approx 500 \, \rm L_\odot / M_\odot$ is a strong indication that radiation pressure is dynamically important \citep{Scoville:17}. 

Possibly, radiation pressure on dust in these extreme environments lies at the origin of some of the AGN-driven outflows now often detected through emission and absorption lines \citep[see][for a recent compilation]{Fiore:17}. Analytic models based on the momentum balance of gravitational and radiation forces have previously shown that AGN are sufficiently luminous to accelerate significant masses of gas to high velocities $> 1000 \, \rm km \, s^{-1}$ at least out of the central regions of massive galaxies \citep{Thompson:15, Ishibashi:15}. Indeed, the prevalence of radiation pressure over, for instance, the thermal pressure of an extremely hot $T \sim 10^9 \, \rm K$ bubble as predicted by some models \citep[e.g.][]{King:03}, has been claimed to be favoured from available constraints based on emission line ratios in systems hosting powerful AGN-driven outflows \citep{Stern:16}.

The possibility that radiation pressure on dust in bright quasars drives large-scale outflows is the subject of this paper.
Direct radiation pressure on dust gives a force $\mathcal{F}_{\rm rad} \, = \,  L_{\rm AGN}/c$, which is lower than the momentum fluxes inferred for a number of observed molecular outflows and is likely insufficient to efficiently regulate black hole accretion and drive powerful large-scale outflows \citep{Debuhr:12, Costa:14}. In configurations in which dusty gas is optically thick also at infrared (IR) frequencies ($\tau_{\rm IR} > 1$), it is possible for the radiation force to exceed $L_{\rm AGN}/c$, though at most\footnote{An exception occurs at extremely high $\tau_{\rm IR}$ if the system is in the `dynamic diffusion' regime \citep[e.g.][]{Krumholz:07, Costa:17}, i.e. when the speed of the flow $v \gg c/\tau_{\rm IR}$. This regime, however, does not apply in the systems we are interested in in this paper, for which $\tau_{\rm IR} < 40$.} by a factor of $\tau_{\rm IR}$.
In this regime, IR photons become trapped within the optically thick gas and must scatter multiple times before escaping \citep[see e.g. Appendix in][]{Costa:17}.

In order for the radiation force to approach $\tau_{\rm IR} L_{\rm AGN}/c$, it is crucial for the radiation to be efficiently confined by optically thick gas.
\citet{Krumholz:09}, for instance, argue that radiation leakage through holes punched through outflowing gas should reduce the IR radiation force to a value comparable to $L_{\rm AGN}/c$.
Using radiation-hydrodynamic simulations, \citet{Krumholz:12, Krumholz:13} find that even when interacting with a homogeneous medium, trapped IR radiation triggers a Rayleigh-Taylor-like instability and is beamed through low-density channels, reducing the radiation force to $\mathcal{F}_{\rm rad} \, \approx \,  L_{\rm AGN}/c$ \citep[see also][]{Rosdahl:15}.
More accurate radiative transfer schemes (e.g. variable Eddington tensor or Monte Carlo methods), however, result in higher (though still reduced) trapping efficiencies \citep{Davis:14, Tsang:15, Zhang:16}.
More recently, \citet{Bieri:17} performed radiation-hydrodynamic simulations to test the efficiency of IR radiation trapping in clumpy galactic discs, showing that, although IR radiation is relatively ineffectively trapped (with efficiency $\eta_{\rm IR} \, \approx \, 0.2 \-- 0.3$), it continues to play a decisive role in accelerating outflows.

\citet{Costa:17} employed the radiation-hydrodynamic code {\sc Ramses-RT} in order to test its ability in reproducing analytic solutions of gas shells driven by radiation pressure out of galactic potentials, as envisaged by \citet{Thompson:15} and \citet{Ishibashi:15}.  
The agreement between numerical simulations and analytic models was found to be excellent for a wide range of AGN luminosities and IR optical depths. 
In particular, IR trapping was seen to be efficient as long as the optically thick gas has a high covering fraction and the IR diffusion times are short in comparison to the hydrodynamic response time.

The feasibility of this scenario has, however, never been tested in cosmological simulations performed with on-the-fly radiative transfer.
The main goal of this paper is to assess whether, from an energetics point of view, radiation pressure on dust can drive large-scale outflows in a cosmological context.
In addition, we investigate the extent to which star formation is inhibited in the presence of single- and multi-scattering radiation pressure from a quasar.
We address this problem for the first time with fully cosmological radiation-hydrodynamic simulations of radiative feedback from a quasar in a massive $z \gtrsim 6$ galaxy, a regime which we have explored with different models for AGN feedback in previous studies \citep[e.g.][]{Costa:14, Costa:15}. 

The paper is organised as follows.
We describe our simulations in Section~\ref{sec_radpresscosmo}. In Section~\ref{sec_results}, we present the results of our radiation-hydrodynamic simulations focussing on the impact of IR multi-scattering in generating large-scale outflows. We discuss the implications of our findings in Section~\ref{sec_discussion} and summarise our conclusions in Section~\ref{sec_conclusions}. We present complementary material in a series of Appendices. In Appendix~\ref{sec_lightcurve}, we present the quasar light-curve used in one of our simulations, in appendices ~\ref{sec_redc},~\ref{sec_convergence} and ~\ref{sec_jeansp}, we test the robustness of our results against changes in the reduced speed of light parameter, numerical resolution and artificial pressure floor, respectively. Finally, in Appendix~\ref{sec_irtrapeff} we quantify the trapping efficiency of IR radiation.

\section{The simulations}
\label{sec_radpresscosmo}

We perform cosmological simulations with the radiation-hydrodynamic code {\sc Ramses-RT} \citep{Rosdahl:13, Rosdahl:15}, the public multi-frequency radiative transfer extension of the adaptive mesh refinement code {\sc Ramses} \citep{Teyssier:02}.
{\sc Ramses} follows the hydrodynamical evolution of gas coupled with the dynamics of dark matter, stellar populations and black holes, under the influence of gravity and radiative cooling.
The Euler hydrodynamic equations are solved using a second-order unsplit Godunov scheme.
The `Local Lax-Friedrichs' (LLF) Riemann solver together with the MinMod Total Variation Diminishing scheme to reconstruct the interpolated variables from their cell-centred values is used to compute fluxes at cell interfaces.
The evolution of the collisionless component (dark matter, stellar populations and black holes) is computed using a particle-mesh method and cloud-in-cell interpolation.

{\sc Ramses-RT} follows the multi-group propagation of radiation, its on-the-fly interaction with hydrogen and helium through photoionisation, heating and momentum transfer (all neglected in this study) as well as its interaction with dust particles embedded within interstellar gas via single- and multi-scattering radiation pressure.
In order to solve radiative transfer, {\sc Ramses-RT} uses a first-order Godunov method with the M1 closure for the Eddington tensor.
We employ the `Global Lax-Friedrichs' (GLF) Riemann solver for the advection of photons between cells.
Since the time-step $\Delta t$, and hence the computational load, scales inversely with the speed of light, we adopt the `reduced speed of light approximation' \citep{Gnedin:01}.
We use a reduced speed of light of $\tilde{c} \, = \, 0.03c$, but also perform simulations at the full speed of light ($\tilde{c} \, = \, c$) in order to assess the convergence of our results. This analysis is presented in Appendix~\ref{sec_redc}.

\subsection{Cosmological initial conditions}

We investigate the ability of AGN radiation pressure to drive outflows in a massive galaxy at $z \,\gtrsim\, 6$.
We assume that the brightest quasars at $z \,=\, 6$ are hosted by dark matter haloes with virial masses of $\gtrsim 2 \times 10^{12} \, \rm M_\odot$ \citep[see][for a discussion on the likely mass of $z \,=\,6$ quasar host haloes]{Costa:14a}.
At $z \,=\,6$, the halo mass function turns over at a mass $\approx 10^{11} \, \rm M_\odot$. With a comoving number density of $\lesssim 10^{-7} \, h^3 \rm \, Mpc^{-3}$, the required haloes become exceedingly rare.
Our simulations have to follow a very large cosmological box in order to include such massive haloes as well as to accurately model the tidal torque field generated by the surrounding large-scale structure.

We generate cosmological `zoom-in' initial conditions with the {\sc MUSIC} code\footnote{The public version we used can be found on the website https://bitbucket.org/ohahn/music.} \citep{Hahn:11} for a cosmological box with a comoving side length $500 \, h^{-1} \, \rm Mpc$.
We adopt the following cosmological parameters: $\Omega_{\rm m} \,=\, 0.3065$, $\Omega_{\rm \Lambda} \,=\, 0.6935$,  $\Omega_{\rm b} \,=\, 0.0483$, $H_{\rm 0} \,=\, 67.9 \, \rm km \, s^{-1} \, Mpc^{-1}$, $\sigma_{\rm 8}\,=\,0.8154$ and $n_{\rm s} \, = \, 0.9681$ in line with the most recent cosmic microwave background constraints \citep{Planck:16}.
Length scales are given in physical units unless stated otherwise.

We first perform a relatively low resolution, purely N-body, simulation on a uniform $1024^3$ grid down to $z\,=\,6$ using {\sc Ramses}.
The low resolution dark matter particles have a mass $m_{\rm DM, 0} \, = \, 1.5 \times 10^{10} \, \rm M_\odot$.
Using the halo finder {\sc AdaptaHop} \citep{Tweed:09}, we identify the location of the most massive dark matter haloes ($M_{\rm vir} \, > \, 10^{12} \, \rm M_\odot$) found in the simulation at $z \,=\,6$. We select the second most massive system, as the most massive halo is found to be undergoing a major merger at $z \,=\, 6$.

We then generate `zoom-in' initial conditions for a roughly spherical region of comoving diameter $\approx 7 \rm \, Mpc$ centred on the selected dark matter halo. 
The radius of our high resolution region is therefore about $5$ times greater than the virial radius ($R_{\rm vir} \, = \, 62.1 \, \rm kpc$) of the targeted halo at $z \,=\, 6$.
Within the high resolution volume, dark matter particles have a mass of $m_{\rm DM, hr} \, = \, 3.5 \, \times \, 10^6 \, \rm M_\odot$.
We verify that the halo of interest is completely free of contamination from low resolution dark matter particles; the nearest low resolution dark matter particle is at a distance of $\approx 8 R_{\rm vir}$ from the central halo at $z \, = \, 6.5$ and $\approx 6 R_{\rm vir}$ at $z \, = \, 6$.
We have listed the main properties of the massive halo in Table~\ref{table1} for various different redshifts, including its virial mass, which, at $z \,=\, 6$, is $M_{\rm vir} \, = \, 2.4 \times 10^{12} \, \rm M_\odot$.  
A visual impression of the cosmological density field at $z \, = \, 6$ in the high resolution region of one of our simulations is provided in Fig.~\ref{fig_cosmofield}.

The grid is dynamically refined during the course of the simulations based on mass density. A cell is refined if $(\rho_{\rm DM} + \frac{\Omega_{\rm DM}}{\Omega_{\rm b}} \rho_{\rm *} + \frac{\Omega_{\rm DM}}{\Omega_{\rm b}} \rho) \Delta x^3 > 8 m_{\rm DM, hr}$, where $\rho_{\rm DM}$, $\rho_{\rm *}$ and $\rho$ are the dark matter, stellar and gas density, respectively.
The minimum cell width achieved in most of our simulations is $\Delta x \,=\, 125 h^{-1} \, \rm pc$. 
At the time at which radiation injection begins, there are about 130,000 such {\sc Ramses} cells within the virial radius, which constitutes $\approx 15 \%$ of the total number of cells (of all sizes) in that region.
We also perform one high resolution simulation (simulation UVIR-4e47-hires in Table~\ref{table_sims}), in which we refine by one additional level.
In this high resolution simulation, the minimum {\sc Ramses} cell width is $\Delta x \,=\, 62.5 h^{-1} \, \rm pc$. The number of high resolution cells within the virial radius of the massive halo is also on the order of 130,000.

\begin{table}
\centering
\begin{tabular}{c*{4}{c}r}
\hline
$z$ & $M_{\rm vir} \, \rm [M_\odot]$ & $R_{\rm vir} \, \rm [kpc]$ & $M_{\rm *} \, \rm [M_\odot]$ & $V_{\rm *}^{\rm max} \, \rm [km \, s^{-1}]$ \\
\hline
$8$ & $3.2 \times 10^{11}$  & $24.5$ & $6.8 \times 10^9$ & $80.7$\\
$7$ & $1.4 \times 10^{12}$  & $44.1$ & $4.1 \times 10^{10}$ & $184.1$\\
$6$ & $2.4 \times 10^{12}$  & $62.2$ & $1.4 \times 10^{11}$ & $585.2$\\
\hline
\end{tabular}
\caption{Main properties of the massive dark matter halo addressed in our various simulations, here shown for simulation noAGN. From left to right, we list the redshift, virial mass, virial radius (in physical units), stellar mass and maximum stellar circular velocity. The halo grows rapidly between $z\,=\,8$ and $z\,=\,6$, increasing by an order of magnitude in mass and a factor of 3 in size. The stellar bulge of the most massive galaxy forms particularly rapidly over this redshift interval and becomes remarkably tightly bound by $z \,=\, 6$ with a peak circular velocity just under $600 \, \rm km \, s^{-1}$.}
\label{table1}
\end{table}

\subsection{Radiation-hydrodynamic simulations}

We use a deliberately simple setup for our numerical experiments.
Our aim here is to isolate the effects of AGN radiation pressure on dust as narrowly as possible, focussing on the conditions under which large-scale outflows can be generated and on the properties of the large-scale outflows that develop.
We do not calibrate the multiple free parameters present in our models or attempt to match observables, for which a range of many additional physical mechanisms (e.g. metal-line and non-equilibrium cooling, supernova-driven outflows) would have to be taken into account.
We do, however, verify how the bulk properties of simulated outflows compare with observations, when possible.  
We describe the physical processes that are included in the following sections.

\subsubsection{Radiative cooling and star formation}

Non-equilibrium hydrogen and helium cooling down to a temperature $T \,=\, 10^4 \, \rm K$ is followed using tabulated cooling rates \citep[as detailed in][]{Rosdahl:13}.
We do not consider feedback from supernovae, since this would itself generate galactic outflows, or interact non-linearly with outflows driven by AGN \citep[e.g.][]{Booth:13, Costa:15}, and thus complicate our task of isolating outflows driven by AGN radiation pressure.

Since we do not follow supernova feedback, we also \emph{do not} include metal-line cooling in our fiducial simulations, as there would be no way of distributing the metals throughout the galactic halo.
While arguably unimportant for average galaxies at $z \,=\,6$, metal-line cooling may already affect the metal-enriched environments of very massive galaxies hosting $z \,>\, 6$ quasars as well as the cooling properties of AGN-driven outflows \citep{Costa:15}. 

We therefore perform a purely hydrodynamic simulation with supernova feedback in order to verify that the gas column density distribution around the AGN in our simulations is not unrealistically high due to the absence of additional feedback mechanisms (see Section~\ref{sec_supernova}).

Star formation is enabled for gas with number density $n_{\rm H} > n_{\rm *} \, = \, 0.13 \, \rm cm^{-3}$ and temperature $T < 2 \times 10^4 \, \rm K$ at a rate
\begin{equation}
\frac{d \rho_{\rm *}}{dt} \, = \, \frac{\rho_{\rm gas}}{t_{\rm *}} \, ,
\label{eq_sfr}
\end{equation} 
where $t_{\rm *} \, = \, t_{\rm 0} \left( n_{\rm H}/n_{\rm *} \right)^{-1/2}$ and $t_{\rm 0} \, = \, 8 \, \rm Gyr$.
Given our star formation density threshold, our choice of $t_{\rm 0}$ corresponds to a star formation efficiency of $\epsilon_{\rm *} \approx 0.02$ per free-fall time.

The number $N_{\rm *}$ of `stellar particles' generated in star forming gas is drawn from a Poisson distribution, i.e. they form with a probability of $P(N_{\rm *}) \, = \, (\lambda^{N_{\rm *}} / N_{\rm *}!) e^{-\lambda}$, where $\lambda \, = \, \epsilon_{\rm *} \left( \rho \Delta x^3 / m_{\rm *, min} \right) \left(\Delta t / t_{\rm ff} \right)$ \citep[see][for details]{Rasera:06}, $\rho$ is the gas density and $\Delta t$ is the time-step. The minimum stellar particle mass is given by $m_{\rm *, min} \, = \, n_{\rm *} \Delta x^3 \, \approx \, 2.3 \times 10^4 \, \rm M_\odot$.

Stellar mass loss, the associated metal loss and enrichment and stellar radiation processes are neglected, but the effect of the latter will be investigated in a future study.

\subsubsection{Polytropic equation of state}
\label{sec_jeanspressure}

Like in many cosmological simulations of galaxy formation, we adopt a polytropic equation of state for dense ISM gas.
The motivation for such an artificial source of pressure support is twofold: (i) the numerical resolution is not sufficient to capture physical processes taking place at ISM scales, including turbulence, thermal conduction and cloud evaporation \citep{McKee:77, Springel:03}, which may generate additional pressure support and (ii) in order to avoid artificial fragmentation in cells in which the Jeans length is not resolved \citep{Truelove:97}.

We recall the definition for the Jeans length 
\begin{equation}
\lambda_{\rm J} \, = \, \left( \frac{\pi c_{\rm s}}{G \rho} \right)^{1/2}\, = \, 16 \left( \frac{T}{1 \mathrm{K}} \right)^{1/2} \left( \frac{n_{\rm H}}{1 \mathrm{cm^{-3}}} \right)^{-1/2} \, \rm pc \, ,
\label{eq_jeanslength}
\end{equation}
where $c_{\rm s} \,=\, (k_{\rm B} T / \gamma)^{1/2}$ is the speed of sound and $\gamma \,=\, 5/3$ is the adiabatic index of a monatomic ideal gas.
Requiring $\lambda_{\rm J}$ to be resolved by $N_{\rm J}$ cells gives a condition for the minimum temperature $T_{\rm J}$ that is necessary to resolve the Jeans length at all times. This is
\begin{equation}
T_{\rm J} \, = \, \left( \frac{n_{\rm H}}{1 \mathrm{cm^{-3}}} \right) \left( \frac{N_{\rm J} \Delta x}{16 \mathrm{pc}} \right)^2 \, \rm K \, ,
\label{eq_tempfloor}
\end{equation}
or, alternatively,
\begin{equation}
T_{\rm J} \, = \, T_{\rm 0}  \left(\frac{n_{\rm H}}{n_{\rm *}} \right) \, ,
\label{eq_efftemp}
\end{equation}
where $T_{\rm 0} \, = \, n_{\rm *} / (1 \mathrm{cm^{-3}})  \left( N_{\rm J} \Delta x / 16 \, \mathrm{pc} \right)^2 \, \rm K $.
In agreement with previous cosmological simulations at comparable resolution and star formation threshold \citep[e.g.][]{Dubois:12a}, we set $T_{\rm 0} \, = \, 10^4 \, \rm K$, which corresponds to $N_{\rm J} \, = \, 24$.
We also perform various simulations for which $T_{\rm 0} \, = \, 10^3 \, \rm K$, i.e. $N_{\rm J} \, = \, 8$ in line with other state-of-the-art cosmological simulations \citep[e.g.][]{Katz:15}.
We find that all main results presented in this paper are \emph{strengthened} by adopting a weaker polytrope with $T_{\rm 0} \, = \, 10^3 \, \rm K$. As a conservative choice, we thus adopt $T_{\rm 0} \, = \, 10^4 \, \rm K$ as our default model and discuss the effect of varying $T_{\rm 0}$ in Appendix~\ref{sec_jeansp}.

We apply an artificial pressure term in terms of the effective temperature function given in Eq.~\ref{eq_efftemp} in all our simulations.
When we quote temperatures in this study, we always refer to the temperature (ignoring the $T_{\rm J}$ term) derived from the thermal pressure, unless stated otherwise.

\subsubsection{Radiation from a bright quasar}
\label{sec_radfromqso}

Following existing analytic models for AGN radiation pressure on dust \citep{Murray:05, Thompson:15, Ishibashi:15, Costa:17}, we consider two radiation frequency bins (optical/UV and IR).
This choice allows us to draw closer comparison with existing literature, for which only one radiation channel is single-scattering and the other is multi-scattering radiation.
In all simulations, AGN radiation is injected in the optical/UV band, but is allowed to be reprocessed into IR.

Radiation injection is performed within a spherical region of radius $\Delta x$ centred on a single massive particle and at every fine time-step.
We employ the {\sc Ramses} clump finder \citep{Bleuler:14, Bleuler:15} to identify gas density peaks considering all {\sc Ramses} cells with a density $80$ times higher than the current background mean cosmic density.
If the saddle point between different density peaks lies at a value higher than $200$ times the background mean gas density, the two peaks are merged and identified as a single clump.
The clump finder is called once every \emph{coarse} time-step\footnote{In the subcycling scheme used in our simulations, two level $l+1$ timesteps are performed for each coarser level $l$ step.} in order to find potential candidate peaks for the formation of a black hole sink particle.
When a clump of mass $3 \times 10^{10} \, \rm M_\odot$ has formed, we create one sink particle with mass $3 \times 10^9 \rm M_\odot$ which is placed at the location of the potential minimum within its parent clump.
In our simulations, the black hole particle is seeded at $z \, = \, 6.8$, but AGN feedback is only allowed to start at $z \, = \, 6.5$.
Note that a high black hole mass is chosen to minimise BH particle wandering and to ensure it remains close to the potential minimum of the halo at all times \citep[see][for a recent physically motivated method to prevent BH scattering]{Biernacki:17}. We also note that, at the scales which are resolved in our simulations, the black hole's gravitational potential is much weaker than those of the stellar and gaseous bulge (see Section~\ref{sec_beforeswitch}).
Only one black hole particle is allowed to form.

The AGN is assumed to radiate at a constant luminosity or to follow a variable light-curve.
For the latter, we consider an irregular light-curve as output by recent cosmological simulations \citep{Costa:15}. 
The AGN light-curve employed in our simulations is shown in Fig.~\ref{fig_lightcurve} in Appendix~\ref{sec_lightcurve}. 
We probe different quasar luminosities in our various simulations, ranging from $2 \times 10^{47} \, \rm erg \, s^{-1}$ to $\gtrsim 10^{48} \rm \, erg \, s^{-1}$.

The luminosities we quote should be understood as the optical/UV luminosity of the quasar, i.e. the portion of the quasar emission that is absorbed by dust grains.
In order to obtain bolometric luminosities from the optical/UV contribution, we must assume a quasar spectrum.
We take the characteristic quasar spectrum of \citet{Sazonov:04} and integrate it between energies $1 \rm \, eV$ and $200 \rm \, eV$, a range which encompasses optical and UV wavelengths and roughly accounts for $56 \%$ of the bolometric luminosity, i.e. $L_{\rm bol} \, \approx \, L_{\rm opt,UV} / 0.56$.

A significant fraction of quasar energy is liberated in the IR band. According to the \citet{Sazonov:04} spectrum, this component amounts to approximately $31 \%$ of the bolometric emission.
However, the origin of this component, which is commonly thought to stem from emission from radiatively heated dust, is complex.
It is often assumed to originate from a pc-scale dusty torus surrounding the AGN (which would not be resolved in our simulations), but could also be generated due to dust absorption at scales we do resolve \citep{Sanders:89}.
Thus, in order to obtain a lower limit on the bolometric luminosity, we also consider a second, modified, spectrum in which the IR contribution is added into that of the optical/UV. 
In this extremal scenario, the bolometric luminosity of the AGN is simply $L_{\rm bol}^\prime \, \approx \, L_{\rm opt,UV} / 0.87$.
When quoting bolometric luminosities, we add an error bar encompassing the range between both limiting cases.

We name our simulations based on the frequency bins included and the adopted quasar luminosity, e.g. `UVIR-4e47' for a simulation in which radiation is injected at a luminosity $4 \times 10^{47} \rm \, erg \, s^{-1}$ in the optical/UV band, but allowed to be reprocessed to the IR.
In one of our simulations, named `noAGN', the opacities are all set to zero, such that the radiation injected by the quasar does not couple to the gas in any way.
We list the main simulations, together with the assumed quasar luminosity in Table~\ref{table_sims}.

\begin{table}
\centering
\begin{tabular}{c*{5}{c}r}
\hline
Name & $L_{\rm opt,UV}$ & $L_{\rm bol}$ & $L_{\rm bol}^\prime$ &IR? \\
 & $\rm [erg \, s^{-1}]$ & $\rm [erg \, s^{-1}]$ & $\rm [erg \, s^{-1}]$ & \\
\hline
UV-4e47       & $4 \times 10^{47}$ & $7.1 \times 10^{47}$ & $4.6 \times 10^{47}$ &  & \\
UV-5e47       & $5 \times 10^{47}$ & $8.9 \times 10^{47}$ & $5.7 \times 10^{47}$ &  & \\
UV-8e47       & $8 \times 10^{47}$ & $1.4 \times 10^{48}$ & $9.2 \times 10^{47}$ &  & \\
UV-1e48       & $1 \times 10^{48}$ & $1.8 \times 10^{48}$ & $1.1 \times 10^{48}$ &  & \\
UV-1.1e48     & $1.1 \times 10^{48}$  & $2 \times 10^{48}$ & $1.3 \times 10^{48}$ &  & \\
UV-1.2e48     & $1.2 \times 10^{48}$  & $2.1 \times 10^{48}$ & $1.4 \times 10^{48}$ &  & \\
UVIR-1e47   & $1 \times 10^{47}$       & $1.8 \times 10^{47}$ & $1.1 \times 10^{47}$ & \checkmark \\
UVIR-2e47    & $2 \times 10^{47}$      & $3.6 \times 10^{47}$ & $2.3 \times 10^{47}$ & \checkmark \\
UVIR-3e47    & $3 \times 10^{47}$      & $5.4 \times 10^{47}$ & $3.4 \times 10^{47}$ & \checkmark \\
UVIR-4e47    & $4 \times 10^{47}$      & $7.1 \times 10^{47}$ & $4.6 \times 10^{47}$ & \checkmark \\
UVIR-5e47    & $5 \times 10^{47}$      & $8.9 \times 10^{47}$ & $5.7 \times 10^{47}$ & \checkmark \\
UVIR-4e47-hires    & $4 \times 10^{47}$  & $7.1 \times 10^{47}$ & $4.6 \times 10^{47}$ & \checkmark \\
UVIR-4e47-dust   & $4 \times 10^{47}$    & $7.1 \times 10^{47}$ & $4.6 \times 10^{47}$ & \checkmark \\
UVIR-4e47-duty   & $4 \times 10^{47}$    & $7.1 \times 10^{47}$ & $4.6 \times 10^{47}$ & \checkmark \\
UVIR-4e47-dustduty   & $4 \times 10^{47}$    & $7.1 \times 10^{47}$ & $4.6 \times 10^{47}$ & \checkmark \\
thermal-4e47   & $4 \times 10^{47}$  & $7.1 \times 10^{47}$ & $4.6 \times 10^{47}$  & &   \\
\hline
\end{tabular}
\caption{The simulations performed and analysed in this study. We probe a wide range of quasar optical/UV luminosities (from $2 \times 10^{47} \, \rm erg \, s^{-1}$ to $1.2 \times 10^{48} \, \rm erg \, s^{-1}$) and explore simulations in which we include only single-scattering radiation pressure (simulations `UV'), while in others we investigate the impact of multi-scattering (IR) radiation (simulations `UVIR'). We also consider a higher resolution simulation (UVIR-4e47-hires), a simulation in which we do not allow radiation to couple to gas hotter than $3 \times 10^4 \, \rm K$ (UVIR-4e47-dust), a simulation in which the AGN light-curve oscillates (UVIR-4e47-duty), a simulation in which both a temperature cut-off for the dust opacity and a variable AGN light-curve is adopted (UVIR-4e47-dustduty) and a simulation in which AGN feedback consists of thermal energy injection instead of radiation (thermal-4e47).}
\label{table_sims}
\end{table}

In this study, we focus on the effects of radiation pressure on dust as an AGN feedback mechanism and do not consider the effects of photoionisation.
We accordingly set all ionisation cross-sections to zero.
We select fixed dust opacities of $\kappa_{\rm UV} \, = \, 1000 \, \rm cm^2 \, g^{-1}$ in the optical/UV and $\kappa_{\rm IR} \, = \, 10 \, \rm cm^2 \, g^{-1}$ in the IR regardless of gas temperature.
These opacities are in line with a dust-to-gas ratio of $f_{\rm D} \, = \, 0.01$ and a metallicity of $Z \, = \, Z_\odot$. 
These are optimistic assumptions, but allow us to link our results to those based on analytic models \citep[e.g.][]{Thompson:15} and cosmological simulations \citep[e.g.][]{Debuhr:11}, where similar assumptions apply, and can be used to place an upper bound on the efficiency of radiation pressure on dust as an AGN feedback mechanism.
We also note that, since there is only one source of radiation in our simulations (the bright quasar) and only the central regions of the massive halo are optically thick in the IR, most of the effect we see is confined to the innermost few kpc of the targeted galaxy.

In one of our simulations (UVIR-4e47-dust), however, we have $\kappa_{\rm UV} \, = \, \kappa_{\rm IR} \, = \, 0 \, \rm cm^2 \, g^{-1}$ for gas with temperature $T \geq 3 \times 10^4 \, \rm K$ in order to mimic a scenario in which dust is destroyed in hot gas. As we show in later sections, introducing this temperature cut-off in the dust opacity mainly limits the development of weak outflows in the simulations at late times, after the obscuring layers initially surrounding the quasar have been ejected. At early times, however, a powerful outflow still develops in this simulation.
We explored also including both a temperature cut-off for the dust opacity and a variable AGN light-curve in one simulation (UVIR-4e47-dustduty).

We finally note that we adopt the `reduced flux approximation' described in the Appendix of \citet{Rosdahl:15b} in order to calculate the direct radiation force of optical/UV photons.
This approximation, which is exact given that we only consider radiation from one source, prevents the radiative force from being artificially suppressed close to the quasar.

\subsubsection{Thermal AGN feedback}

In one of our simulations, we consider also thermal feedback in order to compare its efficiency to that of radiation pressure in driving large-scale outflows.
Thermal energy is injected \emph{continuously}, at every fine time-step, within a spherical region of radius $\Delta x$ at a rate
\begin{equation}
\dot{E}_{\rm th} \, =\, 0.05 \times L_{\rm opt, UV} \, ,
\end{equation}

or, in terms of $L_{\rm bol}$, energy is injected at a rate $\dot{E}_{\rm th} \, =\, 0.028 \times L_{\rm bol}$.
We thus adopt a feedback efficiency of $\approx 3 \%$ in line with that selected in many cosmological simulations that include AGN feedback, which, typically, lies in the range $(1\% \-- 15\%) L_{\rm bol}$ \citep[e.g.][]{DiMatteo:05, Sijacki:07, Booth:09, Biernacki:17}.

Many implementations of AGN thermal feedback cap the temperature to which gas can be heated to $10^9 \, \rm K$ or $5 \times 10^9 \, \rm K$. This procedure ensures that the hydrodynamic time-step does not become too short and prevents gas from entering the relativistic hydrodynamic regime, which is not followed in most simulations.
We, however, find that limiting the temperature of gas heated by AGN feedback to $10^9 \, \rm K$ prevents most of the AGN feedback energy from being injected into the gas. In addition, we find that the gas temperature reaches $10^9 \, \rm K$ in the central regions, while the gas density is reduced as an outflow develops. Since the temperature is not allowed to rise further, while the gas density decreases, the pressure of the central region drops with time unphysically, stalling the outflow.
We thus raise the maximum temperature to which gas can be heated to $10^{10} \, \rm K$ in our simulations with thermal feedback.
In practice, however, the gas mass that gets heated beyond $10^9 \, \rm M_\odot$ in our simulations is very small and on the order of $\sim 10^5 \, \rm M_\odot$, i.e. less than $0.1 \%$ of the total mass in the injection region.
The simulation performed with thermal energy injection, for which $L_{\rm opt,UV} \, = \, 4 \ \times 10^{47} \, \rm erg \, s^{-1}$, is referred to as `thermal-4e47'.
 
\begin{figure*}
\centering 
\includegraphics[scale = 0.5]{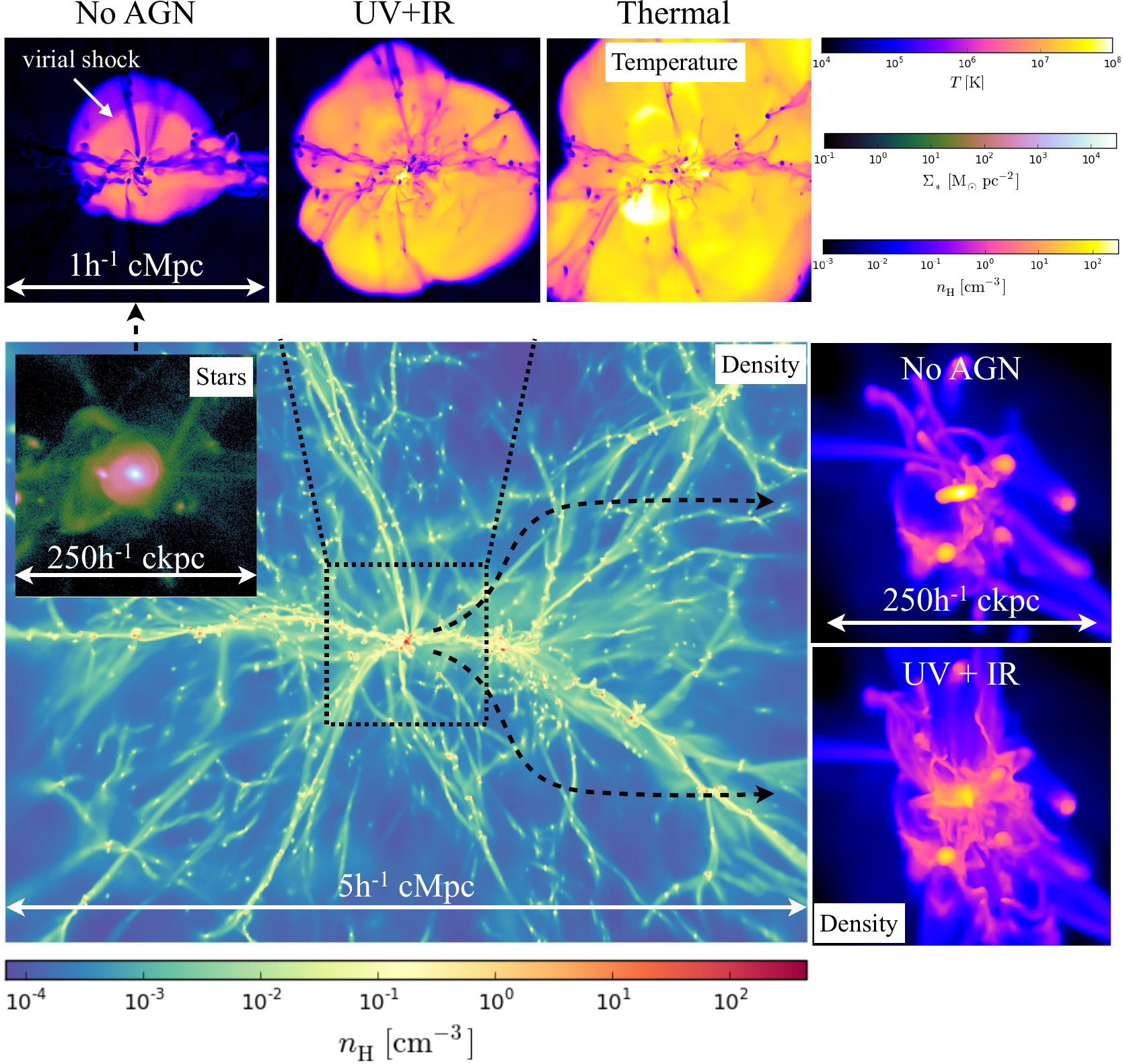}
\caption{The large panel on the bottom left shows the gas density field projected along a slab with thickness $5 h^{-1} \, \rm cMpc$, where $\rm cMpc$ denotes `co-moving Mpc', at $z \, = \, 6$ in simulation noAGN. The quasar host galaxy lies at the intersection of a complex network of gas filaments, which feed the central galaxy at a high rate. At the top, the gas temperature distribution of the central halo is shown at $z \, = \, 6$ for simulations noAGN (left), UVIR-5e47 (middle) and thermal-4e47 (right). The inclusion of AGN feedback drives powerful outflows that transport gas to the outer regions of the halo. On the right, we show density projections for two of our simulations at $z \, = \, 6.3$ (when dense outflows are most prominent); for a simulation without AGN feedback (top panel) and for a simulation with radiative (UV and IR) feedback (UVIR-5e47). Radiation pressure here drives a powerful outflow and disrupts the central galaxy.}
\label{fig_cosmofield}
\end{figure*}
 
\section{IR-driven large-scale outflows}
\label{sec_results}

We now present the main results of our simulations, starting with the evolution of the galaxy down to $z \, = \ 6.5$, the point at which radiation injection begins.

\subsection{Before the quasar is `switched on'}
\label{sec_beforeswitch}

At $z\,=\,6.5$, the targeted galaxy consists of an extended (thick) gaseous disc with diameter $\approx 5 \, \rm kpc$ and a highly concentrated component we shall refer to as a `compact bulge' \citep[see also][]{Dubois:12b}.
Its circular velocity $v_{\rm circ} \,=\, \sqrt{G M(<R) / R}$ peaks at $\approx 440 \, \rm km \, s^{-1}$ for both the stellar and gas components individually, at radial distances of $\approx 600 \, \rm pc$ and $2 \, \rm kpc$, respectively. The dark matter component flattens out at $\approx 430 \, \rm km \, s^{-1}$, though at the much larger radius of $32 \, \rm kpc$.
We find that the total circular velocity peaks at values exceeding $640 \, \rm km \, s^{-1}$ at a radius of $1.5 \, \rm kpc$ \citep[see][for circular velocity profiles for the different components in a similar cosmic environment]{Costa:15}.

Due to the absence of strong stellar feedback, it is likely that this compact bulge is unrealistically tightly bound and over-massive.
The inclusion of supernova and AGN feedback, however, appears not to be able to reduce the total circular velocity much below $\approx 500 \, \rm km \, s^{-1}$ \citep{Costa:15}.
In Section~\ref{sec_supernova}, we present test simulations using strong supernova feedback, finding a moderate \emph{increase} in the peak (total) circular velocity up to $\approx 670 \, \rm km \, s^{-1}$. 
While efficient at suppressing star formation in the low mass progenitors of the central galaxy, we find that the left-over gas is unable to escape the massive halo, where it cools producing an even more compact stellar bulge.
It is worth pointing out that recent observations do suggest that galaxies hosting bright quasars at $z \gtrsim 6$ are, indeed, very compact.
Interferometric observations of emission lines from the cold ISM indicate that $z \gtrsim 5$ quasars are hosted by systems with typical sizes $\approx 2 \-- 5 \, \rm kpc$ \citep[e.g.][]{Wang:13, Wang:16, Trakhtenbrot:17}.   
A recent ALMA measurement of [CII] emission \citep{Venemans:17} from the $z \, =\, 7.1$ quasar J1120 \citep{Mortlock:11} reveals $\approx 80\%$ of the total line flux to be associated with the innermost kpc of the host galaxy.

The formation of such a tightly bound bulge (see inset in Fig.~\ref{fig_cosmofield} for a stellar surface map showing the disc and its central bulge) by $z \,=\,6.5$ appears to be linked to the cosmological environment in which the massive halo forms.
As is illustrated in Fig.~\ref{fig_cosmofield}, the central galaxy lies at the intersection of a large network of filaments. 
These gas streams, which are composed of diffuse cool ($T \approx 10^4 \, \rm K$) gas as well as a large number of satellite galaxies, transport large quantities of gas to the central galaxy.
As they approach the potential minimum, the gas streams accelerate to high speeds $\gtrsim 500 \, \rm km \, s^{-1}$, feeding the central galaxy from different directions.
The environment of the massive galaxy is therefore favourable to angular momentum cancellation; the angular momentum of gas flowing in along the various filaments adds up incoherently \citep{Dubois:12b} and the gas rapidly settles onto a gravitationally unstable disc, which can, in principle, feed the central black hole at a high rate \citep[see e.g.][]{Curtis:16, Prieto:17}.

There is also a more tenuous and spatially extended gas component that surrounds the central galactic disc and is associated with a hot atmosphere that has formed already by $z \,=\, 6.5$ (see top left panel of Fig.~\ref{fig_cosmofield}).
The star formation in the central galaxy remains high ($\approx 500 \, \rm M_\odot \, yr^{-1}$) despite the presence of this hot component.

\begin{figure}
\centering 
\includegraphics[scale = 0.42]{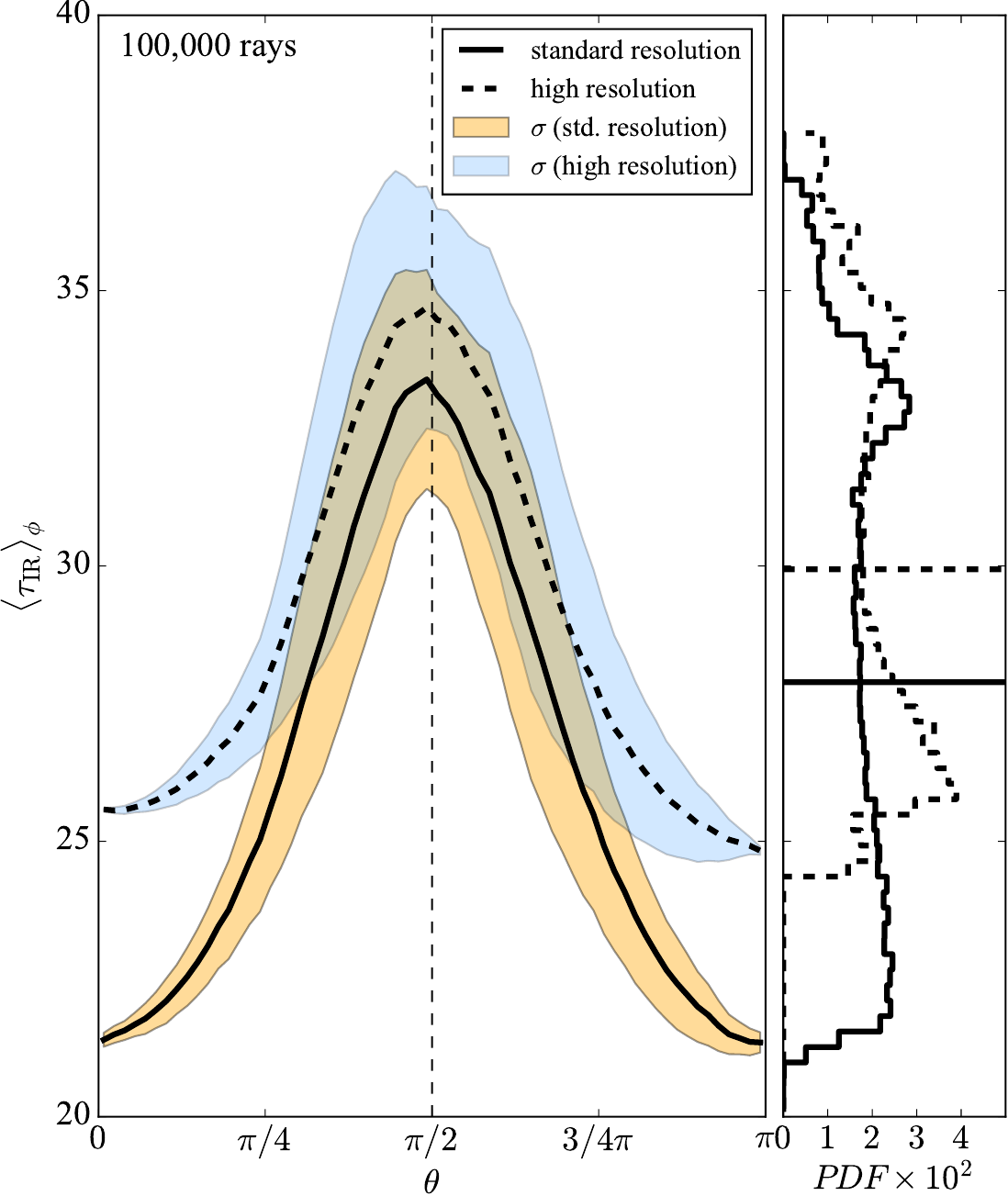}
\caption{IR optical depth distribution at $z \, = \, 6.5$ (at the time at which radiation begins to be injected) as a function of inclination angle $\theta$ (measured from the polar axis in the positive z-direction) obtained by integrating along $10^5$ rays cast in random directions (Michel-Dansac et al., in prep.). We show mean optical depths (black solid line) as well as the $1\sigma$ fluctuation amplitude (orange shades) for one of our standard resolution simulations. The dashed line and the blue shades show results for the high resolution simulation. The vertical dashed line marks the plane of the disc. On the right-hand plot, the histogram shows the optical depth probability distribution for the standard resolution (solid line) and high resolution (dashed line) simulations, with horizontal lines giving the median value in both cases. All lines-of-sight are optically thick in the IR. The optical depths are particularly high along the plane of the central disc, where they nearly reach $\tau_{\rm IR} \, =\, 40$, but lower by about a factor of $2$ along the galactic poles.}
\label{fig_tautheta}
\end{figure}

At $z \, = \, 6.5$, the galactic gas component is so compact that, when integrated from the galactic centre out to a few $\rm kpc$, the hydrogen column densities $N_{\rm H}$ can exceed $10^{24} \, \rm cm^{-2}$. 
We cast $10^5$ rays starting from the location of the black hole particle at $z \, = \, 6.5$ and integrate the gas density in the intercepted {\sc Ramses} cells along the rays out to the virial radius.
The resulting column densities are converted into IR optical depths by computing the product $\tau_{\rm IR} \, = \, \Sigma_{\rm gas} \kappa_{\rm IR}$, where $\Sigma_{\rm gas}$ is the total gas column density obtained by integrating along the various rays.
The solid line in Fig.~\ref{fig_tautheta} shows the mean IR optical depth as a function of inclination angle measured from the polar axis of the galactic disc for one of our standard resolution simulations\footnote{All standard resolution simulations are identical at $z \, = \, 6.5$.}.
We find IR optical depths in the range of $20 \-- 40$, with the highest optical depths occurring at inclination angles of $\theta \approx \pi/2$, i.e. along the disc plane, and the lowest at high inclination angles, i.e. along the poles.
We note that the optical depth distribution seen here converges after only a few kpc from the BH particle and is not sensitive to the upper integration radius as long as it lies beyond the central galaxy.

We ensure that the numerical resolution of our simulations is sufficient to yield a converged column density distribution.
The dashed line in Fig.~\ref{fig_tautheta}, which shows the optical depth distribution obtained in the high resolution simulation, indicates reasonable convergence.
The mean optical depth increases by $\approx 8 \%$, from $\langle \tau_{\rm IR} \rangle \approx 28$, at standard resolution, to $\langle \tau_{\rm IR} \rangle \approx 30$, at high resolution. Fig.~\ref{fig_tautheta} shows that, while the optical depth appears to increase by only a few \% along the disc plane, it increases by $\approx 19\%$ along the polar (rotation) axis.
The normalisation of the column density distribution increases because the galaxy becomes slightly more compact. 
Thus, IR radiation should be slightly more efficiently trapped with increasing numerical resolution. We present a thorough convergence study in Appendix~\ref{sec_convergence}. 

We also show the $1\sigma$ levels associated with fluctuations in the column density surrounding the quasar host galaxy in Fig.~\ref{fig_tautheta}, which are illustrated by the orange shade, for the standard resolution simulation, and by the blue shade, for the high resolution simulation.
We see fluctuations on the order of $15 \%$ around the mean value for both standard- and high resolution simulations.
With increasing resolution and the addition of a cold ISM phase, we should, however, expect the amplitude of these fluctuations to increase and the distribution of gas column densities to become broader, as the gas fragments and clumps into smaller interstellar clouds.  
Some lines-of-sight may then yield optical depths even higher than $\tau_{\rm IR} \, = \, 40$, though other lines-of-sight could be more transparent to IR radiation.

The initial column density distribution is favourable to IR radiation trapping, since optical depths are high in every direction.
At this point the quasar should be completely obscured at optical and UV wavelengths. The quasar is obscured by design in our simulations, since we do not allow any AGN feedback prior to $z \, =\, 6.5$. However, we interpret the initial conditions at $z \, = \, 6.5$ as the possible outcome of a wet merger or accretion episode, which could increase the central gas density in a realistic setting.
We now explore how the onset of quasar radiation shapes the evolution of its host galaxy and the extent to which IR is able to boost an outflow.

\subsection{Galactic-scale outflows}

\begin{figure*}
\centering 
\includegraphics[scale = 0.42]{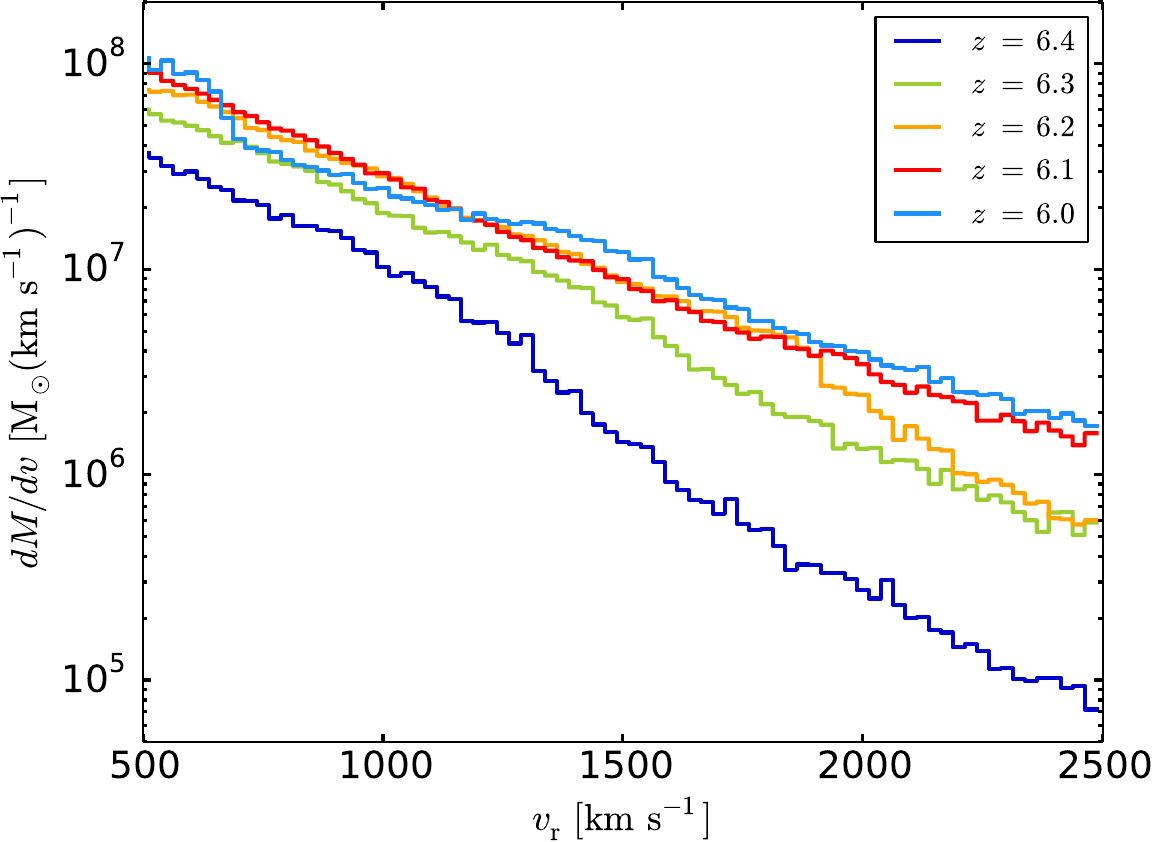}
\includegraphics[scale = 0.42]{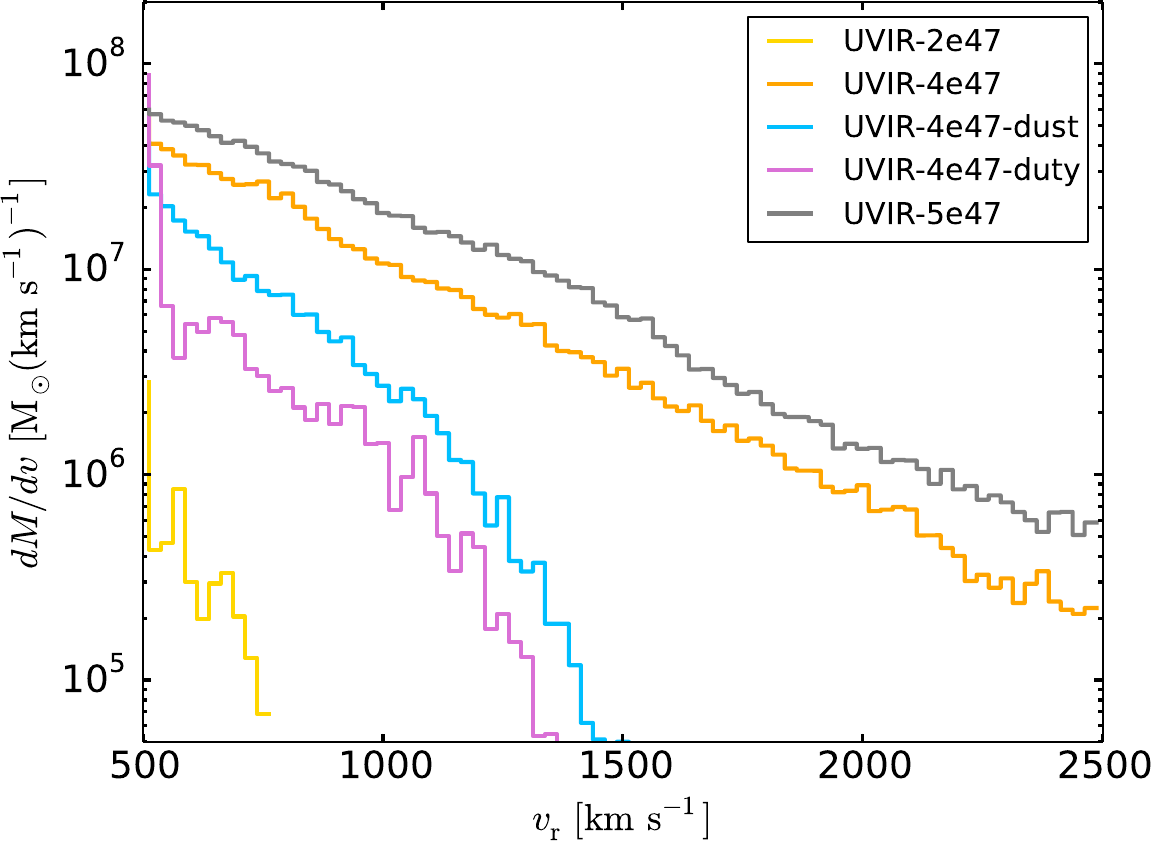}
\caption{\emph{Left:} Mass-weighted gas velocity distribution for gas located within the virial radius of the targeted galaxy in simulation UVIR-5e47 at various redshifts. Radiation pressure here launches outflows with maximum radial velocities exceeding $2,500 \, \rm km \, s^{-1}$. The outflowing mass rises steadily with time as the quasar continues to shine. \emph{Right:} Mass-weighted gas velocity distribution across a sub-set of our simulation sample. Both the outflow mass and the maximum outflow velocity increase with quasar luminosity. Disabling the coupling between radiation and hot gas with $T > 3 \times 10^4 \, \rm K$ (light blue line) or including the AGN lifetime (pink line) lead to lower outflow masses and maximum velocities.}
\label{fig_outflows}
\end{figure*}

Once it is `switched on', quasar radiation pressure on dust accelerates gas outwards.
Large-scale outflows are launched in all but one (UVIR-2e47) of our simulations including IR multi-scattering, but only in the simulations with the highest quasar optical/UV luminosities ($L_{\rm opt,UV} \gtrsim 1.2 \times 10^{48} \, \rm erg \, s^{-1}$) if only single-scattering radiation pressure is included.
We delve deeper into the role played by the quasar luminosity in driving large-scale outflows in Section~\ref{sec_critlum}.
In this section instead, we give a qualitative outline of the main properties of the outflows in simulations in which these exist. 

On the left-hand panel of Fig.~\ref{fig_outflows}, we show the mass-weighted radial velocity distribution for gas located within the virial radius at different times in simulation UVIR-5e47 as an illustration.
We see that substantial amounts of gas ($\gtrsim 10^{10} \, \rm M_\odot$) are accelerated to high velocities ($v_{\rm r} \, > \, 500 \, \rm km \, s^{-1}$).
Due to the continuous injection of radiation at a fixed luminosity, the outflow masses and the maximum outflow velocities steadily rise with time, exceeding $2,500 \, \rm km \, s^{-1}$ by $z \, = \, 6$.

On the right-hand panel of Fig.~\ref{fig_outflows}, we show how the velocity distribution varies across a subset of our simulations at a fixed redshift ($z \, = \, 6.3$). 
We see a clear trend of increasing outflow masses and maximum velocities with quasar luminosity.
The blue and pink lines, which correspond to simulations UVIR-4e47-dust and UVIR-4e47-duty\footnote{In UVIR-4e47-duty, an outflow is not sustained below $z \, = \, 6.2$.}, respectively, show that the outflow mass and maximum velocity can drop drastically with respect to UVIR-4e47, once radiation is not allowed to couple to hot gas and the AGN lifetime is taken into account.
Even in those simulations, however, we find outflow masses on the order of $\sim 5 \times 10^9 \, \rm M_\odot$.
A fast outflow with $v_{\rm max} \, \approx \, 1500 \, \rm km \, s^{-1}$, though carrying  a slightly lower mass of $\sim 10^9 \, \rm M_\odot$, is launched also in UVIR-4e47-dustduty.

The outflow masses we find in our simulations, which reach $\approx 4 \times 10^{10} \, \rm M_\odot$ in one of our most extreme models (UVIR-5e57), are somewhat lower than those found in cosmological simulations following similar haloes with thermal AGN feedback at the same redshift. Using a thermal feedback model, \citet{Costa:15} find outflow masses in the range $5 \-- 8 \times 10^{10} \, \rm M_\odot$, higher by more than an order of magnitude than seen in our more realistic models UVIR-4e47-dust or UVIR-4e47-duty ( $\sim 5 \times 10^9 \, \rm M_\odot$).
It is apparent that the radiatively-driven outflows seen in our simulations, while substantial, are weaker than seen in simulations with (efficient) thermal feedback.
We return to this question in Section~\ref{sec_comparison}.

\begin{figure*}
\centering 
\includegraphics[scale = 0.425]{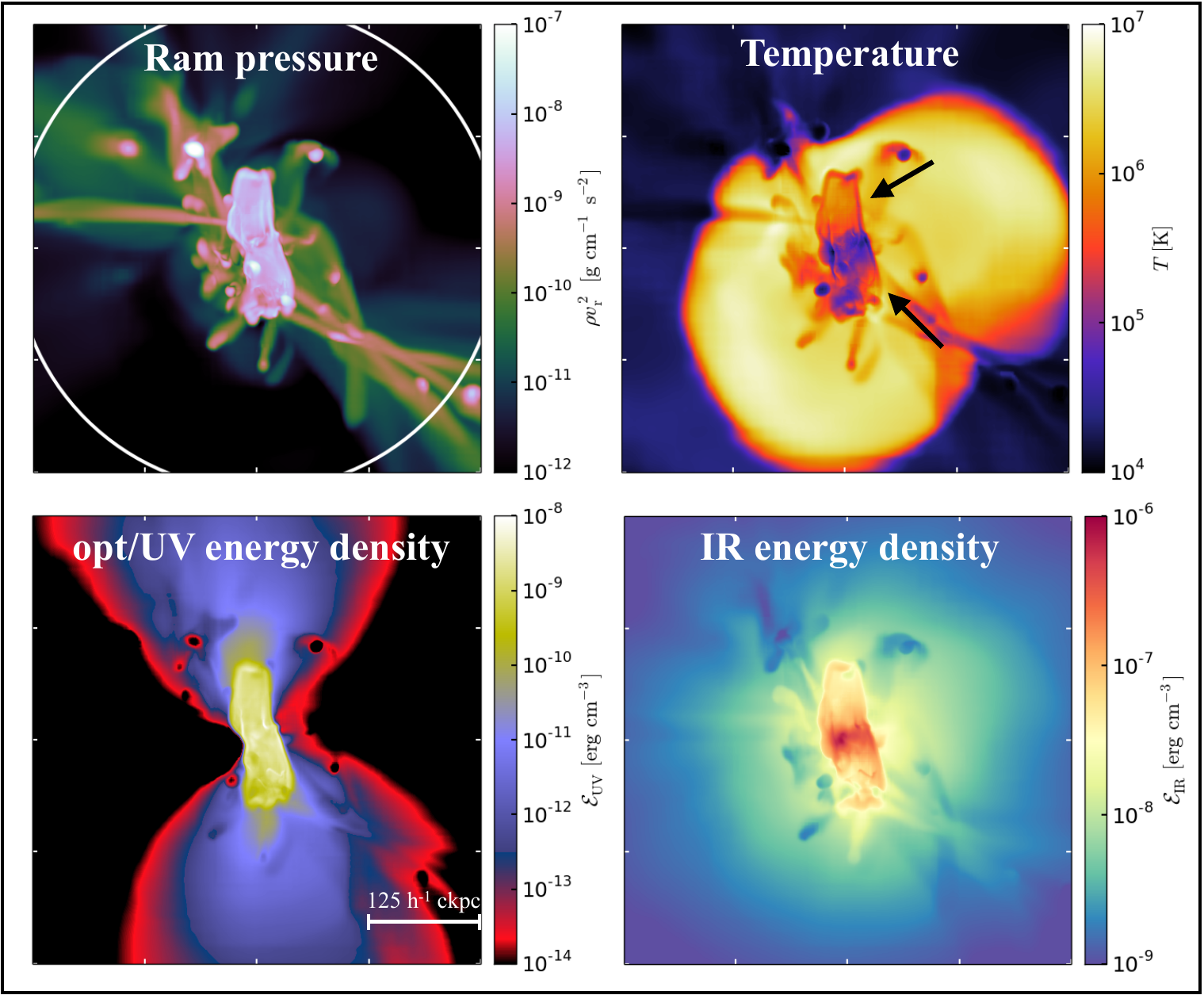}
\includegraphics[scale = 0.425]{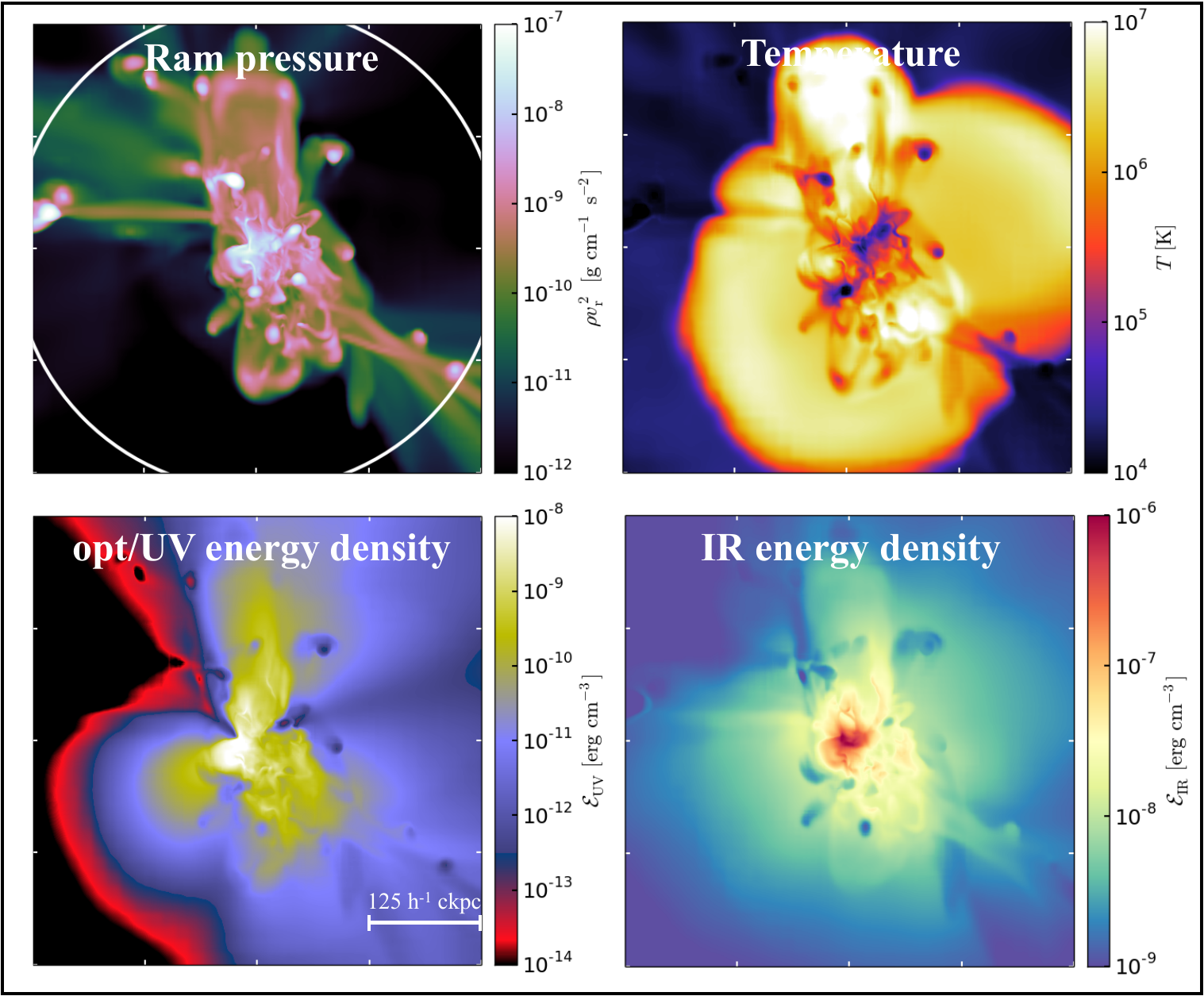}
\caption{Mass-weighted projections of gas ram pressure ($\rho v_{\rm r}^2$), gas temperature, optical/UV and IR energy densities at $z \, = \, 6.4$ (top panels) and $z \, = \, 6.3$ (bottom panels) in a cubic region with size $500 h^{-1} \, \rm kpc$ (comoving) centred on the quasar host galaxy. The white circle in the ram pressure maps gives the virial radius. IR radiation pressure drives a bipolar outflow. The outflow initially consists of shells of cool $T < 10^5 \, \rm K$ gas, which reach out to $\approx 30 \, \rm kpc$ from the BH particle. The optical/UV radiation field is also initially bipolar, but the IR radiation spreads out more isotropically due to the lower dust opacity at this frequency. At later times, the ram pressure of the outflow drops, as the radiation force declines. The outflowing shell fragments and takes on a more filamentary structure. Due to lower optical depths, the optical/UV radiation starts escaping more efficiently, filling up the simulation volume.}
\label{fig_cosmoutflows}
\end{figure*}

We now refer the reader to Fig.~\ref{fig_cosmoutflows}, which shows various radiation and hydrodynamic variables at $z \,=\, 6.4$ (top panel) and at $z \,=\, 6.3$ (bottom panel).
These redshifts correspond to, respectively, $\approx 17 \, \rm Myr$ and $\approx 35 \, \rm Myr$ from the time at which radiation injection begins.

We focus here on simulation UVIR-5e47, which, as we have seen, generates a strong outflow\footnote{Qualitatively, the evolution of the outflow seen in this simulation is similar to that seen in our remaining `UVIR' simulations.}.
In Fig.~\ref{fig_cosmoutflows}, we show gas ram pressure, defined as $\rho v_{\rm r}^2$, gas temperature, and the radiation energy density in the optical/UV and IR bands.
All quantities are projected in a mass-weighted fashion along a slab of thickness $500 h^{-1} \, \rm kpc$ (comoving) oriented in such a way that the disc of the AGN host galaxy is approximately seen edge-on.
In physical units, this length scale corresponds to $\approx 99 \, \rm kpc$ at $z \, = \, 6.4$ and $\approx 100 \, \rm kpc$ at $z \, = \, 6.3$.

The early stages of outflow formation are illustrated in the top panel of Fig.~\ref{fig_cosmoutflows}.
The outflow develops rapidly, breaking out of the galaxy within a few Myr.
At $z \,=\, 6.4$, less than $20 \, \rm Myr$ since the beginning of radiation injection, the outflow already has a size\footnote{In Appendix~\ref{sec_redc}, we show that the spatial extent of the outflow is sensitive to the reduced speed of light approximation only in the first few Myr of radiation injection. At $z \, = \, 6.4$, this is well converged.} of $\approx 30 \, \rm kpc$, from one tip of the cone to the other. 
At this stage, it has a clearly bipolar structure, propagating towards the polar regions of the central disc galaxy, as seen in many previous cosmological simulations \citep[e.g.][]{Costa:14, Curtis:16}; radiatively-driven outflows take paths of least resistance.

The temperature map reveals that a significant portion of outflowing mass is cool, with temperatures $T  < 10^5 \, \rm K$, and is concentrated at the rim of the blast waves generated by radiation pressure.
We have added arrows to the temperature map on the top panel of Fig.~\ref{fig_cosmoutflows} in order to guide the eye of the reader.
The interior of the outflow, instead, appears to contain mostly tenuous gas with temperatures reaching\footnote{We note that such a component exists even in simulations for which radiation is not allowed to couple to gas hotter than $3 \times 10^4 \, \rm K$. This hot component forms as the cool gas pushes against the hot atmosphere of the galactic halo at speeds $\gtrsim 1000 \, \rm km \, s^{-1}$.} $T \approx 10^7 \, \rm K$.
In Section~\ref{sec_thermalprop_outflow}, we show that the cool outflowing gas component forms through radiative cooling of shocked outflowing material. 

The second and fourth rows give the energy densities of optical/UV and IR radiation.
The outflowing cone is filled by optical/UV radiation, which is efficiently absorbed by the dense outer shell of the outflow; the radiation field here also has a clearly bipolar configuration.
Absorption of optical/UV radiation is particularly efficient in directions along the disc plane, such that the quasar remains obscured along this direction.
Due to the lower dust opacity and the fact that it can diffuse through dense gas, IR radiation escapes more easily than optical/UV radiation.
Note, in particular, the higher energy densities seen for the IR radiation component, which exceed those of the optical/UV due to (i) the higher absorption cross-section in the optical/UV and (ii) the fact that IR radiation is reprocessed, while optical/UV is absorbed.
Thus, while most of the volume contains no optical/UV radiation at early times, it is quickly filled by reprocessed IR radiation.
Note that, since we adopt a reduced speed of light $\tilde{c} \, = \, 0.03c$, our simulations inevitably underestimate how long it takes for radiation to propagate across the simulation volume (see Appendix~\ref{sec_redc}).

The coherent structure of the outflow seen on the top panel of Fig.~\ref{fig_cosmoutflows} is, however, short-lived.
By $z \, = \rm 6.3$, the outflowing cold shell has broken up into multiple filaments, many of which remain in the halo and fall back towards the galaxy, and is starting to lose its initial bipolar configuration.
The bottom right panel in Fig.~\ref{fig_cosmofield} gives the density field in the same simulation at the same time.
The galaxy is highly disturbed and surrounded by various shells and filaments of dense ($n_{\rm H} \gtrsim 10 \, \rm cm^{-3}$) gas, which encase a lower density outflow component.
These later stages of the simulation are illustrated on the bottom panel of Fig.~\ref{fig_cosmoutflows}, where we present the same projected maps as shown on the top panel, but generated at $z \, = \, 6.3$.
In particular, the ram pressure of the outflow, i.e. its ability to push gas out from its surroundings, drops as it expands.
Since IR radiation pressure plays a fundamental role in driving the outflow, it is not surprising that the outflow weakens as it moves out to regions in which it becomes optically thin to IR radiation.
In Section~\ref{sec_outflowrate}, we show explicitly that the mass outflow rates drop with radius, and, hence, that the bulk of the mass expelled in the outflow remains within the halo.

At $z \, = \, 6.3$ (and also later on), after $\approx  35 \, \rm Myr$ of continuous radiation injection, the central galaxy, though highly disturbed, remains largely intact.
Even at these high luminosities, radiation pressure is unable to prevent gas from refilling the innermost regions of the halo.
However, the stellar bulge's properties evolve considerably as a consequence of the radiatively-driven outflow.
We find the peak stellar velocity dispersion to have dropped to $\approx 370 \, \rm km \, s^{-1}$ from $\approx 440 \, \rm km \, s^{-1}$ and the effective radius to have increased to $\approx 675 \, \rm pc$ from $\approx 600 \, \rm pc$.
The expansion of the stellar bulge is also important, though somewhat suppressed, in simulations UVIR-4e47-duty and UVIR-4e47-dust, in which the peak circular velocities respectively drop to $\approx 390 \, \rm km \, s^{-1}$ and $\approx 400 \, \rm km \, s^{-1}$ and the effective radius grows to $\approx 700 \, \rm pc$ in both cases.
When performed with inefficient AGN feedback, the peak stellar velocity instead \emph{increases} to $\approx 490 \, \rm km \, s^{-1}$, while the effective radius remains at $\approx 600 \, \rm pc$.
The origin of this size growth (associated with expansion of the stellar component) is likely linked to the rapid modification of the gravitational potential \citep{Mashchenko:06, Pontzen:12, Martizzi:13}.

The picture of radiation pressure-driven outflows suggested by our simulations is in line with the classical scenario \citep{Sanders:88, Hopkins:08, Narayanan:10}.
After an accretion episode, caused by a merger or by a cold flow, the central quasar is buried within a dense gaseous bulge.
The high central gas density should facilitate rapid black hole growth.
The accreting black hole is enshrouded in various layers of dense gas which are thick to the optical/UV radiation of the quasar, which may, at this point, appear as a hot-dust-obscured AGN \citep[e.g.][]{Tsai:15}.
Due to the high binding energy of the gaseous bulge, direct radiation pressure is insufficient to lift off the gas layers surrounding the quasar.
IR radiation pressure, however, is able to drive an outflow, clearing the galactic nucleus (but leaving the gas disc intact) and opening up low density channels through which also the optical/UV radiation field can escape.
At this point, the AGN transitions into an optical quasar.

We examine the properties of the IR-driven outflows in more detail in Section~\ref{sec_outflowrate}, where we show that IR radiation pressure is unlikely to be an efficient mechanism to eject baryons from massive galactic haloes.
Before we return to the properties of the outflows, we first examine the conditions in which they are driven and also show that IR radiation pressure may be extremely important in regulating the star formation rates of quasar host galaxies.

\subsection{A critical luminosity for quasar outflows}
\label{sec_critlum}

Here we study the conditions in which outflows can be launched in our simulations.
We start by investigating the effect of single-scattering radiation only, i.e. in the set of `UV' simulations (see Table~\ref{table_sims}).

\subsubsection{The single-scattering regime}

\begin{figure*}
\centering 
\includegraphics[scale = 0.55]{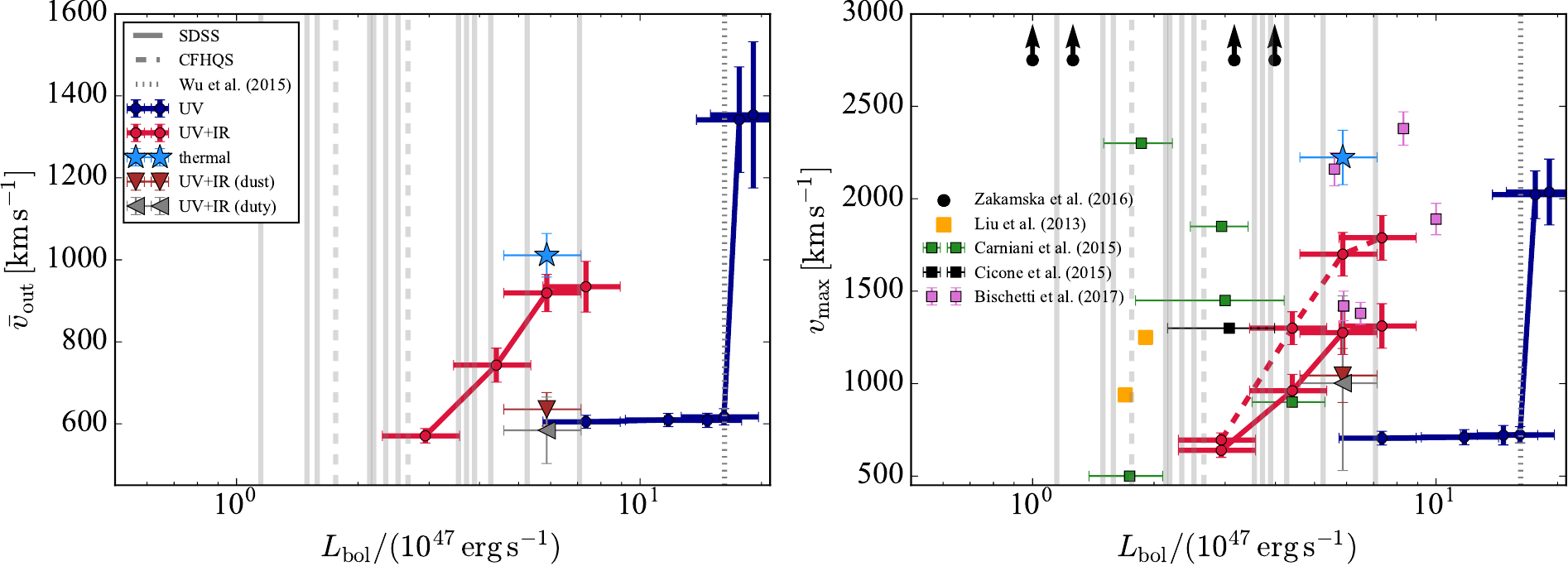}
\caption{Mass-weighted median outflow speed (left-hand panel) and $85^{\rm th}$ radial velocity percentile (right-hand panel) as a function of quasar bolometric luminosity. The dashed line on the right-hand panel gives the $95^{\rm th}$ radial velocity percentile. Blue points refer to simulations with only single-scattering radiation, while red points refer to simulations that also include infrared multi-scattering radiation pressure. The dark red triangle, gray triangle and blue filled star show results for simulations UVIR-4e47-dust, UVIR-4e47-duty and thermal-4e47, respectively. The vertical lines show observational estimates of bolometric luminosities for $z\,\approx\,6$ SDSS quasars (from \citet{Jiang:07}, shown with solid lines), CFHQS (from \citet{Willott:10}, shown with dashed lines) and for J010013 (\citet{Wu:15}, shown with a dotted line). The black data point shows the velocity of the cold outflow detected through [CII] emission in the quasar J1148 at $z \, = \, 6.4$ \citep{Cicone:15}, the green data points show the measured velocities in a sample of ionised outflows detected in $z \,\approx\, 2.4$ quasars \citep{Carniani:15}, the purple data points give results for the WISSH quasar project \citep{Bischetti:17} and orange data points for ionised outflows at $z \,=\, 0.5$ presented in \citet{Liu:13, Liu:13b}. These data are taken from the compilation of \citet{Fiore:17}. \citet{Zakamska:16} estimate outflow velocities $\approx 3,000 \, \rm km \, s^{-1}$. We illustrate these data with upward arrows. There is a critical luminosity $L_{\rm crit}$ above which an outflow is launched in our simulations, though $L_{\rm crit}$ is unrealistically high if only single-scattered radiation is included. Infrared radiation pressure leads to a reduction in $L_{\rm crit}$ by a factor $\approx 4$ to more realistic values and generates outflows with bulk velocities in good agreement with the more moderate of the observed outflows.}
\label{fig_critlum}
\end{figure*}

For simulations performed with single-scattering radiation only, we find that radiation pressure is able to launch a galactic outflow only if the quasar optical/UV luminosity is higher than the critical value $L_{\rm crit} \, = \, 1.2 \times 10^{48} \, \rm erg \, s^{-1}$.
This corresponds to a bolometric luminosity of $L_{\rm crit, bol} \, \approx \, (1.4 \-- 2) \times 10^{48} \, \rm erg \, s^{-1}$ (see Section~\ref{sec_radfromqso}).
The existence of such a critical luminosity is demonstrated in Fig.~\ref{fig_critlum}, where we plot mass-weighted median outflow velocities (on the left-hand panel) and the $85^{\rm th}$ percentile of the velocity distribution (on the right-hand panel) against the quasar bolometric luminosity.
Both quantities are averaged over the period of $\approx 100 \, \rm Myr$ during which the quasar is emitting radiation, with the error bars showing the standard deviation associated with the time-average.
The median is evaluated over all gas located within the virial radius and with an outward radial velocity greater than $500 \, \rm km \, s^{-1}$, which is approximately the maximum speed attained by gas in an identical simulation performed without feedback.
For simulations conducted with single-scattering radiation only (shown with blue symbols), the median outflow speed is constant at $\bar{v}_{\rm out} \, \approx \, 600\, \rm km \, s^{-1}$ up to $L_{\rm opt,UV} \, = \, 1.1 \times 10^{48} \, \rm erg \, s^{-1}$, at which point it increases dramatically to $\bar{v}_{\rm out} \, \approx \, 1400 \, \rm km \, s^{-1}$.
The moderate outflow velocities seen below this critical luminosity are not caused by feedback, but instead represent the high velocity tail of gas undergoing gravitational motions at the centre of the halo.
For this reason, the median and $85^{\rm th}$ percentile values are similar below $L_{\rm crit}$.

If we assume the critical luminosity to be generated by a supermassive black hole accreting at its Eddington limit (Eq.~\ref{eq_eddlimit}), we find a corresponding critical black hole mass of
\begin{equation}
M_{\rm BH}^{\rm crit} \, = \, \frac{L_{\rm crit, bol} \kappa_{\rm T}}{4 \pi G c} \, \approx \, 1.1 \times 10^{10} \left( \frac{L_{\rm crit, bol}}{1.4 \times 10^{48} \, \mathrm{erg \, s^{-1}}} \right) \, \rm M_\odot \, .
\end{equation}
The required quasar luminosities and the corresponding black hole masses are, in this case, implausibly high.
The total stellar mass within the halo is $M_{\rm *} \, = \, 1.1 \times 10^{11} \, \rm M_\odot$ at $z \, = \, 6.5$, such that the required black hole mass would amount to $\gtrsim 10\%$ of the stellar mass. 
Estimated bolometric luminosities for the majority of $z \, = \, 6$ quasars are lower than $L_{\rm crit}$ by factors $\gtrsim 2 \-- 3$, as can be seen from comparison with the observational data shown in Fig.~\ref{fig_critlum} with gray, vertical lines.
These observational data give the estimated bolometric luminosities of SDSS and CFHQS quasars at $z \gtrsim 6$ and are taken from \citet{Jiang:07} and \citet{Willott:10}, respectively.
In the probably rare case of J010013, a hyperluminous quasar at $z \, = \, 6.3$ identified by \citet{Wu:15}, the estimated bolometric luminosity is higher than $1.5 \times 10^{48} \, \rm erg \, s^{-1}$ and would therefore compare well with $L_{\rm crit}$ as obtained in our simulations.
It is important to note, however, that surveys such SDSS or 2MASS probe cosmological volumes much larger than that of our simulations.
Associating such a bright quasar to every $\approx (2 \-- 3) \times 10^{12} \, \rm M_\odot$ halo found in our cosmological volume would lead us to greatly overestimate their number density\footnote{This argument assumes a quasar duty cycle of order unity. For the brightest quasars at $z \,=\, 6$, this condition is likely required in order to enable supermassive black holes to form rapidly, as shown in e.g. Fig.~\ref{fig_lightcurve}.}.

A critical quasar luminosity for outflows is, in fact, a key prediction of analytic models based on feedback through radiation pressure \citep[e.g.][]{Fabian:99, Murray:05, Heckman:16, Costa:17}.
According to these models, the outward radiation pressure acceleration $\sim L/c$ eventually exceeds the inward gravitational acceleration $G M_{\rm gas}(R) M_{\rm tot}(R)/R^2$, where $M_{\rm gas}(R)$ and $M_{\rm tot}(R)$ are enclosed gas and total masses\footnote{There are inconsistencies in the literature regarding the computation of $M_{\rm tot}$ and on whether and how the gas mass (and its self-gravity) is accounted for. We calculate $M_{\rm tot} \, = \, M_{\rm col} + M_{\rm gas}$, where $M_{\rm col}$ is the mass of the collisionless component (dark matter, stars, black holes), in agreement with e.g. \citet[][]{Zubovas:14b}. Note that e.g. \citet[][]{Thompson:15} instead take $M_{\rm tot} \, = \, M_{\rm col} + \frac{1}{2} M_{\rm gas}$. Our computation of $L_{\rm eq}$ (Eq.~\ref{eq_leq}), however, drops only by a factor $\approx 1.3$ if we take the latter. This does not affect our conclusions.}, respectively.
The critical quasar luminosity at which both forces balance is hence
\begin{equation}
L_{\rm eq} \, \sim \, \frac{G M_{\rm gas}(R)M_{\rm tot}(R) c}{R^2} \, = \, \frac{ v_{\rm circ, gas}^2 v_{\rm circ, tot}^2 c}{G} \, .
\label{eq_leq}
\end{equation}
The right-hand side of Eq.~\ref{eq_leq} varies with radius, but a sufficient condition for an outflow is that $L_{\rm eq}$ is greater than the maximum value of this expression.
We evaluate $v_{\rm circ, gas}^2 v_{\rm circ, tot}^2$ at $z \, = \, 6.5$ and, using Eq.~\ref{eq_leq}, find that $L_{\rm eq} \, = \, 3.4 \times 10^{48} \, \rm erg \, s^{-1}$, higher than obtained in the simulations by a factor $\approx 2.8$.

That a significant outflow is launched at a somewhat lower luminosity is not surprising, since the simple analytic argument presented above (i) requires all gas to be ejected from within the radius corresponding to the peak of the circular velocity curve, and this does not happen at the luminosities probed by our simulations\footnote{We have verified that even if the luminosity is set to $L_{\rm eq}$, it is still not possible to eject all gas from the halo. The reason for this is that Eq.~\ref{eq_leq} also ignores density anisotropy and ram pressure from infalling gas.}, and (ii) it neglects the fact that the galaxy is not spherically symmetric, but elongated along the disc plane (see Fig.~\ref{fig_tautheta}). 
In reality, radiation preferentially accelerates gas located towards the galactic poles, such that material driven out at $L_{\rm opt,UV} < L_{\rm eq}$ corresponds to the low-end tail of the gas surface density distribution.

This result indicates that the critical luminosity for outflows in the single-scattering regime seen in our simulations is in reasonable agreement with analytic predictions based on force balance between gravity and radiation pressure.
However, we show that large-scale outflows can be launched already at $L_{\rm opt,UV} < L_{\rm eq}$, such that it is likely that black hole growth would self-regulate before the associated quasar reaches the luminosity required to expel all gas from the galaxy \citep[see also][]{Hartwig:17}.
This finding also agrees with observations, which show high redshift quasars to lie within extended gaseous discs \citep[e.g.][]{Wang:16} and with previous simulations, which show central outflows to pave paths of least resistance without interacting with most of the mass in the galaxy \citep{Gabor:14, Bourne:14, Curtis:16}.

Finally, note that our conclusion that too high UV/optical luminosities are required to launch outflows with single-scattering radiation pressure only is not caused by numerical underestimation of the radial photon flux \citep[see Appendix B in][]{Rosdahl:15b}, which has been corrected for in our simulations. We verified that if we did not adopt the `reduced flux approximation' described in Section~\ref{sec_radfromqso}, we would require even higher quasar luminosities in order to drive outflows with single-scattering radiation pressure.

\subsubsection{How IR radiation pressure boosts galactic outflows}

The inclusion of IR radiation pressure reduces $L_{\rm crit}$ by a factor of $\approx 4$, as can be seen from the red data points in Fig.~\ref{fig_critlum}.
The resulting $L_{\rm crit} \, = \, (2 \-- 3) \times 10^{47} \rm erg \, s^{-1}$, which corresponds to $L_{\rm crit, bol} \, = \, (3.4 \-- 5.4) \times 10^{47} \rm erg \, s^{-1}$, appears to be in better agreement with the luminosities of quasars observed at $z \, = \, 6$.
The required bolometric luminosities, if achieved through Eddington limited accretion, now correspond to black hole masses of $(2.7 \-- 4.3) \times 10^9 \, \rm M_\odot$, in good agreement with the black hole masses predicted by cosmological simulations that model black hole growth self-consistently in galactic haloes of similar mass at $z \, = \, 6$ \citep[e.g.][]{Sijacki:09, DiMatteo:12, Costa:14a}. 
The required black hole mass is now $\lesssim 3\%$ of the total stellar mass within the halo.
Thus, we confirm that, as in energy-driven outflows \citep{Costa:14}, trapped IR radiation pressure has the ability to generate large-scale outflows at lower luminosities.
In Appendix~\ref{sec_irtrapeff}, we calculate the IR radiation force in simulation UVIR-4e47 explicitly and find this to be $\sim 10 L_{\rm opt,UV} / c$ at the earliest times.

For optical/UV luminosities $L_{\rm opt,UV} \, \geq \, 3 \times 10^{47} \, \rm erg \, s^{-1}$, the resulting IR-driven outflows have median speeds on the order of $1000 \, \rm km \, s^{-1}$, with the fastest component reaching velocities of nearly $1500 \, \rm km \, s^{-1}$. 
In particular, we find the median and maximum outflow luminosities to increase with $L_{\rm opt,UV}$, though not linearly.
The difference between simulations performed at different quasar luminosities is that, with increasing $L_{\rm opt,UV}$, outflows start sweeping up more mass such that the mean and maximum outflow velocity are prevented from increasing linearly.

Note that the outflow velocities drop significantly in simulations UVIR-4e47-dust, for which the dust opacity is set to zero at high gas temperatures (dark red triangle in Fig.~\ref{fig_critlum}) and UVIR-4e47-duty, for which the AGN light-curve fluctuates.
In UVIR-4e47-dust, outflow velocities decrease, because gas which is shock-heated as it pushes into the low-density halo gas can no longer couple to radiation, while in UVIR-4e47-duty the quasar luminosity drops by an order of magnitude after $z \, \approx \, 6.3$ and is unable to sustain an outflow.
However, even in these simulations, IR radiation pressure is able to drive outflows for which the fastest component has $v_{\rm max} \gtrsim 1000 \, \rm km \, s^{-1}$.

\subsubsection{Comparison with observed outflows}

Powerful quasar outflows are observed at quasar luminosities in the range explored by our simulations.
However, comparison with observations is challenging, since (i) observations usually probe only one phase of outflows, e.g. the cold phase as detected through [CII] or CO emission \citep[e.g.][]{Maiolino:12, Cicone:15} or the warm ionised phase as detected through high ionisation metal lines \citep[e.g.][]{Carniani:15, Zakamska:16, Bischetti:17} and (ii) various different definitions of outflow velocity are adopted in the literature.
Thus we merely verify whether the bulk velocity of the outflows generated in our simulations compares well with available observational constraints.

On the right-hand panel of Fig.~\ref{fig_critlum}, we show data points for observationally measured outflow velocities in high-redshift quasars with estimated bolometric luminosities of $> 10^{47} \, \rm erg \, s^{-1}$, such as those probed in our simulations.
We compare our simulations to results from five different data sets, including (i) the powerful, spatially extended outflow seen in [CII] emission around the $z \, =\, 6.4$ quasar J1148 \citep{Cicone:15} and (ii) the ionised outflows from the $z \approx 0.5$ obscured quasar sample presented in \citet{Liu:13, Liu:13b}, (iii) the ionised outflows from a sample of $z \, \approx \, 2.4$ quasars \citep{Carniani:15}, (iv) the outflows from the WISSH quasars project \citep{Bischetti:17} and (v) the extreme outflows in reddened quasars at $z \, = \, 2.5$ of \citet{Zakamska:16}, all detected in [OIII] emission.
We take the bolometric luminosity\footnote{We take the bolometric luminosities provided in \citet{Fiore:17}, but include an error bar if the bolometric luminosity is also quantified in the original papers and this does not match the value listed in \citet{Fiore:17}. In this case, the error bar encompasses the range of bolometric luminosities for a given system.} and maximum velocity of outflows $v_{\rm max}$ from Table B.1 from \citet{Fiore:17} for the outflows investigated by \citet{Liu:13, Liu:13b, Cicone:15, Carniani:15} and \citet{Bischetti:17}, while \citet{Zakamska:16} give $v_{\rm max} \, = \, 3,000 \, \rm km \, s^{-1}$ as the likely velocity of their quasar-driven outflows (see their Section 4.2).

We find that the maximum outflow velocities of the simulated outflows are in much better agreement with observations if IR multi-scattering is included.
At luminosities at which a force $L_{\rm opt, UV} / c$ is insufficient, trapped IR radiation ensures that an outflow is driven.
Interestingly though, even under optimistic assumptions of high and uniform dust-to-gas ratio, IR radiation pressure is not able to generate sufficient gas at velocities much exceeding $1500 \, \rm km \, s^{-1}$, as shown on the right-hand panel of Fig.~\ref{fig_critlum}.
When compared to the observational data points, the IR-driven outflows appear to have velocities which are somewhat too low.
We have verified that these conclusions do not change if we define the maximum velocity as the $95^{\rm th}$ radial velocity percentile, in which case the maximum speed attained by IR-driven outflows is on the order of $1700 \, \rm km \, s^{-1}$ (see dashed red line on the right-hand panel of Fig.~\ref{fig_critlum}).
Some observed outflows at similar quasar luminosities have outflow speeds exceeding $2,500 \, \rm km \, s^{-1}$ \citep[e.g.][]{Zakamska:16, Bischetti:17}.
The filled blue star in Fig.~\ref{fig_critlum}, which shows results for the thermally-driven outflow, is in much better agreement with these extreme outflows.

At face value, our simulations suggest that, while IR-driven outflows have bulk speeds in agreement with some of the more moderate AGN-driven outflows, energy-driving may be required to generate large-scale winds moving faster than $2,000 \, \rm km \, s^{-1}$.
Possible alternatives would require (i) the galaxies hosting the quasars in the samples of e.g. \citet{Zakamska:16} or \citet{Bischetti:17} to be significantly less compact than in our simulations, (ii) that the IR optical depths are even higher than seen in our simulations (see Appendix~\ref{sec_jeansp}), (iii) that additional feedback mechanisms such as supernova feedback or cosmic rays boost the AGN radiatively-driven outflow \citep[e.g.][]{Costa:15} or (iv) that the observed outflows exist on scales that are too small to be resolved accurately by our simulations. The latter is unlikely since, based on photoionisation modelling, \citet{Zakamska:16} constrain the spatial extension of their extreme outflows to be on the order of a few kpc, a scale which is well resolved in our simulations.
Performing higher resolution and more self-consistent simulations following a statistically significant number of quasars at lower redshift will be crucial to generalise the findings presented here.

\subsection{Regulating star formation}
\label{sec_regulation}

\begin{figure*}
\centering 
\includegraphics[scale = 0.44]{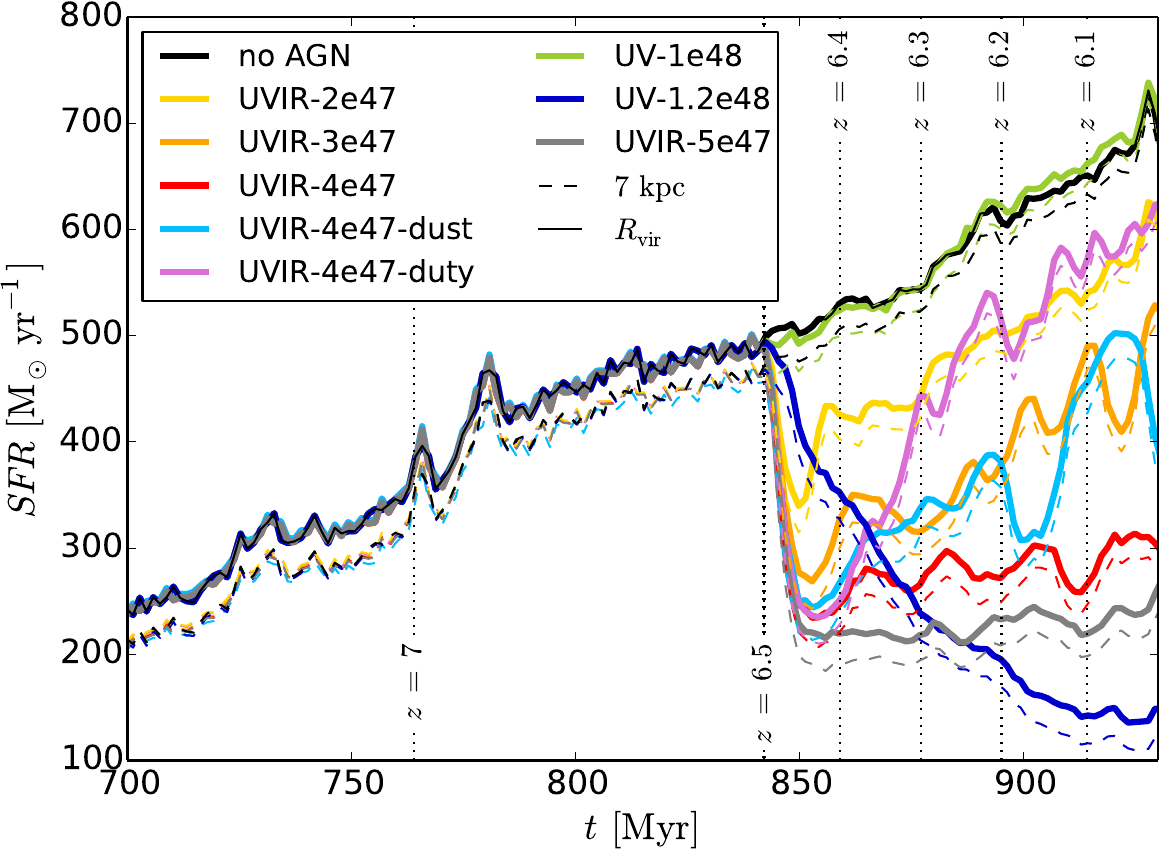}
\includegraphics[scale = 0.44]{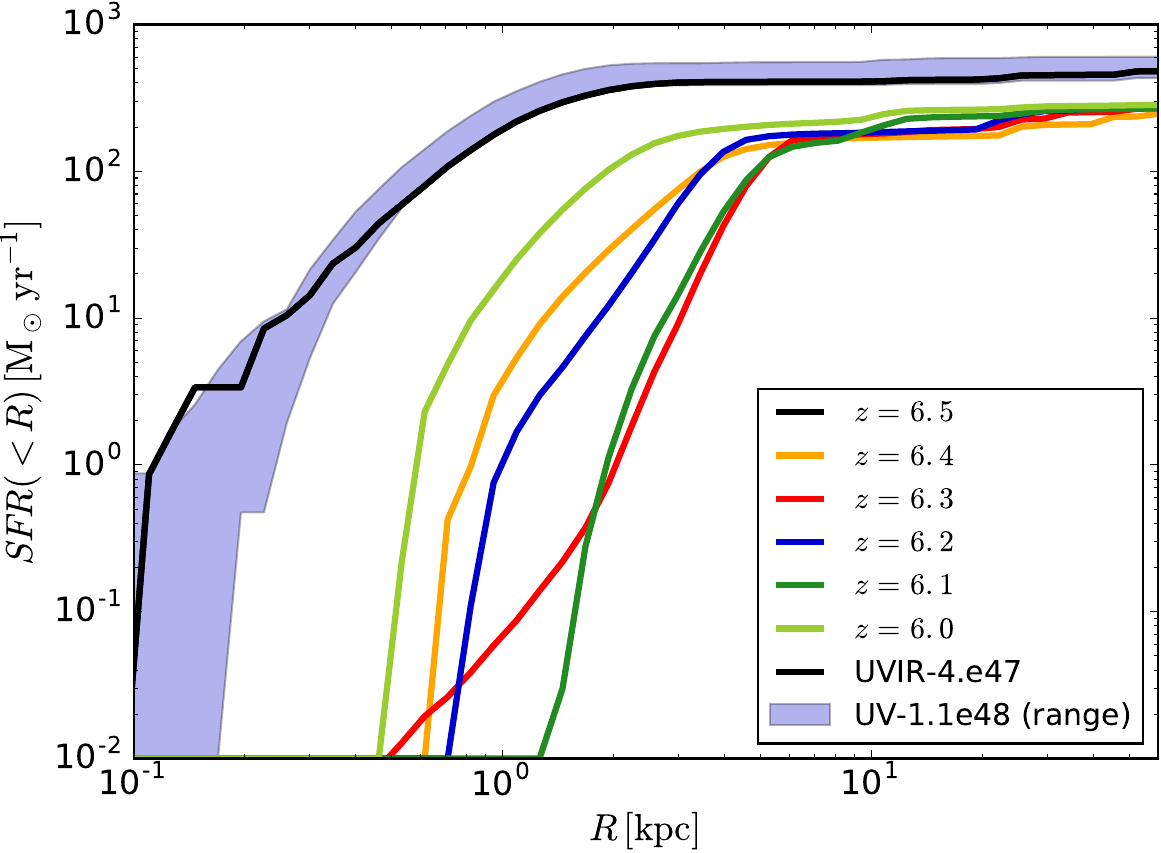}
\caption{\emph{Left:} Star formation rate within the virial radius (solid lines) and the central $7 \, \rm kpc$ (dashed lines). In our simulation with no radiative coupling (noAGN, black lines), star formation rises steadily, exceeding $700 \, \rm M_\odot \, yr^{-1}$ by $z \, \approx \, 6.0$. The bulk of this star formation ($> 80 \%$) occurs within the central 10 kpc. In all our simulations with radiative feedback, radiation injection starts at $t \, \approx \, 841 \, \rm Myr$ ($z \, = \, 6.5$), a time which is marked with a dotted vertical line. If only single-scattering radiation is included, the star formation history of the quasar host galaxy is not affected unless the quasar luminosity is very high ($\gtrsim 1.2 \times 10^{48} \, \rm erg \, s^{-1}$), as shown by the dark blue line. If IR radiation pressure is included, the star formation rate is suppressed already at UV/optical luminosities of $2 \times 10^{47} \, \rm erg \, s^{-1}$ (yellow line). The suppression is more significant for higher quasar luminosities (e.g. orange and gray lines), as expected. \emph{Right:} Star formation rate profiles for simulations UV-1.1e48, shown as blue band encompassing the range of profiles obtained from $z \, = \, 6.5$ to $z \, = \, 6.0$, and UVIR-4e47, shown with solid lines for different times. The addition of IR radiation pressure results in a significant, or total, reduction in star formation within the innermost $1 \, \rm \, kpc$.}
\label{fig_sfr}
\end{figure*}

We now address the effect of radiation pressure on star formation in the quasar host galaxy.
We select all stellar particles within $R_{\rm vir}$ and reconstruct the star formation history using their birth times.
The star formation rate as a function of time in some of our simulations is shown on the left-hand panel of Fig.~\ref{fig_sfr}.
Star formation rates within spheres with radii of $\approx 7 \, \rm kpc$ and $R_{\rm vir}$ are shown by dashed and solid lines, respectively.
These star formation rates account for both in- and ex situ contributions.
Since the central galaxy forms stars at a rapid rate at $z \, = \, 6$ due to efficient cooling, however, most of the star formation occurs in situ, as is shown below.

\begin{figure}
\centering 
\includegraphics[scale = 0.44]{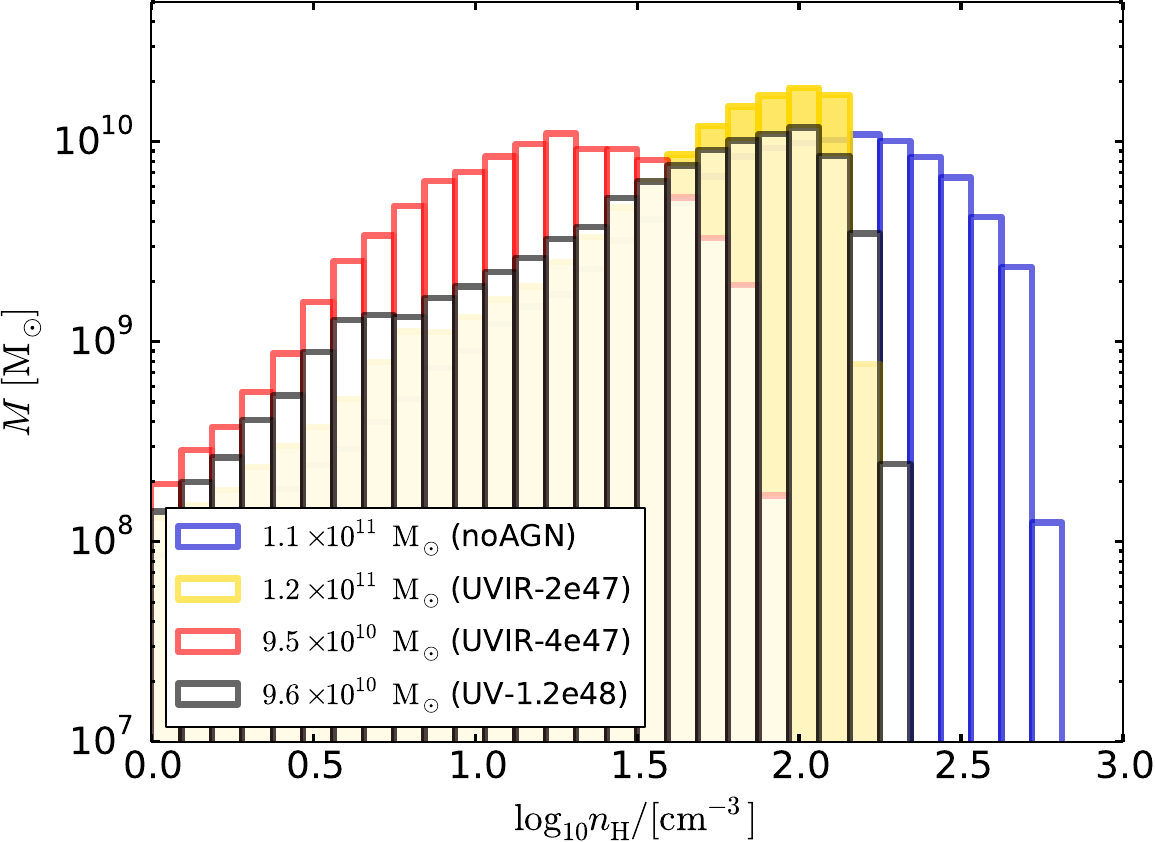}
\caption{Density probability distribution within $5 \rm kpc$ of the quasar for gas with absolute radial velocities $< 300 \, \rm km \, s^{-1}$ at $z \, = \, 6.4$ in a representative sample of our simulations. The mass associated with this gas component is given in the legend. Radiative feedback reduces the central gas density; optical/UV radiation pressure does so by ejecting gas (though very high quasar luminosities are required). IR radiation, however, can reduce the gas density without ejecting it, as can be seen from comparing the gas masses with those of simulation noAGN. Thus, one of the channels by which trapped IR radiation quenches star formation is through local reduction of gas density.}
\label{fig_densmax}
\end{figure}

For the simulation in which the opacity is zero (noAGN), the star formation rate increases from $500 \, \rm M_\odot \, yr^{-1}$ to over $700 \, \rm M_\odot \, yr^{-1}$ within the time span of $\approx 100 \, \rm Myr$ during which radiation injection is allowed to take place. 
Most of this star formation occurs within the innermost $7 \, \rm kpc$ of the galaxy, as can be seen by comparing dashed and solid black lines. 

If IR radiation pressure is neglected and only single-scattering radiation is included, unrealistically high luminosities are required to affect the star formation evolution.
We find that the quasar optical/UV luminosity would have to be increased to about $1.2 \times 10^{48} \, \rm erg \, s^{-1}$, roughly the luminosity output of a $\gtrsim 10^{10} \, \rm M_\odot$ black hole accreting at its Eddington limit, in order for the star formation to drop with respect to simulation noAGN.
This luminosity is the same as the threshold value $L_{\rm crit}$ for which an outflow can be generated, as we have seen in the previous section.

Remarkably, IR radiation pressure reduces the required quasar luminosity to values lower by a factor $\approx 6$. 
The luminosity required to suppress star formation ($L_{\rm opt,UV} \, = \, 2 \times 10^{47} \, \rm erg \, s^{-1}$) in the host galaxy is somewhat lower than that required to launch a large-scale outflow ($L_{\rm opt,UV} \, = \, 3 \times 10^{47} \, \rm erg \, s^{-1}$), indicating that trapped IR radiation can reduce the star formation rate without ejecting gas (yellow line).
Star formation reduces even further as the quasar luminosity increases (see e.g. red line for $L_{\rm opt,UV} \, = \, 4 \times 10^{47} \, \rm erg \, s^{-1}$).

In the more restrictive simulations in which radiation cannot couple to hot gas (UVIR-4e47-dust) and the quasar lightcurve fluctuates (UVIR-4e47-duty), shown by light blue and pink curves, respectively, star formation is efficiently suppressed in the first $\sim 10 \, \rm Myr$.
However, at later times, the inability of radiation to couple to hot gas (in the case of UVIR-4e47-dust) and the lower quasar luminosities (in the case of UVIR-4e47-duty) result in star formation rates up to a factor $2$ higher than seen in UVIR-4e47, failing to quench star formation in the long term.
The effect is similar to that of reducing the quasar luminosity, which can be seen from comparison with the yellow and orange curves in Fig.~\ref{fig_sfr}.
For UVIR-4e47-dustyduty, not shown here, the star formation history closely traces that of UVIR-4e47-dust in the first $10 \, \rm Myr$ and, as the quasar luminosity falls, increases thereafter to values $\approx 10 \-- 20 \%$ higher than in UVIR-4e47-duty.
Overall, however, due to early gas ejection, some degree of long term star formation suppression occurs in all simulations including IR radiation pressure.

The star formation rates evaluated within spheres of radii $7 \, \rm kpc$ never drop below $100 \, \rm M_\odot \, \rm yr^{-1}$.
It is clear that, in our simulations with IR radiation pressure, only the central few kpc of the quasar host galaxies are affected.
In the right-hand panel of Fig.~\ref{fig_sfr}, we show the enclosed star formation rate as a function of radial distance from the BH particle in simulation UV-1.1e48, illustrated with a blue band encompassing the range of profiles obtained between $z \, = \, 6.5$ and $z \, = \, 6$, and UVIR-4e47 simulations, shown with solid coloured lines at different redshifts.
Here, star formation rates are computed directly by applying Eq.~\ref{eq_sfr} and therefore include only in situ contributions\footnote{We have also verified that if we remove the temperature criterion for star formation, the star formation rates remain similar to those shown on the right-hand panel of Fig.~\ref{fig_sfr}}.
Although the total star formation rates, evaluated within $R_{\rm vir}$, drop only by factors $2 \-- 3$, central star formation is quenched efficiently.
In particular, for simulation UVIR-4e47, the region within which star formation is completely terminated has a radius $\sim 1 \, \rm kpc$.
We investigated how the size of the region within which star formation is suppressed scales with quasar optical/UV luminosity in our simulations.
We find that in simulations which include IR multi-scattering, star formation is suppressed efficiently within $\approx 300 \, \rm pc$ for simulation UVIR-2e47 to $\approx 2 \, \rm kpc$ for UVIR-5e47. 
In simulations UVIR-4e47-dust and UVIR-4e47-duty, star formation is efficiently suppressed in the innermost $\approx 800 \, \rm pc$ within $\sim 10 \, \rm Myr$ of radiation injection. At later times, however, star formation, though still reduced, is no longer suppressed as efficiently.

Star formation suppression occurs due to a drop in gas density (see Section~\ref{sec_limit} for a discussion of how missing physics and insufficient resolution may affect these findings).
We show the density probability distribution in Fig.~\ref{fig_densmax} for gas in the central regions of the halo (within $5 \, \rm kpc$) and with low absolute radial velocities ($|v_{\rm r}| < 300 \, \rm km \, s^{-1}$) for a representative sample of our simulations at $z \, = \, 6.4$.
For simulation noAGN (shown in blue), the gas density can reach $\approx 500 \, \rm cm^{-3}$.
The peak of the density distribution drops by about $0.25 \, \rm dex$ in simulation UVIR-2.e47 (yellow histograms).
As we have seen in Section~\ref{sec_critlum}, there is no large-scale outflow in this simulation. Together with the fact that the mass at low velocity remains unchanged (see legend in Fig.~\ref{fig_densmax}), the shift in the density probability distribution must be interpreted as the consequence of local suppression of gas density as opposed to gas ejection.

\begin{figure*}
\centering 
\includegraphics[scale = 0.44]{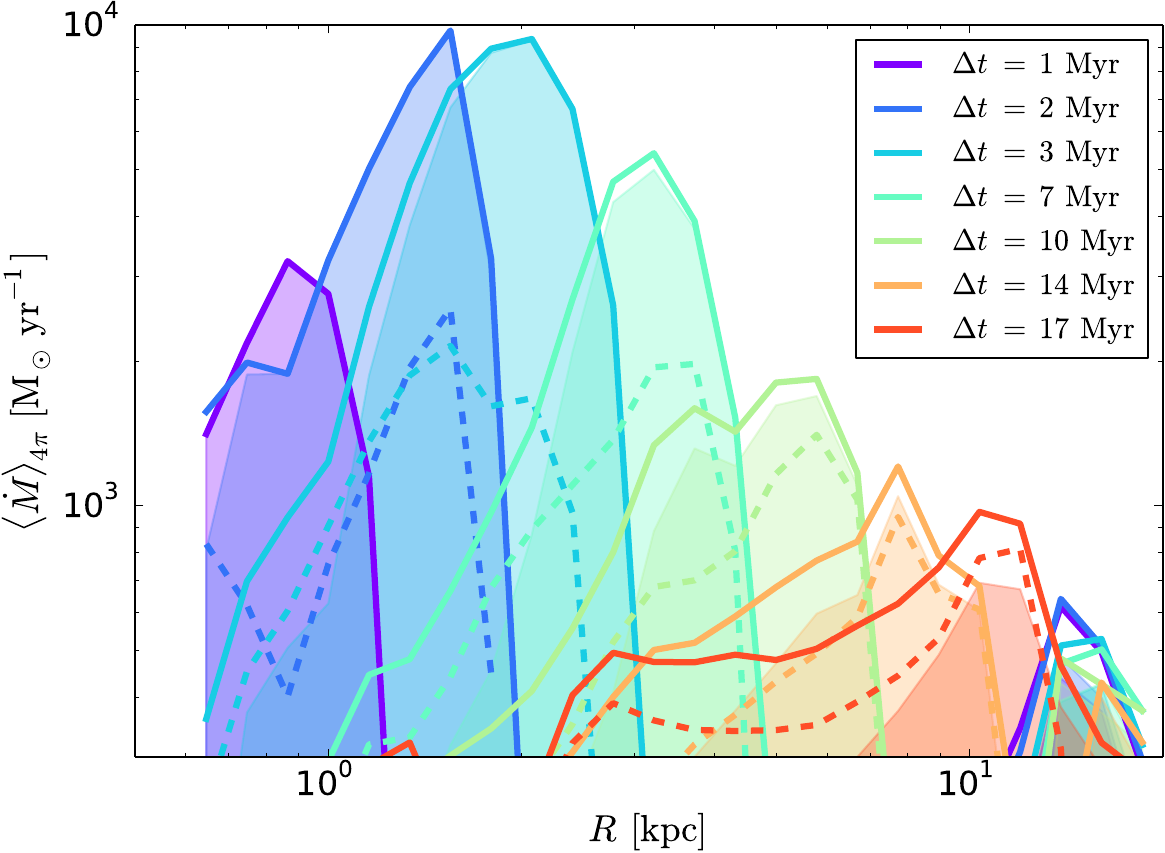}
\includegraphics[scale = 0.44]{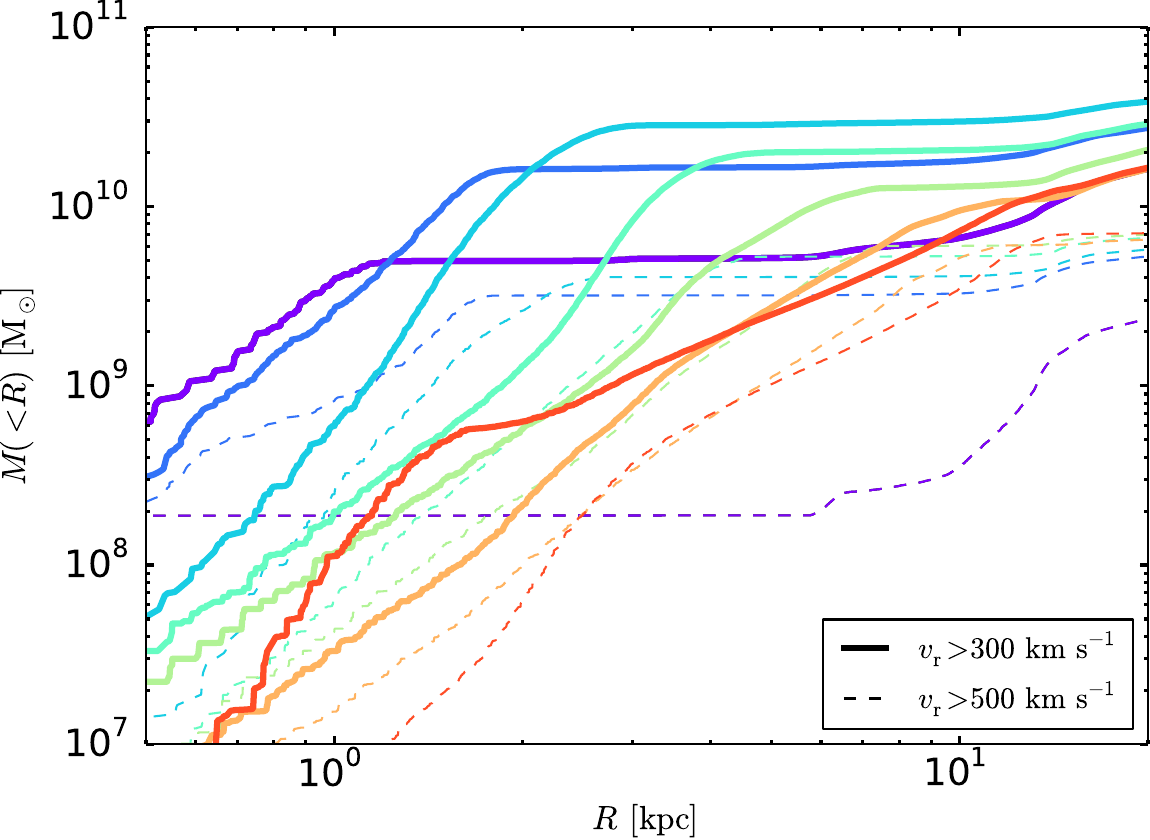}
\caption{\emph{Left:} spherically averaged mass outflow rate in simulation UVIR-4e47 within the first $20 \, \rm Myr$ of quasar radiation injection. Solid (dashed) lines give \emph{total} mass outflow rates considering gas with $v_{\rm r} > 300 \, \rm km \, s^{-1}$ ($v_{\rm r} > 500 \, \rm km \, s^{-1}$). Shaded regions give outflow rates only for gas with temperature $T < 3 \times 10^4 \, \rm K$ and $v_{\rm r} > 300 \, \rm km \, s^{-1}$. \emph{Right:} cumulative outflow mass profile as a function of radius at various times after the onset of radiation injection. After sweeping up $\approx 3 \times 10^{10} \, \rm M_\odot$, mass is gradually subtracted from the outflow.}
\label{fig_outflowrate}
\end{figure*}

The ability of IR radiation pressure to suppress the local gas density becomes clearer once we consider somewhat higher quasar optical/UV luminosities.
 We show the density probability distribution in Fig.~\ref{fig_densmax} for simulation UVIR-4e47 (in light red). The mass of low velocity gas drops to $9.5 \times 10^{10} \, \rm M_\odot$ (less than a factor 2 decrease), but the peak of the probability distribution shifts by almost an order of magnitude.
The mass of low velocity gas is comparable to that in simulation UV-1.2e48 (which has single-scattering radiation only, but a luminosity 3 times higher), but gas density is not suppressed nearly as efficiently in the latter.
We have also verified that suppression in gas density occurs also in simulations UVIR-4e47-duty and UVIR-4e47-dust, suggesting that our conclusions are unlikely to change as long as IR radiation can couple efficiently to interstellar gas.

The suppression of star formation seen in our simulations that follow IR multi-scattering has two main origins: (i) gas ejection in an outflow and (ii) local suppression of gas density due to internal pressurisation by trapped IR radiation. The second mechanism has the potential to lead to low star formation efficiencies without necessarily removing the gas.

\subsection{Outflow properties: mass outflow rates, kinetic luminosity and momentum flux}
\label{sec_outflowrate}

Key outflow observables include the mass outflow rate, kinetic luminosity and momentum flux.
For outflows driven by AGN, the highest mass outflow rates can reach $10^3 \-- 10^4 \, \rm M_\odot \, yr^{-1}$, though a wide span of outflow rates, which can be as low as $1 \-- 10 \, \rm M_\odot \, yr^{-1}$, is seen even at high AGN luminosities \citep{Fiore:17}. 

In observational studies, kinetic luminosities and momentum fluxes are derived from the mass outflow rate and outflow velocity estimates as $\dot{E}_{\rm kin} \, = \, 1/2 \dot{M}_{\rm out} v^2_{\rm out}$ and $\dot{P} \, = \, \dot{M}_{\rm out} v_{\rm out}$, respectively.
The highest kinetic luminosities appear to reach a few percent of the AGN bolometric luminosity \citep[e.g.][]{Cicone:14, Zakamska:16, Bischetti:17, Bae:17, Fiore:17}, though significantly lower values on the order of $10^{-4}\-- 10^{-3} L_{\rm bol}$ are also often measured \citep[e.g.][]{Husemann:16}.
Similarly, momentum fluxes as high as $\gtrsim 20 L_{\rm bol} / c$ have been reported \citep[e.g.][]{Cicone:14, Tombesi:15, Rupke:17}, though, in other cases, values on the order of $L_{\rm bol} / c$ or lower are inferred \citep[e.g.][]{Cicone:15, Veilleux:17}.

In this section, we measure these various outflow properties in our simulations of radiation pressure-driven outflows.

\subsubsection{Mass outflow rates}
 
There are multiple ways in which outflow and momentum rates as well as kinetic luminosities are evaluated in the existing literature.
Strictly speaking, given a flux $\mathbf{\mathcal{F}} \, = \, Q \mathbf{v}$ for some density $Q \, = \, dq/dV$, e.g. the gas density, the rate at which $q$ flows across some surface $\mathcal{S}$ is $\int_{\mathcal{S}}{Q \mathbf{v}} \cdot d\mathbf{A}$, where $\mathbf{v}$ is the flow velocity and $d \mathbf{A}$ is the vector area.
For a spherically symmetric flow bounded by a spherical surface with radius $R$, this would be $4 \pi R^2 Q v_{\rm r} (R)$.
However, it is common in both observational and theoretical studies to evaluate outflow rates in an integrated sense.
For instance, \citet{Cicone:15} and \citet{Carniani:15} identify individual outflowing clouds, estimate their mass $M$, velocity $v$, and (projected) radial distance to the quasar $R$ and measure the outflow rate by summing $M v / R$, where $v/R$ is the (inverse) flow time, over all such clouds.
In this study, we apply and compare both methods.

We begin by examining the spherically averaged outflow rate $\langle \dot{M} \rangle_{\rm 4\pi}$.  
We generate a radial grid centred on the halo centre, with logarithmic radial bins spaced at about 0.06 dex in the range $(0.6 \-- 20) \, \rm kpc$.
For each shell, we compute $\langle \rho v_{\rm r} \rangle_{\rm 4\pi} \, = \, V_{\rm shell}^{-1} \sum_{i} m_{i} v_{i, \rm rad}$, where $V_{\rm shell}$ is the volume of the shell, $m_{i}$ the cell gas mass and $v_{i, \rm rad}$ the gas cell radial velocity within the shell, respectively, and obtain the spherically averaged mass outflow rate as  $\langle \dot{M} \rangle_{\rm 4\pi} \, = \, 4 \pi \langle \rho v_{\rm r} \rangle_{\rm 4\pi} R^2$, where $R$ is the radius of the radial bin.
The sum is performed over all gas elements with radial velocities $> 300 \, \rm km \, s^{-1}$, but we also investigate the effect of raising this threshold to a radial velocity $> 500 \, \rm km \, s^{-1}$.

We show the resulting outflow rate profiles on the left-hand panel of Fig.~\ref{fig_outflowrate} for simulation UVIR-4e47 at various times, using solid lines for the velocity threshold of $300 \, \rm km \, s^{-1}$ and dashed lines for $500 \, \rm km \, s^{-1}$.
We focus on the first $17 \, \rm Myr$ of outflow evolution, in the regime in which our simulations are least sensitive to the assumption of long quasar lifetimes and constant dust-to-gas ratio\footnote{We recall that both the simulation performed with a variable quasar lightcurve and with a temperature cut-off for the dust opacity yield similar results during the first few Myr.}. 

Considering our lower velocity cut (solid lines), the outflow rate $\langle \dot{M} \rangle_{\rm 4\pi}$ initially rises, peaking at about $10^4 \, \rm M_\odot \, yr^{-1}$ after $2 \, \rm Myr$ of AGN activity, when the shell-like outflow\footnote{The peak at higher radius is associated with a satellite fly-by.} has a radius of $1 \-- 2 \, \rm kpc$.
Thereafter, the outflow rate declines rapidly within short timescales of $\sim 1 \, \rm Myr$.
We first see the outflow rate drop to $\lesssim 5000 \, \rm M_\odot \, yr^{-1}$ at radii $R \approx 3 \, \rm kpc$ and, shortly thereafter, down to $\sim 1000 \, \rm M_\odot \, yr^{-1}$ at radii $R \approx 5 \-- 10\, \rm kpc$.
This trend in fact persists out to later times in this simulation. We find outflow rates on the order of $\sim 100 \, \rm M_\odot \, yr^{-1}$ at scales of $\approx 40 \, \rm kpc$ at $z \, = \, 6.2$ (not shown in Fig.~\ref{fig_outflowrate}).
It follows that only a few percent of the gas ejected from the central galaxy actually reaches the virial radius.

The outflow rates are lower if we consider a higher radial velocity threshold of $> 500 \, \rm km \, s^{-1}$, as shown by the dashed lines on the left-hand panel of Fig.~\ref{fig_outflowrate}. The peak outflow rate drops to $\approx 2600 \, \rm M_\odot \, yr ^{-1}$, suggesting that, at $\rm kpc$ scales, most of the mass is driven out at a relatively slow speed in the range $300 \-- 500 \, \rm km \, s^{-1}$. Most of the outflow that makes it out of the central few kpc is fast and therefore, the outflow rates for high and low velocity thresholds converge at at later times ($\Delta t \, \approx \, 10 \, \rm Myr$). From the right-hand panel of Fig.~\ref{fig_outflowrate}, it is also clear that the mass in fast-outflowing material does not decrease as sharply as for outflow gas with $v_{\rm r} < 300 \, \rm km \, s^{-1}$, as it has a higher chance of moving out to large radii.

The shaded regions in Fig.~\ref{fig_outflowrate} give the mass outflow rates of gas with temperature $T < 3 \times 10^4 \, \rm K$.
The cool outflow component can here be seen to dominate the mass budget of the outflow, though it starts to become less important at large radii $\approx 7 \, \rm kpc$, when it moves to lower density regions. 
In Section~\ref{sec_thermalprop_outflow}, we show that the cool phase forms through a thermal instability in the shocked outflowing medium.

\begin{figure*}
\centering 
\includegraphics[scale = 0.43]{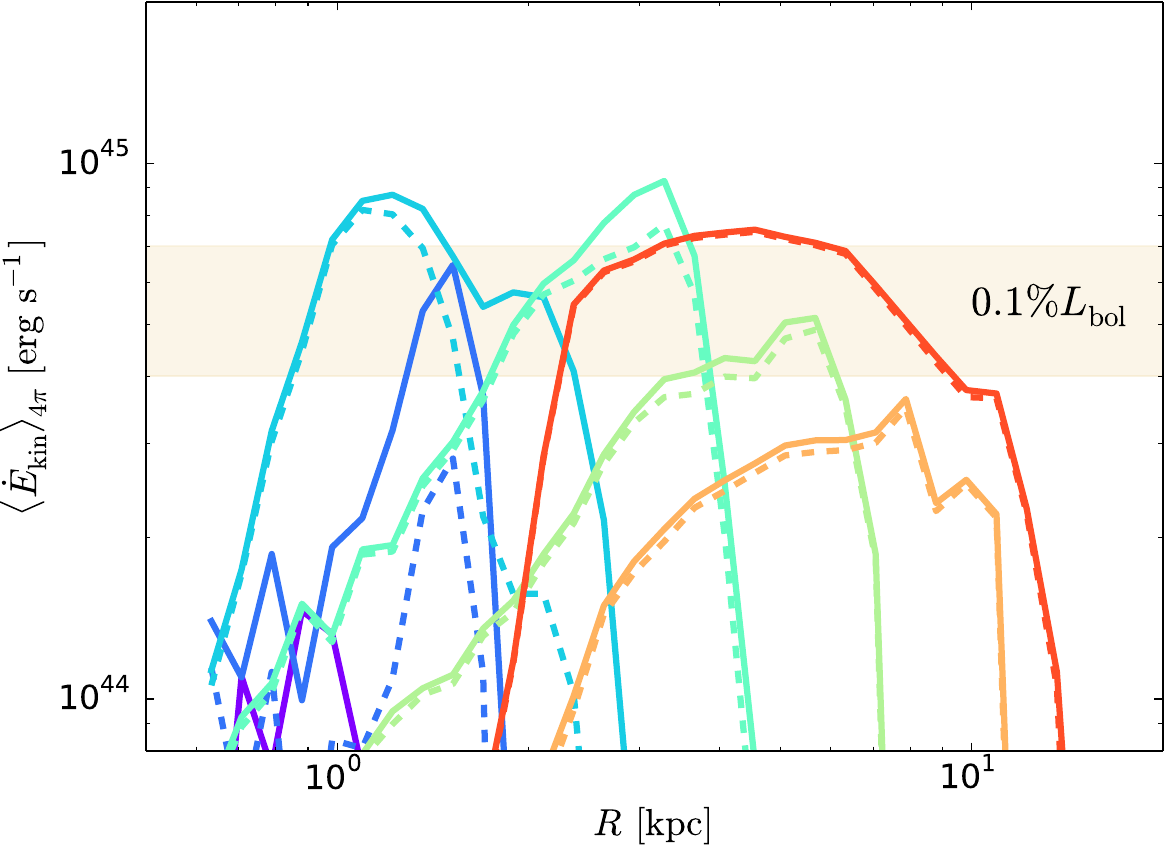}
\includegraphics[scale = 0.43]{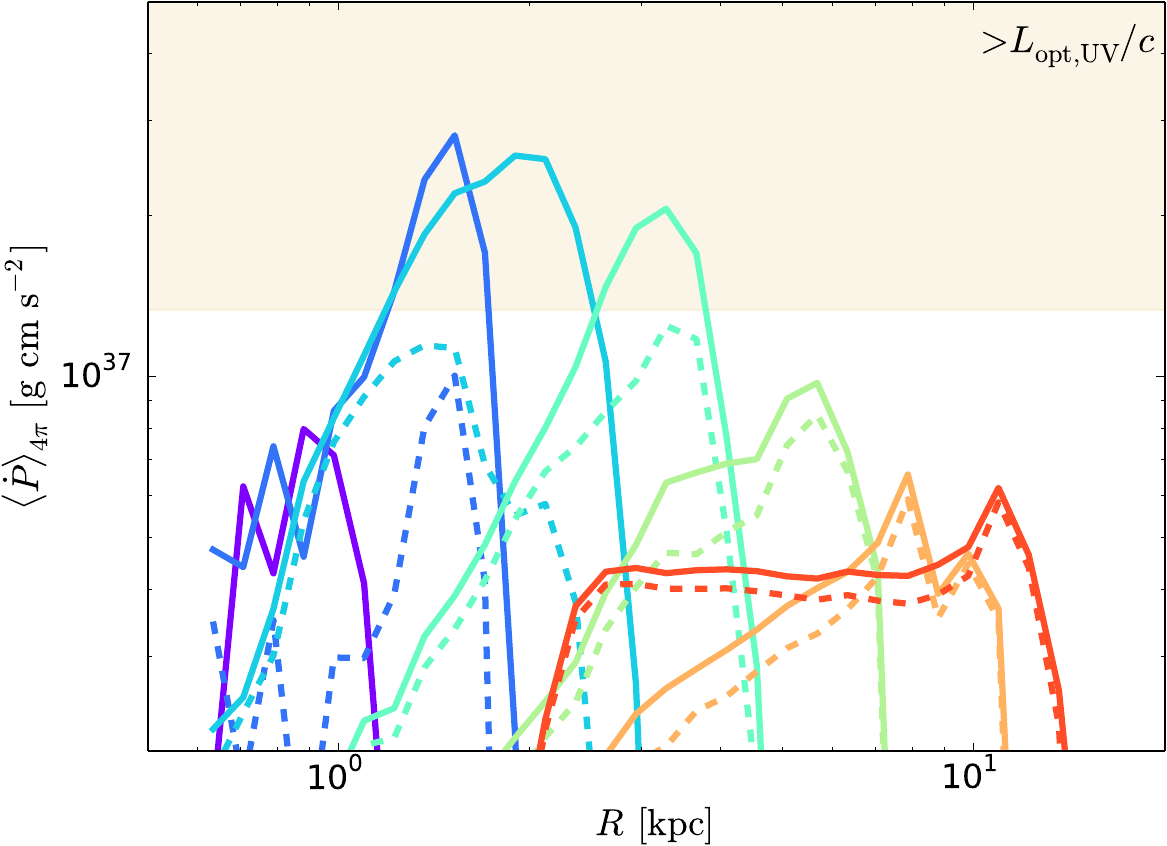}
\caption{\emph{Left:} spherically averaged kinetic luminosity in simulation UVIR-4e47 within the first $20 \, \rm Myr$ of quasar radiation injection. The shaded region gives the level corresponding to $0.1 \%$ conversion efficiency between quasar- and kinetic luminosity. \emph{Right:} spherically averaged `momentum flux' as a function of radius. The shaded region has $\langle \dot{P} \rangle_{\rm 4 \pi} > L_{\rm UV,opt}/c$ and indicates that the outflow is moderately boosted for a duration of $\approx 5 \, \rm Myr$.}
\label{fig_outflowekin}
\end{figure*}

The declining mass outflow rates indicate that the radiatively-driven outflow is not mass-conserving.
The left-hand panel of Fig.~\ref{fig_outflowrate} shows that, besides declining with radius, the mass outflow rate profiles become broader, as some of the expelled gas lags behind the fastest outflow component. 
On the right-hand panel, we show the enclosed outflowing mass (with $v_{\rm out} > 300 \, \rm km \, s^{-1}$) as a function of radius for the same simulation times as shown on the left-hand panel.
The outflowing mass initially increases, as mass is swept-up, reaching its maximum of $\approx 3 \times 10^{10} \, \rm M_\odot$ when the outflow is located at a radius $\approx 2 \, \rm kpc$.
The enclosed mass, however, drops as the outflow moves out to larger radii, implying that a significant mass ($\gtrsim 10^{10} \, \rm M_\odot$) of outflowing gas slows down to radial velocities lower than $300 \, \rm km \, s^{-1}$.
The outflow rates therefore drop with radius (and time) as high velocity gas is subtracted from the outflow.

We have also checked how the outflow rates seen here vary with quasar luminosity.
We did not perform simulations with outputs spaced at $\sim \, \rm Myr$ intervals as discussed in this section for all our models and we therefore focus on results at $z \, = \, 6.4$.
The peak outflow rate scales with quasar luminosity and ranges from $<550 \, \rm M_\odot \, yr^{-1}$ in simulation UVIR-2e47 to $\gtrsim 1500 \, \rm M_\odot \, yr^{-1}$ in UVIR-5e47. The peak outflow rate drops to $\approx 660 \, \rm M_\odot \, yr^{-1}$ in simulation UVIR-4e47-dust, though higher outflow rates are seen at earlier times, but they rise to $\approx 1390 \, \rm M_\odot \, yr^{-1}$ in thermal-4e47.

\subsubsection{Kinetic luminosity and momentum flux}

We now measure the kinetic luminosity and momentum flux of the simulated radiation pressure-driven outflows.
Specifically, we compute $4 \pi \langle \frac{1}{2} \rho v_{\rm r}^3 \rangle_{\rm 4\pi} R^2 \, = \, 2 \pi V_{\rm shell}^{-1} \sum_{i} m_{i} v_{i, \rm rad}^3$ for the kinetic luminosity and $4 \pi \langle \rho v_{\rm r}^2 \rangle_{\rm 4\pi} R^2 \, = \, 4 \pi V_{\rm shell}^{-1} \sum_{i} m_{i} v_{i, \rm rad}^2$ for the momentum flux.
We plot the radial profiles of these two quantities in Fig.~\ref{fig_outflowekin}, again considering the radial velocity cuts of $v_{\rm r} > 300 \, \rm km \, s^{-1}$ (solid lines) and $v_{\rm r} > 500 \, \rm km \, s^{-1}$ (dashed lines) when performing the sum.

We find maximum kinetic luminosities on the order of $10^{45} \, \rm erg \, s^{-1}$.
These are weakly sensitive to the velocity cut we consider due to the steep scaling of kinetic luminosity with velocity $\propto v^{3}_{\rm r}$, except at early times, when most of the outflow is still relatively slow.
Given the quasar optical/UV luminosity of $4 \times 10^{47} \, \rm erg \, s^{-1}$, or $L_{\rm bol} \approx 7 \times 10^{47} \, \rm erg \, s^{-1}$, the kinetic luminosities measured here imply an energy conversion efficiency of $\approx 0.14\% L_{\rm bol}$, comparable to observational constraints of quasar outflows \citep[e.g.][]{Rupke:17}, though lower by about an order of magnitude than the most powerful AGN-driven outflows.
Similarly, we find moderately boosted (maximal) momentum fluxes on the order of $\approx 3 L_{\rm opt,UV} / c \, \approx \, 2 L_{\rm bol} / c$ comparable to what is found in some AGN-driven outflows \citep[e.g.][]{Veilleux:17}, but certainly lower than the high values $\approx 20 L_{\rm bol}/c$ sometimes measured \citep[e.g.][]{Cicone:14}. 
The momentum boosts we measure are lower than $L_{\rm opt,UV} / c$ if we consider the fastest portion of the radiatively-driven outflow (dashed lines), indicating that this quantity is strongly dependent on the precise velocity cut taken when defining an outflow. We would be led to conclude that IR trapping does not generate a momentum boost if we were to only consider the fastest outflowing component, whereas we would conclude the opposite if we were to include the slower component into our consideration. Based on this analysis, we suggest that a substantial fraction of the AGN radiation momentum is transferred to a slower outflow component with a speed which may make it difficult to distinguish from gas undergoing random gravitational motion, greatly complicating the task of observationally evaluating the coupling efficiency of AGN feedback.

\begin{table}
\centering
\begin{tabular}{c*{5}{c}r}
\hline
$\Delta t$    & $\dot{M}$ & $\dot{P}_{\rm out} / (L_{\rm opt,UV} / c)$ & $\dot{E}_{\rm kin} / L_{\rm opt,UV}$\\
$\rm [Myr]$ & $\rm [M_\odot \, yr^{-1}]$ &                                       &  \\
\hline
$1$   & $3394$  & $0.80$ & $0.10 \%$\\
$2$   & $6709$  & $2.23$ & $1.08 \%$\\
$3$   & $7223$  & $1.72$ & $0.22 \%$\\
$7$   & $3870$  & $1.17$ & $0.21 \%$\\
$10$ & $2259$  & $0.78$ & $0.15 \%$\\
$14$ & $1557$  & $0.56$ & $0.12 \%$\\
$17$ & $1643$  & $0.74$ & $0.28 \%$\\
\hline
\end{tabular}
\caption{Many observational studies compute outflow rates, momentum fluxes and kinetic luminosities in an integrated sense, by adding up e.g. for the mass outflow rate $\sum_{i} M_i v_i / R_i$ for all outflowing components (e.g. different outflowing clouds). We show the result of this calculation applied to our simulations in this table. The values listed here are comparable to the maxima of the curves shown in Figs.~\ref{fig_outflowrate} and~\ref{fig_outflowekin}.}
\end{table}
\label{table2}

High outflow kinetic luminosities and momentum fluxes, however, last only in the $\lesssim 10 \, \rm Myr$ phase in which the outflow sweeps up mass at a high rate. The timescale during which the outflow has high kinetic luminosity $\gtrsim 0.1 \% L_{\rm bol}$ and a `momentum boost' $> L_{\rm bol}/c$ is only on the order of $5 \, \rm Myr$.
After the radiatively-driven outflow breaks out of the quasar hosting galaxy, at scales of a few kpc, the kinetic luminosity drops to values below $0.1 \% L_{\rm bol}$ and the momentum flux declines to $< L_{\rm bol} / c$. 
Based on our simulations, we thus suggest that the most extended radiatively-driven outflows typically have $\dot{E}_{\rm kin} \lesssim 0.1 \% L_{\rm bol}$ and $\dot{P}_{\rm out} \lesssim L_{\rm bol} / c$.
One of the most spatially extended outflows detected to date is the [CII] emitting cold wind seen by \citet{Maiolino:12} and \citet{Cicone:15}, where the emission associated with fast moving gas is shown to be distributed over some $30 \, \rm kpc$. Intriguingly, \citet{Cicone:15} measure a (projected) momentum flux of $\dot{P}_{\rm out} \, \approx \, L_{\rm bol}/c$ and a kinetic luminosity of $\dot{E}_{\rm kin} \, \approx \, 0.16 \% L_{\rm bol}$. Both these values, as well as the outflow rate $\approx \, 1400 \, \rm M_\odot \, yr^{-1}$ are in good agreement with the values measured in the later stages of our simulations, when the outflow is most extended.
We have also verified that the high mass outflow rates, momentum fluxes and kinetic luminosities obtained in UVIR-4e47 persist in simulation UVIR-4e47-dust during the first $\approx 20 \, \rm Myr$ of radiation injection. At this time, there is a large mass of cold gas surrounding the black hole particle and radiation couples efficiently (see also Fig.~\ref{fig_sfr}). It is at late times, when the cold gas reservoir has been cleared out that differences between UVIR-4e47 and UVIR-4e47-dust start becoming relevant.

For UVIR-4e47-duty, the quasar luminosity is high and close to the maximum value during the first $20 \, \rm Myr$ of radiation injection, i.e. that simulation is essentially identical to UVIR-4e47 during the time period investigated in this section. It is interesting that at later times, when the outflow still propagates through the halo but the quasar luminosity has dropped to $\approx 0.02$ its initial value, the peak kinetic luminosities and momentum fluxes appear to be high when normalised to the instantaneous quasar luminosity. At $z \, \approx \, 6.3$, we find a peak kinetic luminosity of $\approx 0.2 \% L_{\rm opt, UV}$ and momentum flux $\approx L_{\rm opt, UV}/c$ higher than in UVIR-4e47 at the same redshift. Such `fossil outflows' misleadingly appear to retain more of the quasar's energy and momentum just by virtue of quasar variability.

We conclude this section by comparing the outflow rates, the momentum fluxes and kinetic luminosities seen in the radial profiles of Figs.~\ref{fig_outflowrate} and ~\ref{fig_outflowekin} with values obtained by computing the sum $\sum_i{Q_i / t_{i \, \rm flow}}$, where $Q \, = \, M \, , Mv_{\rm r}\, , 1/2 M v_{\rm r}^2$, respectively. We perform the sum over all gas cells within the virial radius\footnote{We verify that the results are insensitive to the choice of the outer radius as long as it encloses all of the outflow.} and with a velocity\footnote{The values shown in Table~\ref{table2} increase only by $15 \%$ and only in the first Myr if a this condition was changed to $v_{\rm r} > 100 \, \rm km \, s^{-1}$.} $v_{\rm r} > 300 \, \rm km \, s^{-1}$ and list the results in Table~\ref{table2}. 
The resulting mass outflow rates, momentum fluxes and kinetic luminosities are comparable to the maxima of the radial profiles shown in Figs.~\ref{fig_outflowrate} and ~\ref{fig_outflowekin} and their trend with time is the same.
Since our mass outflow profiles are sharply peaked, it is no surprise that both methods return comparable values.
However, for flatter outflow rate profiles, adding up the contribution from outflowing components at different radii should bias the resulting rates to unphysically high values.

\subsection{Multi-phase structure of radiation pressure-driven outflows}
\label{sec_thermalprop_outflow}

The properties of observed AGN-driven outflows appear to be very diverse.
Such outflows have been observed to transport hot, warm ionised, neutral as well as cold atomic and molecular gas components \citep[see e.g.][]{Fiore:17}.
The origin of this multi-phase structure and particularly the cold component, however, remains highly debated.
There are two prevailing scenarios for the existence of fast cold gas at kpc scales: (1) cold clouds are ejected from the galaxy through ram pressure from a hot wind or directly through radiation pressure \citep{Klein:94, Murray:11, Scannapieco:15, McCourt:15, Schneider:17} and (2) the shocked outflowing component cools down radiatively and forms cold clouds in situ \citep[see e.g.][]{Zubovas:14, Costa:15, Thompson:16, Ferrara:16, Scannapieco:17, Richings:17}.
Our simulations are too simplistic to test either of these scenarios in great detail due to resolution limitations, the absence of a cold ISM phase with $T < 10^4 \, \rm K$ and simplified cooling physics (i.e. no metal-lines).
However, since we have seen that the simulated outflows do appear to contain gas at a range of temperatures, it is useful to investigate how this multi-phase structure comes about.

\subsubsection{Characteristic temperature scales}

\begin{figure*}
\centering 
\includegraphics[scale = 0.32]{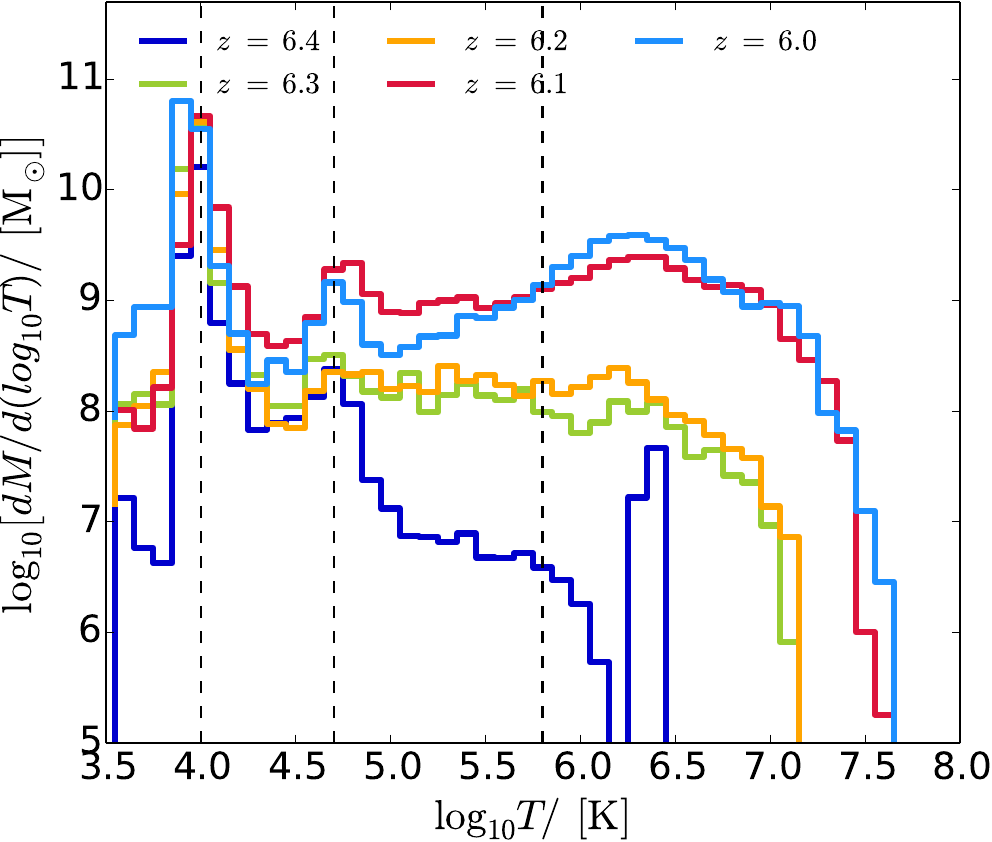}
\includegraphics[scale = 0.36]{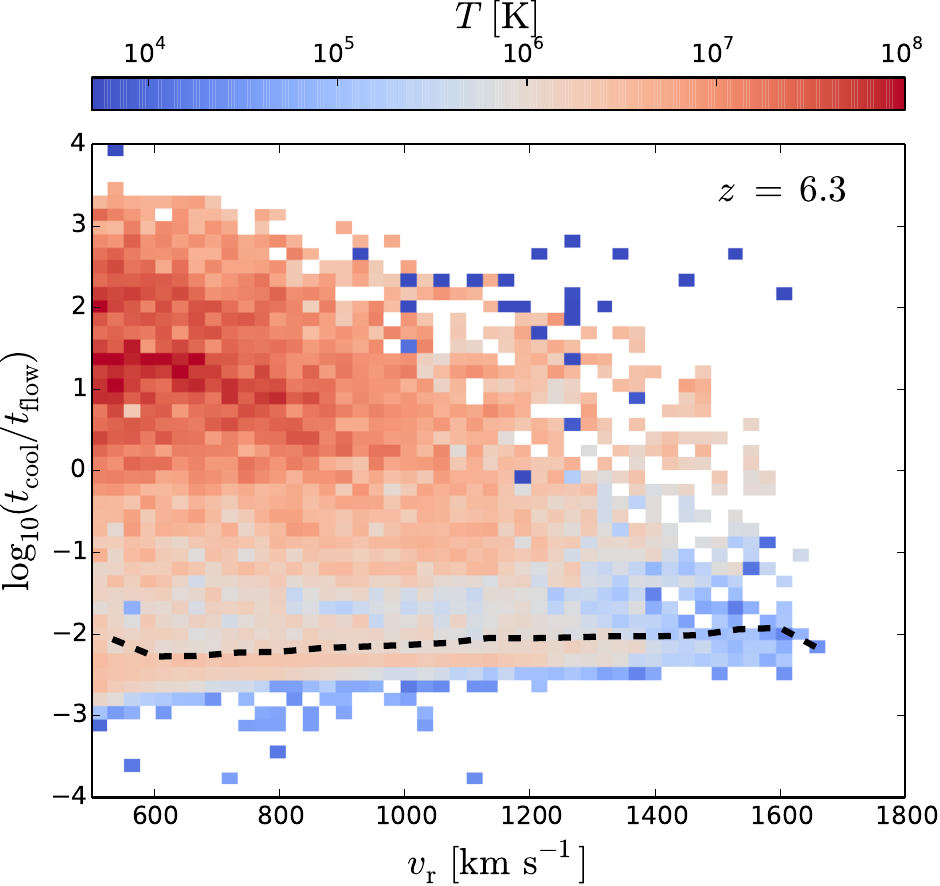}
\includegraphics[scale = 0.32]{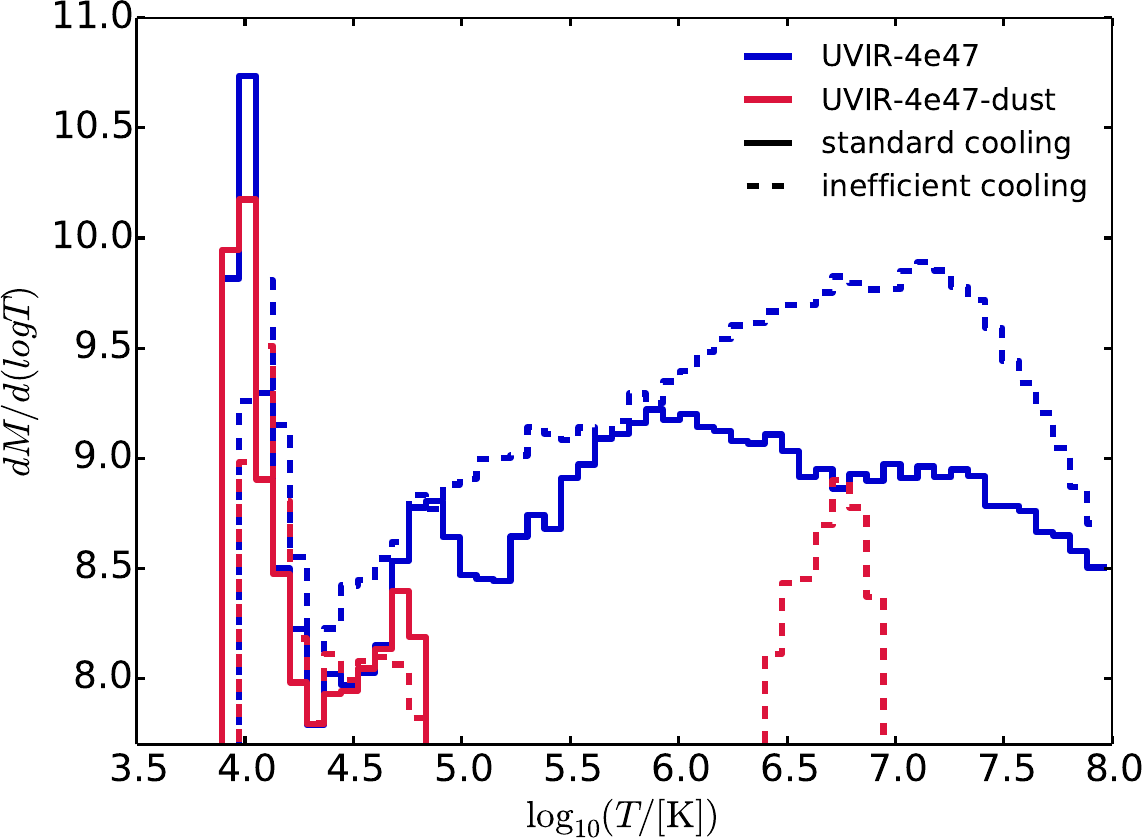}
\caption{\emph{Left:} The mass distribution of the outflow shown as a function of temperature. Only gas with radial velocity $> 500 \, \rm km \, s^{-1}$ is considered. The bulk of the outflowing gas is in the `cool phase' with temperatures $\lesssim 3 \times 10^4 \, \rm K$. There are two characteristic temperatures at which the outflowing mass accumulates. These characteristic temperatures correspond to local minima (vertical dashed lines) in the radiative cooling function. \emph{Middle:} Logarithmic ratio of cooling to outflow expansion times (`flow times') as a function of outflow radial velocity, colour-coded according to mean temperature per bin, at $z \, = \, 6.3$. The dashed line gives the mass-weighted median relation. The bulk of the outflowing material radiates away its thermal energy and becomes cold within a flow time. \emph{Right:} The relative mass contribution of different gas phases and $v_{\rm r} > 500 \, \rm km \, s^{-1}$ as a function of gas temperature in our UVIR-4e47 (blue lines) and UVIR-4e47-dust (red lines) simulations as well as in identical simulations performed with an inefficient cooling function in which the cooling rates were uniformly decreased by a factor of 10 (dashed curves), all at $z \, = \, 6.4$.}
\label{fig_outflowprops}
\end{figure*}

We start by taking a closer look at the temperature distribution of the outflows generated in our simulations.
We focus on simulation UVIR-4e47-dust, where the dust opacity temperature cut-off treats the inefficient coupling between radiation and hot, dust-free gas more realistically.
The temperature distribution of the outflows launched in simulation UVIR-4e47-hires is explored in Appendix~\ref{sec_convergence}.

In Fig.~\ref{fig_outflowprops}, we plot the outflow mass distribution as a function of gas temperature (left-hand panel).
In producing these plots, we select only gas with radial velocity $> 500 \, \rm km \, s^{-1}$, which is approximately the maximum radial velocity reached by gas in an identical simulation performed without AGN feedback, and gas within the virial radius of the galactic halo.
The dominant outflowing component has $T \approx 10^4 \, \rm K$, placing it at the bottom of the adopted cooling function.
Since we have neglected metal-line and molecular cooling, it is possible that this outflow component could be even more massive and colder than seen in our simulations.
In Appendix~\ref{sec_convergence}, we show the temperature distribution of the outflow seen here to be robust with increasing numerical resolution.

Apart from the dominant cold component at $T \approx 10^4 \, \rm K$, there are two other characteristic outflow temperatures, seen as local maxima on the left-hand panel of Fig.~\ref{fig_outflowprops}.
A significant portion of the outflow has $T \approx 10^{4.7} \, \rm K$, corresponding to the dip in the cooling function just below the temperature at which Helium cooling is efficient, while another important component has  $T \approx 10^6 \, \rm K$, though this becomes significant only at late times.
The black dashed lines on the left-hand panel of Fig.~\ref{fig_outflowprops} show the temperatures at which the cooling function has local minima.
The three different outflow components we have identified trace these lines very closely, suggesting that the outflowing gas preferentially has a temperature such that radiative cooling proceeds slowly.
For temperatures $\approx 10^6 \, \rm K$, note that (i) the associated dip in the cooling function is very broad and (2) the cooling times become comparable to the characteristic outflow times, such that the agreement between the temperature peak seen in this simulation and the right-most dashed line is less close.
Cooling of outflowing material can only be expected if the characteristic outflow expansion time (`flow time', in short) is longer than the cooling time of the gas.
In the following, we show that this condition is indeed satisfied for the bulk of the outflowing material.

\subsubsection{Cooling- vs. outflow expansion timescales}

Na\"ively, outflowing gas is expected to cool radiatively if its cooling time is short compared to its characteristic outflow expansion time \citep[see e.g.][for recent studies]{Zubovas:14, Costa:15}.
Here, we define the cooling time as 
\begin{equation}
t_{\rm cool} \, = \, \frac{3/2 n k_{\rm B} T}{n_{\rm H}^2 \Lambda (T)} \, ,
\end{equation}
where $\Lambda (T)$ is the primordial, non-equilibrium cooling rate adopted in our simulations.
The outflow expansion time, or simply the `flow time', is defined as 
\begin{equation}
t_{\rm flow} \, = \, \frac{R}{v_{\rm r}} \, .
\end{equation}

In the central panel of Fig.~\ref{fig_outflowprops}, we present a 2-dimensional histogram for the ratio $\log_{10}{ \left( t_{\rm cool} / t_{\rm flow} \right)}$ as a function of radial velocity at $z \,=\, 6.3$.
The colour shows the mean gas temperature per bin, as indicated by the colour bars at the top, and the dashed lines show the mass-weighted median values as a function of outflow radial velocity.
We find that there is little evolution in the redshift range\footnote{At early times $\Delta t \lesssim 10 \, \rm Myr$, however, the cooling vs. flow time ratios vary rapidly, as the gas is shock-heated and the cooling times are long.} for which radiation is emitted by the AGN, with the exception of the formation of the hot component with long cooling times, and we therefore choose to show results only for $z \, = \, 6.3$. 

In agreement with our expectations, as shown by the dashed curve in the central panel of Fig.~\ref{fig_outflowprops}, most outflowing gas has $t_{\rm cool} / t_{\rm flow} \ll 1$.
The bulk of the outflow should therefore be cooling efficiently.
The cooling portion of the outflow, in fact, accounts for $80\% \-- 97\%$ of the outflow mass and has a temperature in the range $\approx 10^{4} \-- 10^{6} \, \rm K$.
There is also a weakly mass-loaded hot component with temperatures $T \gtrsim 10^6 \, \rm K$, which remains hot and tenuous.
This component, however, amounts to, at most, $\approx 3\% \-- 20\%$ of the total outflowing mass, where $20\%$ applies at $z \, = \, 6$.
The cool gas with $T \approx 10^{4} \, \rm K$ seen at the top has long cooling times because its temperature has reached the bottom of the cooling function and is thus not allowed to cool radiatively.
We note that the temperature of $T \approx 10^{4} \, \rm K$ still corresponds to the maximum of the temperature probability distribution even if we do not introduce an opacity cut-off at high temperatures (see Fig.~\ref{fig_dmdt_res}), though the hot phase can then amount to $\approx 30 \-- 80 \%$ of the total outflow mass in the simulation, where $80\%$ applies at $z \, = \, 6$.

The presence of the hot weakly mass-loaded component may appear surprising given that radiation couples only to gas with  $T < 3 \times 10^4 \, \rm K$ in UVIR-4e47-dust.
This gas component is, indeed, cool and dense when it is initially accelerated, but is shock-heated.
When its temperature increases, it stops coupling to quasar radiation, which is why this phase generally traces only relatively low velocity $v_{\rm r} \lesssim 800 \, \rm km \, s^{-1}$ gas (see central panel of Fig.~\ref{fig_outflowprops}).

\subsubsection{The consequences of inefficient cooling}

If in-shock radiative cooling is the dominant mechanism for the production of fast outflowing gas in our simulations, then suppressing cooling should decrease the mass of cold outflowing material.
In this section we perform this second test.
We re-run simulations UVIR-4e47 and UVIR-4e47-dust starting from $z \,= \, 6.5$, when radiation injection begins, with a modified cooling function for which the normalisation of the total cooling rate is reduced by a factor of $10$.

In the right-hand panel of Fig.~\ref{fig_outflowprops} we plot the temperature distribution of the resulting outflows at $z \, = \, 6.4$.
As before, we select only gas within the virial radius of the parent galactic halo.
We find that reducing the cooling rate has a very strong impact on the temperature distribution of the outflow.
For UVIR-4e47 (shown with blue lines), we see that the outflow mass at $T \, \approx\, 10^4 \, \rm K$ drops by over one order of magnitude.
Conversely, the mass of hot gas, at  $T \, \approx\, 10^7 \, \rm K$ increases dramatically.

For UVIR-4e47-dust (shown with red lines), there is virtually no hot component (at this time) if cooling is efficient.
Inefficient cooling, however, prevents the shock-heated gas from radiating away its thermal energy, such that a significant portion of the outflow remains at high temperatures.
The cool outflow component that forms in this simulation is also less significant.

\begin{figure*}
\centering 
\includegraphics[scale = 0.3]{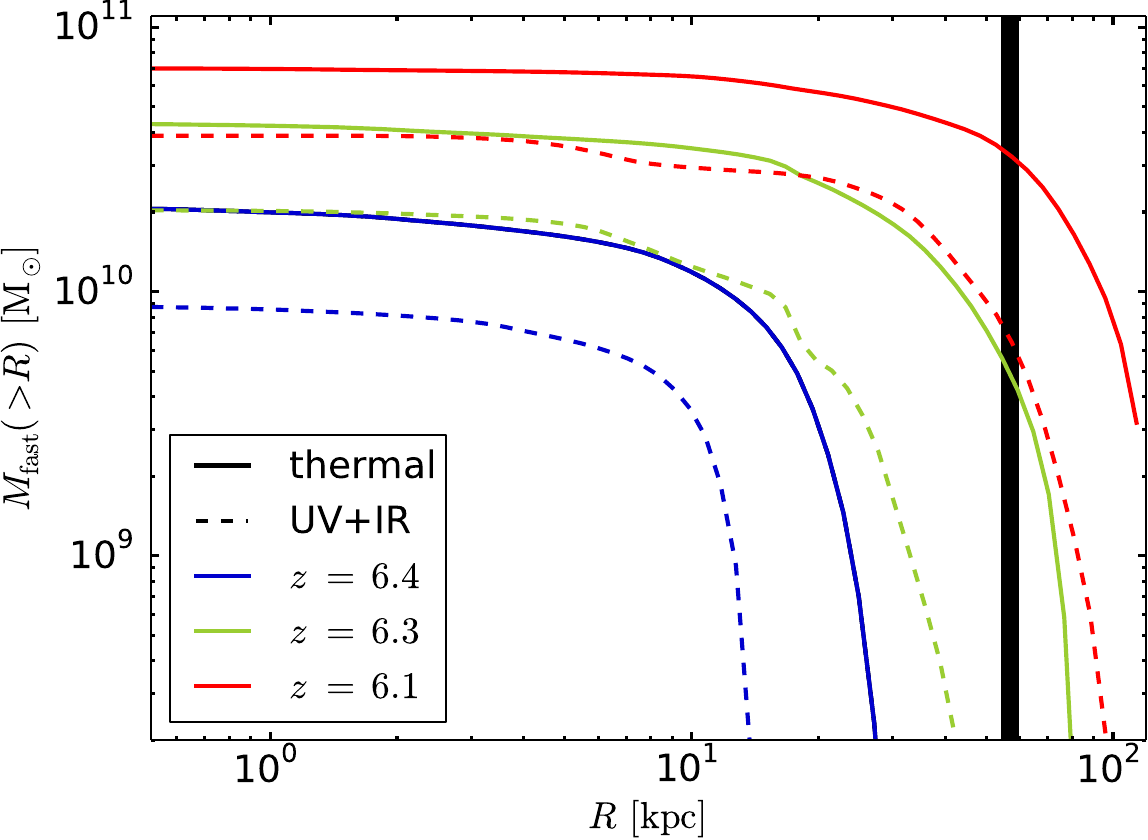}
\includegraphics[scale = 0.3]{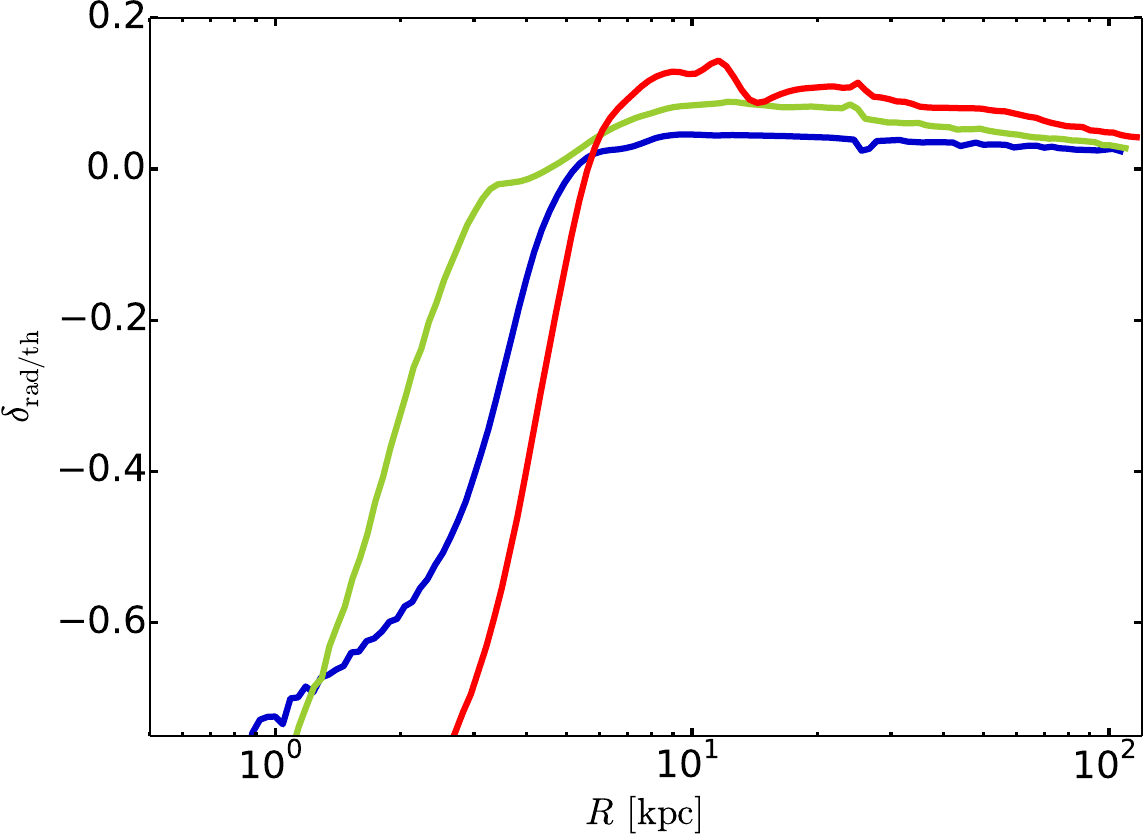}
\includegraphics[scale = 0.3]{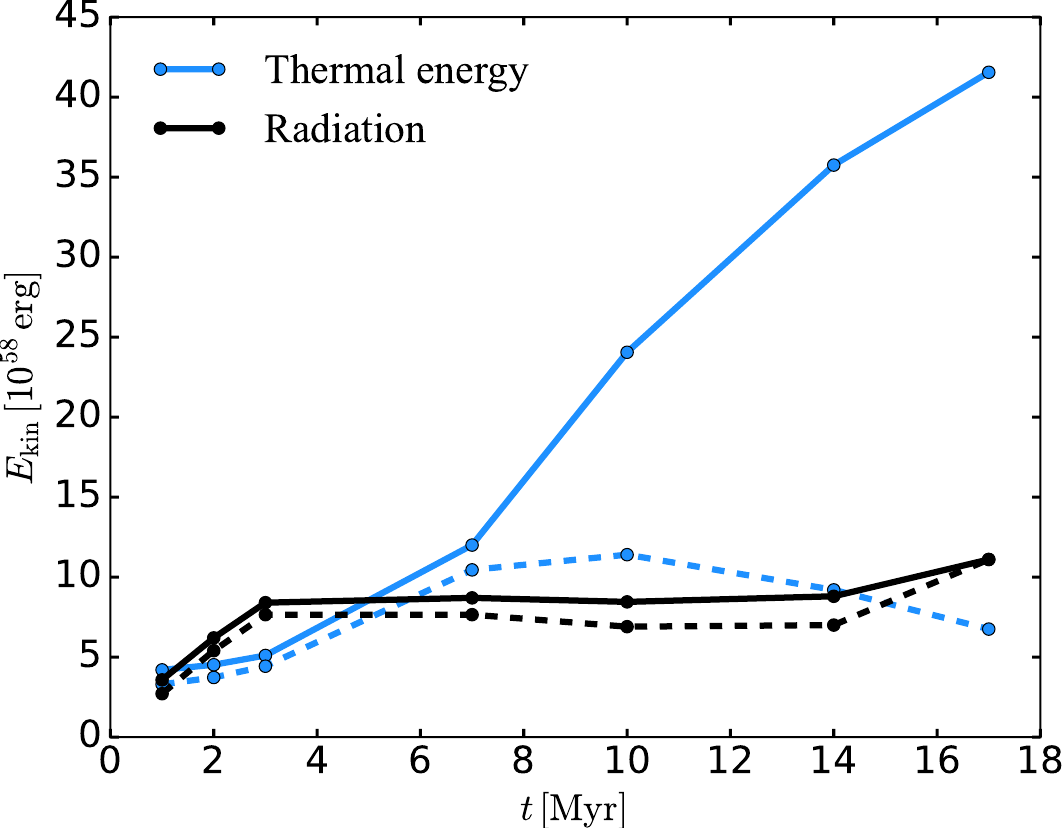}
\caption{\emph{Left:} Cumulative mass distribution for gas with $v_{\rm r} \geq 500 \, \rm km \, s^{-1}$ in the simulation with thermal feedback (solid lines) and the simulation with radiative feedback (dashed lines), plotted at different times. The vertical shade gives the range of the time-varying virial radius. \emph{Middle:} Difference in enclosed low velocity ($v_{\rm r} < 500 \, \rm km \, s^{-1}$) gas mass between the simulation with radiative feedback and that with thermal feedback. \emph{Right:} Kinetic energy of outflowing material with $v_{\rm r} > 300 \, \rm km \, s^{-1}$ in the simulation with thermal feedback (blue lines) and radiative feedback (black lines). The dashed lines give the kinetic energy if the low-density hot component is excluded (see text). At matching quasar luminosity, thermally-driven outflows are the most powerful, because they are faster and sweep-up larger amounts of mass. However, thermally-driven outflows preferentially sweep-up gas at larger radii, i.e. at $R \gtrsim 5 \, \rm kpc$, while radiation pressure is more efficient at removing gas from the central regions. Quenching operates differently in these two mechanisms: thermal energy expels gas from the halo, whereas multi-scattering radiation pressure launches dense gas from the galactic nucleus and suppresses the galaxy's density globally.}
\label{fig_comp}
\end{figure*}

We are forced to conclude that radiative cooling plays a fundamental role in regulating the amount of cool gas in radiation pressure-driven outflows.
It also plays a role in modulating the mass of inflowing material, which in turn affects how much ram pressure from inflowing gas the outflow needs to overcome \citep[see also][]{Costa:14}.
Since the cooling function we adopt in this study assumes primordial radiative cooling, we should think of the cool outflowing gas masses seen in our simulations as a lower limit of what might be achievable in the presence of metal-line cooling.
We also have neglected the ionisation field of the quasar. While we should expect our outflowing cool phase to become photoionised, its temperature should not much exceed $\approx 10^4 \, \rm K$. We have performed separate simulations in which we included photoionisation from a quasar using 5 radiation frequencies bins for IR, optical and UV radiation (three bands), finding no significant difference in the mass of `cool gas' as defined in this paper.
Quasar radiation may, however, have a significant effect on the metal-line cooling rates and on the ionisation of metals in the CGM  even after it has turned off \citep[e.g.][]{Segers:17}.

\subsection{Radiatively- vs. thermally-driven outflows}
\label{sec_comparison}

In this section, we compare the properties of outflows driven by radiation pressure to those launched through thermal energy injection.
We refer to the simulation with thermal feedback as `thermal-4e47' and compare it against simulation UVIR-4e47. 

In our simulations, continuous injection of thermal energy results in the formation of a hot, $T \, \approx \, 5 \times 10^9 \, \rm K$, bubble.
As it expands, this bubble initially pushes the surrounding gas approximately isotropically.
However, when its radius becomes comparable to the height of the quasar host disk, the bubble breaks out and escapes, propagating at speeds $> 1000 \, \rm km \, s^{-1}$ and taking on an increasingly anisotropic morphology \citep[see also][]{Costa:14}.
As it pushes into ambient halo gas, the expanding bubble drives a shock that sweeps across the circumgalactic medium, ultimately filling the halo (and its outer regions) with hot and tenuous gas at temperatures  $10^7 \-- 10^8 \, \rm K$, as has been illustrated in Fig.~\ref{fig_cosmofield} \citep[see also][]{Gilli:17}.

\begin{figure*}
\centering 
\includegraphics[scale = 0.35]{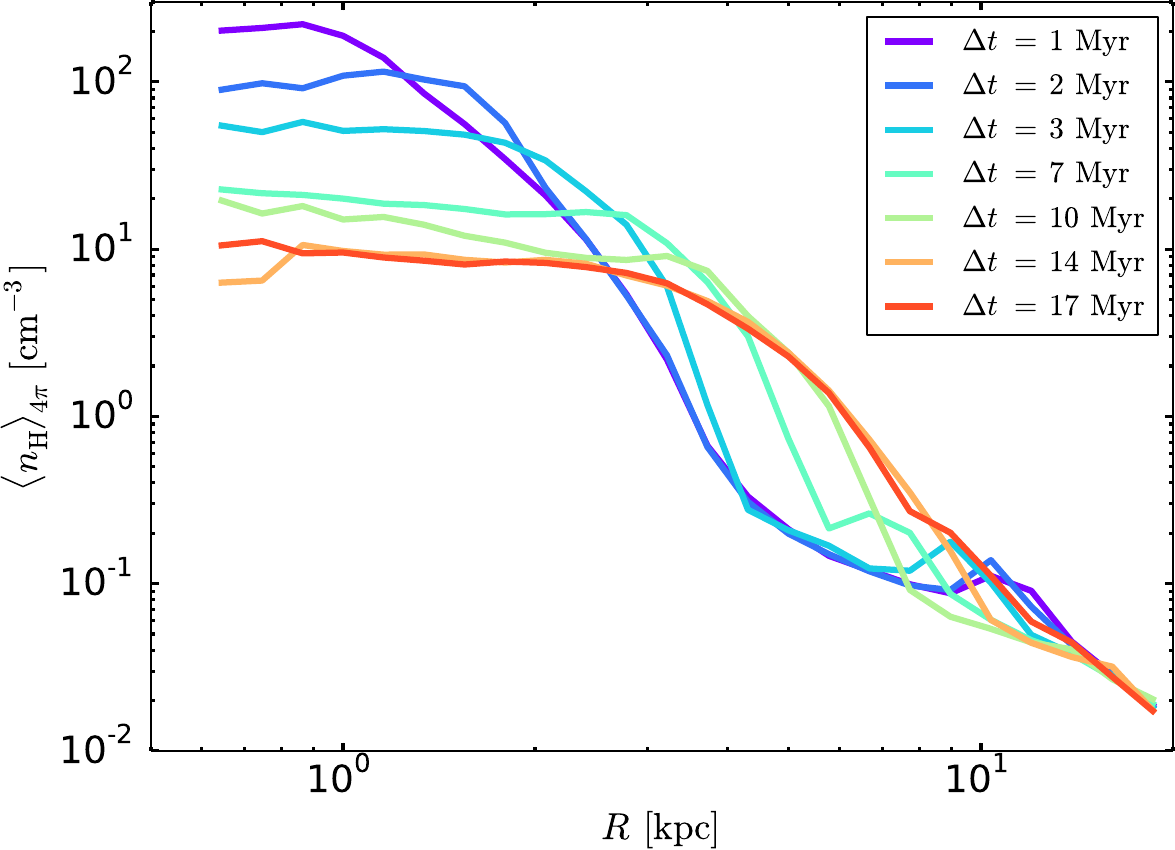}
\includegraphics[scale = 0.35]{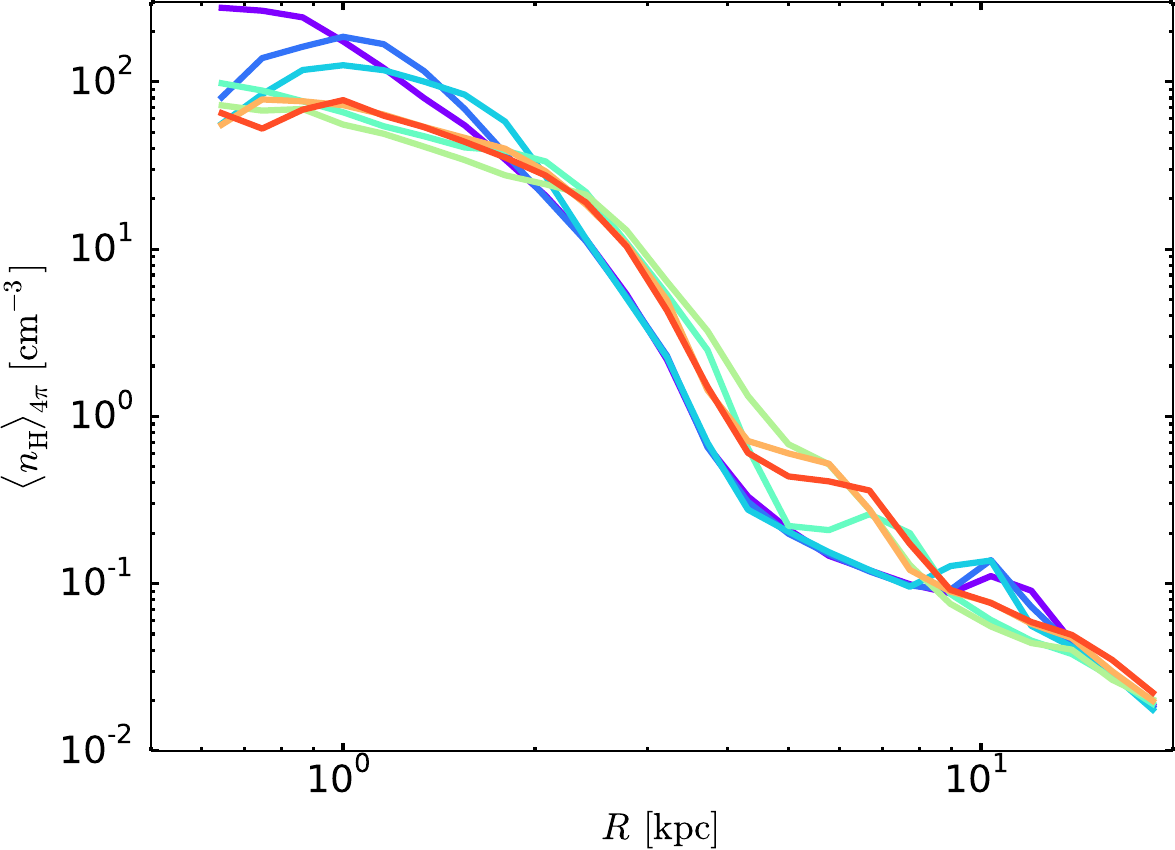}
\caption{Gas density profiles in the early stages of outflow evolution in the simulation with purely radiative feedback (left-hand panel) and with purely thermal feedback (right-hand panel). Radiation pressure suppresses the central gas density more efficiently than thermal energy, even if the latter results in the formation of hot bubbles with $T \, \approx \, 5 \times 10^9 \, \rm K$. While IR radiation diffuses and hence penetrates dense gas, thermal energy vents out of the galaxy. Thus, even if thermally-driven outflows are more energetic, they produce a weaker immediate impact on the host galaxy than IR radiation pressure.}
\label{fig_dens_comp}
\end{figure*}

The outflow propagates more quickly in the case of thermal energy injection than for radiation pressure\footnote{However, note that this is true given the feedback efficiency of $\epsilon_{\rm f} \, = \, 0.028$ adopted here. For lower feedback efficiencies, both results may become more similar.}, as is shown on the left-hand panel of Fig.~\ref{fig_comp}, where we plot the cumulative mass distribution, as $M(>R)$, of fast outflowing gas with $v_{\rm vrad} > 500 \, \rm km \, s^{-1}$ as a function of radius.
At any given time, thermally-driven outflows have propagated out to a larger distance and swept-up more mass than in the case of radiation pressure.
For instance, we see gas masses flowing out faster than $500 \, \rm km \, s^{-1}$ of $\approx (2 \-- 7) \times 10^{10} \, \rm M_\odot$ (within $2 \times R_{\rm vir}$) in the simulation with thermal feedback (solid lines), while we find $8 \times 10^{9} \-- 4 \times 10^{10} \, \rm M_\odot$ in the simulation with radiative feedback (dashed lines). 
It is also clear from Fig.~\ref{fig_comp} that some outflowing gas does reach the virial radius (which grows with time and is shown with a black shade). 
At $L_{\rm opt, UV} \, = \, 4 \times 10^{47} \, \rm erg \, s^{-1}$, this amounts to $\lesssim 3 \times 10^{10} \, \rm M_\odot$ in the simulation with thermal feedback and $\lesssim 5 \times 10^9 \, \rm M_\odot$ in the simulation with radiative feedback, less than $10 \%$ of the ejected gas\footnote{We caution, however, that if we perform our simulations for a longer time, it is possible for these values to rise as some outflowing material coasts through the halo.}.

While thermally-driven outflows displace the largest masses, the scales within which gas is removed differ between both feedback mechanisms.
The central panel of Fig.~\ref{fig_comp} shows the difference between the enclosed mass of low velocity gas as a function of radius in both simulations, at different redshifts.
Specifically, we plot the quantity 
\begin{equation}
\delta_{\rm rad/th} \, = \, \frac{M_{\rm rad}(<R) - M_{\rm th}(<R)}{M_{\rm rad}(<R) + M_{\rm th}(<R)} \, ,
\end{equation}
where $M_{\rm rad}(<R)$ is the cumulative mass in simulation UVIR-4e47 and $M_{\rm th}(<R)$ is the cumulative mass in simulation thermal-4e47. 
This quantity is computed for gas with an absolute radial velocity $|v_{\rm r}| < 500 \, \rm km \, s^{-1}$, in order to probe the presence of low velocity gas.
If $\delta > 0$, gas is more efficiently accelerated in simulation thermal-4e47; if  $\delta < 0$, gas is more efficiently accelerated in simulation UVIR-4e47.
There is a deficit of low velocity gas mass in the simulation with radiative feedback in the central $\approx 3 \, \rm kpc$ when compared to that performed with thermal feedback, i.e. radiation pressure accelerates gas in the central regions more efficiently than thermal feedback.
We also show radial density profiles in Fig.~\ref{fig_dens_comp} for UVIR-4e47 (on the left-hand panel) and thermal-4e47 (on the right-hand panel) within the first few $17 \, \rm Myr$ of AGN feedback. 
The density profiles show that the central density is indeed more efficiently suppressed with radiation pressure than with thermal energy injection.
Conversely, at large radii, thermal feedback is the more efficient mechanism when it comes to expelling material, which can be seen from the excess of low velocity gas ($\delta > 0$) in the simulation with radiative feedback in the central panel of Fig.~\ref{fig_comp}.
Finally note that the total mass of low velocity gas is higher in the simulation with feedback through radiation pressure, as can be noticed from its higher cumulative mass at high radii.
We conclude that, overall, thermal feedback ejects more material from the halo than radiation pressure.

We show the kinetic energy of outflowing gas (with $v_{\rm r} > 300 \, \rm km \, s^{-1}$) in the first $17 \, \rm Myr$ of AGN feedback on the right-hand panel for UVIR-4e47 (black lines) and thermal-4e47 (blue lines). Since they are fastest and most massive, outflows driven through thermal energy injection are the most energetic.
Note, however, that the kinetic energy of outflowing gas in thermal-4e47 only overtakes that of UVIR-4e47 after $\approx 5 \, \rm Myr$ of AGN feedback, when the hot bubble begins to expand into low-density material, converting its thermal energy into kinetic energy.
Conversely, in the case of radiation pressure, the kinetic energy of outflowing gas stalls after the initial outflow is launched and $\tau_{\rm IR} \approx 1$ (see Fig.~\ref{fig_tau_eff}).
If we exclude hot $T > 10^6 \, \rm K$, low-density $n_{\rm H} < 1 \, \rm cm^{-3}$ gas from our computation (dashed lines), we find that, while the kinetic energy drops only marginally in UVIR-4e47, it falls by factors $\gtrsim 2 \-- 4$ in thermal-4e47, indicating that the bulk of energy in thermally-driven outflows is contained in relatively low density gas. 

Intuitively, these findings make sense; IR radiation can only operate in IR thick, and hence highly dense material, and should become ineffective after the obscuring layers are removed and when propagating through low density gas in the halo.
However, since it can diffuse and penetrate dense gas, IR radiation has the ability to couple to a higher volume of the galactic gas.
The bubble generated by thermal energy deposition, on the other hand, leaks out, coupling to the lower density halo gas outside of the central galaxy instead\footnote{If thermal conduction operates, which may be expected given the steep temperature gradient at the intersection between the bubble's outer surface and the colder ISM, it may be possible for thermal feedback to couple to a larger volume than seen here.}.
Thus, thermally-driven outflows impact the halo environment more profoundly, while the innermost regions are most efficiently regulated by radiation pressure.

\section{Discussion}
\label{sec_discussion}

We now consider the implications of radiation pressure  on dust as an AGN feedback mechanism and discuss the physical and numerical limitations of our work.

\subsection{AGN radiation pressure as a feedback mechanism}

Radiation pressure on dust from a quasar appears to be an important feedback mechanism, which can drive large-scale outflows and reduce star formation in massive galaxies by factors $\gtrsim 3$.
In most simulations, star formation is suppressed efficiently only in the first $\sim 10 \, \rm Myr$ from the onset of radiation injection, i.e. when gas is still optically thick in the IR. Though the initial removal of gas as well as the expansion of the ISM of the quasar host galaxy lead to long term star formation suppression in all simulations, star formation tends to increase at late times.
In our simulations, which follow the evolution of a massive galaxy capable of hosting a bright quasar at $z \gtrsim 6$, IR radiation pressure is essential to ensure feedback operates at reasonable quasar luminosities.
Single-scattering is ineffective because the potential well is too deep and the galaxy is fed by gas inflows at very high rates.
While single-scattering radiation is inefficient in galaxies which are very compact, high central densities favour high optical depths in the IR, which could lead to efficient multi-scattering radiation pressure.

In this study, we have shown that \emph{from an energetics point of view}, IR radiation pressure can be a significant AGN feedback mechanism.
It leads to the generation of mass-loaded outflows with peak outflow rates $10^3 \-- 10^4 \, \rm M_\odot \, yr^{-1}$, displacing masses $\sim 10^{10} \, \rm M_\odot$ at peak velocities $\gtrsim 1000 \, \rm km \, s^{-1}$.
However, the outflows generally have only moderate kinetic luminosities $\sim 0.1 \% L_{\rm bol}$, which are considerably lower than observational reports of $\gtrsim 1 \% L_{\rm bol}$ \citep[e.g.][]{Cicone:14, Zakamska:16}.
In addition, outflow rates, as well as any quantity derived from them, vary rapidly on $\sim \, \rm Myr$ scales as the amount of mass which can be entrained changes and different portions of the outflow stall; the same outflow may appear differently mass-loaded depending on when and where it is observed, even for a constant quasar light-curve.
The non-steady nature of the outflows generated in our simulations could partially account for the scatter of measured outflow rates at fixed AGN luminosity \citep[e.g.][]{Fiore:17, Rupke:17}.

While the outflows launched by IR radiation pressure are only moderately powerful, particularly when compared to outflows driven by thermal energy injection, the impact of IR radiation on the star formation rate is considerable.
We have shown that star formation can be completely quenched within the innermost kpc at reasonable quasar luminosities. 
Quenching here happens through a combination of gas removal and through the local suppression of gas density in the central galaxy.
The latter, which does not occur in simulations with single-scattering only radiation or thermal energy injection, is caused by internal pressurisation of trapped IR radiation.
This process could potentially result in quenching without gas ejection in massive high-z galaxies.

Throughout this study, we have taken a number of simplifying assumptions that favour radiation pressure on dust as an AGN feedback mechanism.
We have assumed dust opacities in line with metal-rich and hot dust, high quasar luminosities, a spatially constant dust-to-gas ratio and, in many of our simulations, we have neglected the possibility that dust is destroyed in the outflow through shocks and thermal sputtering.
The quasar host galaxy is, by construction, initially highly obscured, which enhances the impact of IR radiation pressure. Supernova feedback appears not to reduce the column densities substantially (see Section~\ref{sec_supernova}), but even stronger feedback or earlier AGN-driven outflows may further reduce the IR optical depths.

Despite these assumptions, we have found that radiation pressure on dust from a quasar is able to generate only relatively weak outflows with masses not greater than $\sim 10^{10} \, \rm M_\odot$, amounting to just a few percent of the total halo gas mass.
In some of our simulations, we have adopted a simple model for dust destruction, by setting the dust opacities to zero for gas hotter than $T \, = \, 3 \times 10^4 \, \rm K$, finding outflow velocities lower by only a small factor $\sim 1.5$, and a weaker impact on star formation rate; the results here become comparable to those seen for our simulations with slightly lower quasar optical/UV luminosities, i.e. star formation is suppressed efficiently within the first $10 \, \rm Myr$ and then starts to rise, approaching the result of the `noAGN' simulation.

In principle, the outflow could be boosted if supernovae inject energy at a very high rate, which might be expected given the high star formation rates  of $\gtrsim 500 \, \rm M_\odot \, yr^{-1}$ present in the galactic disc, which is less affected than the nucleus, or if the contribution from radiation pressure from the stellar radiation field is also significant.
Missing physical ingredients, such as cosmic rays excited in the shock fronts developed in AGN-driven outflows or supernova explosions, may also play a role in generating more powerful outflows.

Nevertheless, one should question whether AGN-driven outflows are able, by themselves, to lead to rapid quenching in massive galaxies.
We have shown that, even in circumstances in which the cold and dense gas component has a high covering fraction, outflows pave paths of least resistance, whether driven by radiation pressure or through a hot bubble.
The outflows then merely vent out, as has been seen in many previous simulations \citep[e.g.][]{Costa:14, Costa:14a, Gabor:14, Bourne:14, Curtis:16, Bieri:17}.
The cold gas that is seen through molecular line emission in observations may arise from a thermal instability in the shocked outflowing medium and not necessarily indicate a thorough depletion of the molecular reservoir.
Moreover, we calculate inflow rates on the order of $100 \-- 1000 \, \rm M_\odot \, \rm yr^{-1}$, despite the presence of a hot atmosphere, even when this is boosted by thermal feedback, in the quasar host halo.
These high inflow rates are certainly more persistent than the only temporarily high outflow rates seen in our simulations and continue to replenish the central galaxy down to (at least) $z \, = \rm 6$.

Based on our findings, we suggest that non-ejective modes of AGN feedback may play a complementary, if not a more dominant, role in quenching massive galaxies.
Recent far-infrared observations of high redshift post-starburst galaxies support such a picture. \citet{Suess:17}, for instance, find significant molecular gas reservoirs ($M_{\rm gas} \, \approx \, 6 \times 10^9 \-- 3 \times 10^{10} \, \rm M_\odot$) in two quenched post-starburst massive galaxies at $z \,\approx\,0.7$, while \citet{Schreiber:17} present possible evidence for quenching without complete mass ejection in an extremely obscured galaxy at $z \, \approx \, 4$.
In our simulations, trapped IR radiation leads to the `inflation' of the massive galaxy as a whole, pushing its star formation rate to lower levels, but other mechanisms could also be at play (see Section~\ref{sec_limit}).

\subsection{Are the required quasar luminosities realistic?}
\label{sec_supernova}

Observed bright quasars at $z \gtrsim 6$ have estimated bolometric luminosities of $10^{47} \-- 5 \times 10^{47} \, \rm erg \, s^{-1}$ \citep[e.g.][]{DeRosa:14, Reed:17}.
In this regard, the luminosity of $3 \-- 4 \times 10^{47} \, \rm erg \, s^{-1}$ for which IR radiation pressure is able to generate an outflow in our simulations is consistent with that implied by observations, \emph{assuming that the dark matter halo followed in our simulations should host such a bright quasar.}
The brightest quasars detected at $z \gtrsim 6$ have an estimated comoving number density of $\approx 1 \, \rm Gpc^{-3}$ \citep{Fan:06}, such that even the $500 h^{-1} \, \rm Mpc$ cosmological box employed in our simulations is arguably too small to capture their likely host haloes.
Since the binding energy of the central galaxy should increase for higher mass dark matter haloes, even the `momentum boost' provided by IR radiation might prove too weak to efficiently regulate stellar mass and black hole growth.
Note, however, that the enhanced IR optical depths could compensate by boosting radiation pressure.

We have neglected feedback from supernova explosions in this study.
While unlikely to unbind the very tightly bound gaseous bulge \citep{Costa:15}, stellar feedback could plausibly decrease its binding energy, leading to a drop in the critical quasar luminosity required to launch an outflow. 
In such conditions, we might expect $L_{\rm crit} < 3 \-- 4 \times 10^{47} \, \rm erg \, s^{-1}$, or a  $< 2 \-- 3 \times 10^9 \, \rm M_\odot$ SMBH radiating at its Eddington limit, for our $2.4 \times 10^{12} M_\odot$ halo.
Supernova feedback may, however, produce the competing effect of reducing the IR optical depths to much lower values, lowering the expected IR momentum boost.

\begin{figure}
\centering 
\includegraphics[scale = 0.44]{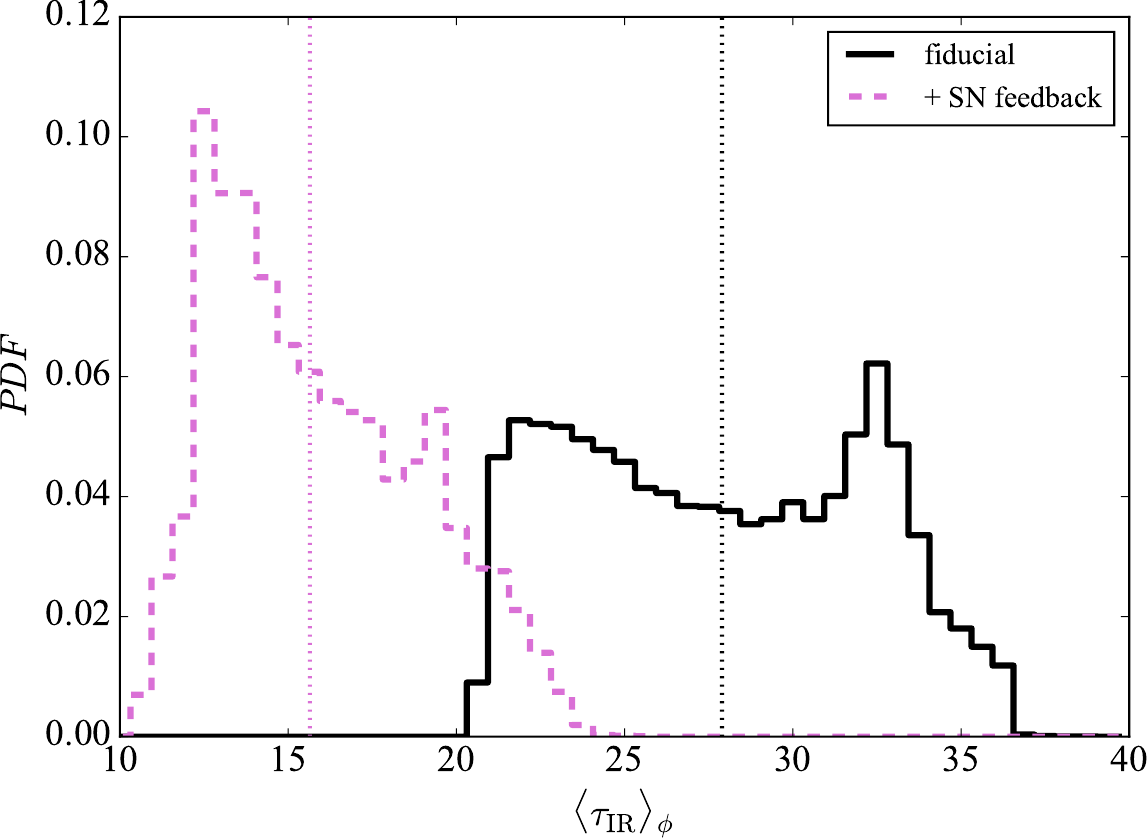}
\caption{The IR optical depth distribution around the BH particle in our fiducial simulation with no radiative feedback (black line) and in our simulation with SN feedback modelled via delayed cooling (lilac dashed line). Vertical dotted lines given the median values of the distributions. The median IR optical depth decreases by less than a factor 2 from $\tau_{\rm IR} \, \approx \, 28$ to $\tau_{\rm IR} \, \approx \, 16$. Crucially, all lines-of-sight remain IR optically thick.}
\label{fig_taudist}
\end{figure}

In order to verify that the IR optical depths obtained in this study do not depend strongly on the absence of supernova-feedback before the onset of AGN feedback in our simulations, we perform a new simulation, identical to noAGN, but following feedback from supernovae. 
There are various ways in which supernova feedback is modelled in {\sc Ramses} and it is beyond the scope of this study to investigate how the various models affect the structure of massive $z > 6$ galaxies.
Instead, we select a particularly strong implementation of supernova feedback to probe how the column density distribution responds to violent supernova feedback. 

We adopt the `delayed cooling' prescription described in \citet{Teyssier:13}, whereby radiative cooling is shut down in gas which has been recently heated by nearby supernova events.
We assume that each stellar particle of our simulations samples a single stellar population, each of which injects a supernova (thermal) energy $E_{\rm SN}$ of
\begin{equation}
E_{\rm SN} \, = \, 10^{51} \eta_{\rm SN} \frac{m_{\rm *}}{m_{\rm SN}} \, \rm erg \, ,
\end{equation}
where the fraction of mass per stellar particle $\eta_{\rm SN}$ recycled into supernova ejecta is set to $\eta_{\rm SN} \, = \, 0.2$ and the average stellar mass of a Type II SN progenitor $m_{\rm SN}$ is set to $m_{\rm SN} \, = \, 10 \, \rm M_\odot$, following the assumption that each stellar population is approximately described by a Chabrier initial mass function (IMF).
Each SN injection episode is associated with mass injection at a rate 
\begin{equation}
m_{\rm ej} \, = \, \eta_{\rm SN} m_{\rm *} \, , 
\end{equation}
though the associated metal injection is neglected in our simulations. 
Energy is simultaneously added in the form of a passive scalar to a non-thermal energy variable $\rho \epsilon_{\rm turb}$, where $\epsilon_{\rm turb}$ is a specific energy tracer typically associated with unresolved turbulent energy \citep[see][]{Teyssier:13}.
The non-thermal passive scalar advects with the gas and is assumed to decay on a timescale $t_{\rm delay}$ \citep[see e.g.][for details]{Rosdahl:17}.
In order to maximise the effect of SN feedback, we set $t_{\rm delay} \, =\, 20 \, \rm Myr$, twice the value expected given our simulation parameters \citep[see Eq. A8 in][]{Dubois:15}.

Supernova feedback in our simulations reduces the optical depths by about a factor $\sim 2$.
In Fig.~\ref{fig_taudist} we plot the IR optical depth distribution around the BH particle at $z \,=\, 6.5$, when radiation injection begins in our simulations with AGN feedback, showing results for noAGN with a solid black line and for our simulation with stellar feedback with a dashed lilac line.
The dotted lines show the median values of each distribution.
The highest optical depths drop from $\approx 37$ to $\approx 25$, the minimum optical depths drop from $\approx 22$ to $\approx 10$, while the median drops from $\approx 28$ to $\approx 16$.
The central regions of the galaxy thus remain optically thick to IR radiation in every direction.

Interestingly, the potential well deepens in the presence of SN feedback and the peak of the total circular velocity increases from $640 \, \rm km \, s^{-1}$ to $670 \, \rm km \, s^{-1}$ at $z \, = \, 6.5$. 
The larger gas mass present in the latter, which is caused by inefficient star formation in the progenitors of the massive galaxy, means that the critical luminosity, as obtained by Eq.~\ref{eq_leq} also increases.
We estimate that $L_{\rm eq} \approx 5.3 \times 10^{48} \, \rm erg \, s^{-1}$, a factor $\approx 1.6$ increase. 
Thus, supernova feedback could, in principle, create the requirement of even higher quasar luminosities in order for radiation pressure to drive outflows.
However, the absolute value of this threshold luminosity should depend on the supernova model and also on the time at which rapid black hole accretion occurs.

\subsection{Numerical and physical limitations}
\label{sec_limit}

\subsubsection{M1 method}
\label{sec_m1method}

{\sc Ramses-RT} solves radiative transfer using a moment method, i.e. conservation laws are derived for the radiation energy density and flux \citep[see][for details]{Rosdahl:13}.
The resulting set of partial differential equations is closed by assuming the M1 closure for the Eddington tensor.
While fast and reasonably accurate \citep[see e.g.][]{Costa:17}, the moment method is less accurate and more diffusive than other techniques available.
As a consequence, {\sc Ramses-RT} tends to underestimate the trapping efficiency of IR radiation when compared to ray-tracing or Monte Carlo methods \citep[see e.g.][]{Davis:14, Tsang:15}. It is therefore possible that radiation pressure is even more efficient than seen in our simulations (given our setup).
In practical terms, however, due to simplifications of dust physics and the choice of high dust opacities, we consider a more efficient scenario to be implausible.

\subsubsection{Dust physics}

We have used a very simplified treatment of dust physics throughout the paper.
In order to bridge the gap with existing analytic models of radiation pressure-driven outflows \citep[e.g.][]{Murray:05, Thompson:15, Ishibashi:15}, we have assumed the Rosseland mean opacity to be fixed at all temperatures, or to drop to zero for gas above a temperature of $T \, = \, 3 \times 10^4 \, \rm K$.
In reality, the opacity in the optically thick limit $\tau_{\rm IR} > 1$ scales with the dust temperature as $\propto T^2$ for $T \lesssim 150 \, \rm K$ \citep[e.g.][]{Semenov:03}.
In particular, the value of $\kappa_{\rm IR} \, = \, 10 \, \rm cm^2 g^{-1}$ assumed here is valid for hot dust with $T \approx 150 \, \rm K$ and for a dust-to-gas ratio of $f_{\rm D} \, =\, 0.01$.
In addition, the dust-to-gas ratio can vary spatially and with redshift, which we have ignored.

We have also shown that some of our results are sensitive to whether and for how long dust is able to survive in the outflowing material.
While radiation was found to couple to cool and dense gas in our simulations even without explicitly reducing the opacities at temperatures above $T \approx 10^4 \, \rm K$, the long term impact on the star formation rate is found to be equivalent to a moderate reduction in the quasar luminosity. 
While star formation is efficiently suppressed in the first few Myr, it rises after the cold dense material has been expelled from the host galaxy and the remaining gas is  too hot to couple to radiation.
We have seen that radiative cooling is able to maintain a large fraction of gas at $T \approx 10^4 \, \rm K$, which may permit some of the dust grains to survive if the cooling times are sufficiently short, but a portion of the dust should nevertheless be destroyed through grain-grain collisions, shocks and the ambient radiation field, as well as thermal and non-thermal sputtering \citep{Draine:79, McKee:89, Jones:96, Hensley:14}. It is likely that all these effects would further decrease the ability of trapped IR radiation to drive large-scale outflows.
More sophisticated dust formation and destruction models are now starting to be incorporated into simulations of galaxy formation \citep[e.g.][]{Zhukovska:16, McKinnon:17}. 
A crucial test will be to couple AGN and stellar radiation to such a dust model in order to probe the extent to which the various physical processes outlined here affect this AGN feedback scenario.

Another simplifying assumption we have made is that gas and dust are always coupled hydrodynamically.
The coupling between gas and dust in reality depends on the size of the dust grain \citep[e.g.][]{Booth:15, Hopkins:16}.
Moreover, in the presence of strong shocks, dust grains may become decoupled from the gas phase, further limiting the net momentum imparted on the gas.
Clearly, all these questions require further exploration in future work.

\subsubsection{Black hole accretion}

We have not modelled black hole accretion self-consistently and have therefore not captured the effect of black hole accretion self-regulation.
Instead, we have assumed the quasar to release energy at a constant luminosity or according to a predefined light-curve (Fig.~\ref{fig_lightcurve}). 
Unlike we have assumed here, the AGN light-curve should be, at some level, correlated to the black hole accretion rate. Due to insufficient resolution, however, it is unclear to what extent the AGN luminosity should correlate with the accretion rate, when evaluated at $\sim 100 \, \rm pc$ scales; some of the gas may not accrete onto the black hole or do so only after a considerable time delay.
It is also unclear whether tying the AGN luminosity to the black hole accretion would result in less efficient IR radiation pressure feedback. 
It is possible that, as the AGN fades, the dense nucleus of the galaxy reforms, boosting the IR radiation force. 
This situation is, for instance, difficult to achieve in our fiducial simulations, in which the AGN luminosity is fixed at a constant value, preventing the regrowth of a central density peak.

On the other hand, this simplification allows us to study the properties of radiation pressure-driven outflows cleanly without invoking a `subgrid model' for black hole accretion.
Black hole accretion models are notoriously uncertain, mainly because the relevant physics is poorly understood and the spatial (e.g. the accretion disc) and time scales are not resolved.
Most models rely on prescriptions for Bondi-Hoyle-type accretion flows and often make the additional, simplifying, assumption that AGN feedback occurs instantaneously.
Choices have to be made regarding the spatial scale within which the Bondi parameters (gas density and speed of sound) are to be measured and how the average is to be performed.
\citet{Curtis:15, Negri:17}, for instance, find discrepancies of up to two orders of magnitude in the black hole mass required for self-regulated growth just due to different choices of resolution, averaging method, the spatial scale within which the Bondi parameters are evaluated as well as the feedback implementation.

Our results, however, should hold in general as long as the quasar luminosity $L$ is allowed to reach the high values probed here.
In particular, outflows should be launched as long as $L \sim L_{\rm crit}$. 

\subsubsection{Star formation criterion}

One of our key findings is the ability of trapped IR radiation pressure to reduce the star formation rate in massive compact galaxies.
We identify two quenching channels: (i) gas removal through outflows and (ii) density suppression due to internal pressurisation by trapped IR radiation.
These results depend on our star formation model, which only allows star formation to occur in dense and cold ($T < 2 \times 10^4 \, \rm K$) gas.
In reality, star formation likely depends also on the dynamical and turbulent state of dense gas \citep[e.g.][]{Semenov:16} and also on the ability of dust to shield star forming regions from hard radiation \citep[e.g.][]{Draine:96}.
Both physical ingredients should increase the complexity of how radiation pressure on dust may affect star formation.
For instance, variations in dust-to-gas ratio induced by the removal of dust grains via radiation pressure could conceivably facilitate the destruction of molecular gas or iron out density inhomogeneities \citep[see e.g.][]{Rosdahl:15b}, preventing or delaying their collapse.
In order to gain further insight into the possibility of non-ejective quenching, future investigations should address the interaction of AGN radiation with the interstellar medium at $\sim \, \rm pc$ scales.

\subsubsection{Other missing physics}

We have focussed on radiation pressure on dust as an AGN feedback mechanism.
There are, however, multiple ways in which AGN radiation may affect the interstellar medium of the host galaxy.
Photoionisation and photo-heating of gas may, for instance, contribute to suppressing star formation in the AGN host galaxy \citep[see][for the effect of stellar photoionisation]{Rosdahl:15b}. Radiation pressure from the stellar radiation field could both aid the propagation of a large-scale outflow as well as reduce the density of interstellar gas, further limiting star formation.

An additional effect of the AGN radiation field is Compton heating of gas with $T \lesssim 10^7 \, \rm K$ in the inner regions of the galaxy \citep{Gan:14, Xie:17}.
This process may drive slow outflows from the nuclear regions and may help maintaining the black hole accretion rate of the host galaxy at a low level \citep[e.g.][]{Ciotti:97}. 

Importantly, a cold ISM phase is missing from our simulations. 
Cold gas has a much lower volume filling fraction \citep[though see][]{McCourt:16}, is inhomogeneous\footnote{Though note that lack of confinement possibly resulting by density inhomogeneity should also limit the efficiency of feedback generated by hot over-pressurised bubbles and not just radiation.} and could therefore be less efficient at trapping IR radiation \citep[e.g.][]{Bieri:17}.
We therefore underscore the need to revisit the problem at higher resolution and in the presence of a cold ISM phase.

Additional AGN feedback processes (such as nuclear winds or jets) could also affect the properties of AGN host galaxy.
In our simulations, AGN feedback at $z > 6.5$, before we include radiation injection, could potentially have cleared out low density channels, limiting the trapping efficiency of radiation subsequently. 

Finally, we highlight thermal conduction as a potentially important physical process.
In the context of energy-driven outflows, this mechanism could be an efficient way of coupling the thermal energy contained in the central hot bubble to a larger volume of the quasar host galaxy, analogously with trapped IR radiation.

\section{Conclusions}
\label{sec_conclusions}

We have presented a suite of cosmological radiation-hydrodynamic simulations performed with the code {\sc Ramses-RT} in order to test the scenario in which quasars launch large-scale outflows via radiation pressure on dust.
We have found that single-scattering radiation pressure (from UV/optical radiation) on its own leads to galactic outflows only at very high quasar luminosities $L \gtrsim 1.1 \times 10^{48} \, \rm erg \, s^{-1}$. Given that the brightest $z \, = \, 6$ quasars plausibly reside in haloes even more massive than considered in this study, we consider such high quasar luminosities to be unrealistic. This conclusion applies even in the presence of strong supernova feedback. 
A momentum input rate of $> L/c$ thus appears to be required to launch large-scale outflows and affect star formation in the galaxy as a whole \citep[see also][]{Debuhr:11, Costa:14}.

Multi-scattering of reprocessed IR radiation leads to large-scale outflows at luminosities a factor $\approx 4$ lower than in the single-scattering case.
In our simulations, IR multi-scattering is efficient because the gas configuration is initially optically thick in the IR along virtually every line-of-sight.
The `boost factor' over which the radiation force is amplified compared to the single-scattering scenario is $\approx 10 L/c$.
The coupling, however, quickly becomes inefficient as the dense gas envelope is cleared out and low density channels are created.
Accordingly, radiation pressure on dust is only efficient in the initial `obscured' phase, when the quasar is enshrouded by dense gas.

Even though radiation pressure on dust appears to be strong enough to launch a galactic outflow, its effects are limited to the innermost regions of the galactic halo.
We find a net suppression in star formation rate in the galactic halo by a factor of $\approx 3$.
The star formation in the innermost few kpc, however, drops to zero, suggesting that IR radiation pressure might be very efficient when it comes to regulating the star formation rate in compact starburst galaxies at high redshift. 
The inclusion of a temperature cut-off in the dust opacity, mimicking dust destruction in hot gas, or AGN variability both decrease the efficiency of radiation pressure as a feedback mechanism, leading to outflows with properties similar to those obtained by reducing the quasar luminosity. While the effect of IR radiation pressure is still not negligible in periods of high quasar luminosity and high central densities in these simulations, it is clear that the ability of dust to survive in outflowing material and the duration of quasar outbursts are key to the efficiency of radiation pressure on dust as a feedback mechanism.

The way in which IR radiation pressure affects the quasar host galaxy is twofold: (i) gas is expelled from the nuclear region, reducing the density and hence star formation rate locally, (ii) the gas density is suppressed globally since a portion of IR radiation can diffuse out to dense regions at large radii.
The last mechanism is particularly promising as it could open up a new avenue for suppressing star formation in high-redshift galaxies.
Given its potential, this scenario deserves to be analysed at higher resolution in more realistic simulations.

The radiation pressure-driven outflows generated in our simulations initially have an approximately symmetric conic structure and peak mass outflow rates of $(10^3 \-- 10^4) \, \rm M_\odot \, yr^{-1}$.
The outflows are not mass-conserving and only a small fraction of the gas mass is accelerated beyond $\approx 10 \, \rm kpc$.
Quantities such as mass outflow rates, momentum fluxes and kinetic luminosities are sensitive to \emph{when} and \emph{where} they are measured, fluctuating by large factors over timescales of $\sim 1 \, \rm Myr$. 
Note that we measure kinetic luminosities on the order of $0.1 \% L$, which agree with various observational measurements \citep[e.g.][]{Cicone:15, Carniani:15, Veilleux:17}, but are about an order of magnitude lower than the canonical $5 \% L$ often claimed to be required for efficient feedback.

The simulated outflows are naturally multi-phase and have a temperature structure which is set by radiative cooling. The bulk of outflowing mass has very short cooling times and cools radiatively over a flow time.
Using simulations with lower radiative cooling efficiencies, we explicitly find that the cool $10^4 \, \rm K$ component found in our simulations forms via in-shock radiative cooling.

There are interesting differences between the effects of IR radiation pressure feedback and thermal feedback as commonly implemented in simulations of galaxy formation with AGN feedback. We find that, in the regime in which the gas density in the host galaxy is very high and the quasar is likely obscured, IR radiation pressure is more efficient at regulating star formation globally due to its ability to diffuse. Instead, thermal energy vents out of the galaxy as soon as the outflow has cleared out low density channels.
After the thick gas layers around the quasar are expelled, however, radiation pressure becomes inefficient because the gas is mostly optically thin (in the IR).
The hot pressurised bubbles generated by the thermal energy injection, however, accelerate lower density gas in the galactic halo and lead to a slower starvation of the gas reservoir in the central galaxy as the cooling times are prolonged and diffuse halo gas is pushed out.

Our findings open up many intriguing questions concerning the nature of AGN feedback and galactic outflows.
Mainly, they highlight the many channels in which so called `quasar mode' feedback may affect the evolution of massive galaxies, which include gas heating, expulsion as well as the gentler effect of pressurisation by trapped IR radiation.
 
Our results are subject to various simplifications regarding dust physics, including the constant dust-to-gas ratio and high dust opacities.
What they do suggest is that, from an energetics point of view, IR radiation pressure appears to be dynamically significant not only in simple spherical models \citep[e.g.][]{Thompson:15, Ishibashi:15, Costa:17} but also  in a more realistic cosmological setting.
In order to further scrutinise the ability of IR radiation pressure to affect star formation in galaxies, future studies must turn to dust physics and include a more realistic treatment of the cold interstellar medium and its interaction with a quasar radiation field.

\section{Acknowledgements}

We thank the referee for a thorough and useful report.
TC is grateful to Arif Babul, Pawe\l{} Biernacki, Andy Fabian, Harley Katz, Sowgat Muzahid, Joop Schaye and Corentin Schreiber for their interest and for helpful discussions. We thank L\'eo Michel-Dansac and J\'er\'emy Blaizot for sharing the code {\sc Rascas}, which we have used to cast rays through the {\sc Ramses} grid, Pawe\l{} Biernacki for making his {\sc Ramses} movie routine public and Romain Teyssier for developing the code {\sc Ramses} and for making it publicly available. TC is supported by a NOVA Fellowship. JR acknowledges support from the ORAGE project from the Agence Nationale de la Recherche under grand ANR-14-CE33-0016-03 and from the European Research Council under the European Unions Seventh Framework Programme (FP7/2007-2013) / ERC Grant agreement 278594-GasAroundGalaxies. The authors further acknowledge support by the ERC ADVANCED Grant 320596 `The Emergence of Structure during the epoch of Reionization' (PI: M. G. Haehnelt) and the ERC Starting Grant 638707 `Black holes and their host galaxies: coevolution across cosmic time' (PI: D. Sijacki). All simulations were performed on the following: the Darwin Supercomputer of the University of Cambridge High Performance Computing Service (http://www.hpc.cam.ac.uk/), provided by Dell Inc. using Strategic Research Infrastructure Funding from the Higher Education Funding Council for England and funding from the Science and Technology Facilities Council and the COSMA Data Centric system based at Durham University, operated by the Institute for Computational Cosmology on behalf of the STFC DiRAC HPC Facility (www.dirac.ac.uk). This equipment was funded by a BIS National E-infrastructure capital grant ST/K00042X/1, STFC capital grant ST/K00087X/1, DiRAC Operations grant ST/K003267/1 and Durham University. DiRAC is part of the National E-Infrastructure. Various simulations presented here were also performed on the Dutch national e-infrastructure with the support of SURF Cooperative.

\bibliographystyle{mn2e} 
\bibliography{references}

\appendix

\section{Adopted variable AGN light-curve}
\label{sec_lightcurve}

\begin{figure}
\centering 
\includegraphics[scale = 0.425]{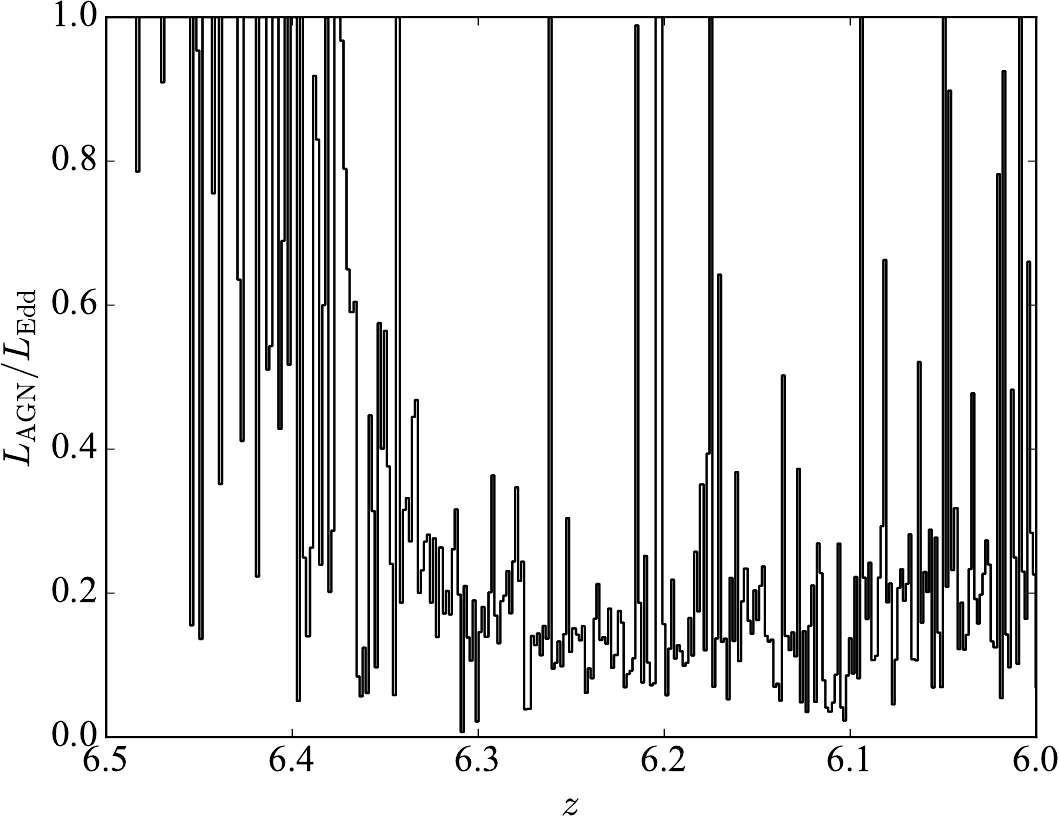}
\caption{AGN light-curve adopted in simulation UVIR-4e47-duty shown in terms of the Eddington ratio as a function of redshift. The Eddington ratio is close to unity until $z \approx 6.4$ (about 17 Myr since radiation begins to be injected), but it oscillates between $L_{\rm AGN} / L_{\rm Edd} \,\approx \, 1$ and $L_{\rm AGN} / L_{\rm Edd} \,\approx \, 0.1$ thereafter.}
\label{fig_lightcurve}
\end{figure}

In one of our simulations (UVIR-4e47-duty), we explore varying the AGN light-curve in order to illustrate a more realistic scenario in which radiation is injected at a time-varying rate.
We show the light-curve we adopt in UVIR-4e47-duty in Fig.~\ref{fig_lightcurve}, where we plot the Eddington ratio $L_{\rm AGN} / L_{\rm Edd}$ as a function of redshift. 
The light-curve is taken from one of the cosmological `zoom-in' simulations of \citet{Costa:15}, where black hole growth is investigated self-consistently in the presence of AGN and supernova feedback. 
The Eddington ratio is close to unity until $z \approx 6.4$ (about 17 Myr since radiation begins to be injected in the simulations presented in this paper), but it oscillates between $L_{\rm AGN} / L_{\rm Edd} \,\approx \, 1$ and $L_{\rm AGN} / L_{\rm Edd} \,\approx \, 0.1$ thereafter.
Note that $L_{\rm AGN}$ is set to the optical/UV luminosity $L_{\rm opt,UV}$ in UVIR-4e47-duty.

\section{Reduced speed of light approximation}
\label{sec_redc}

\begin{figure*}
\centering 
\includegraphics[scale = 0.425]{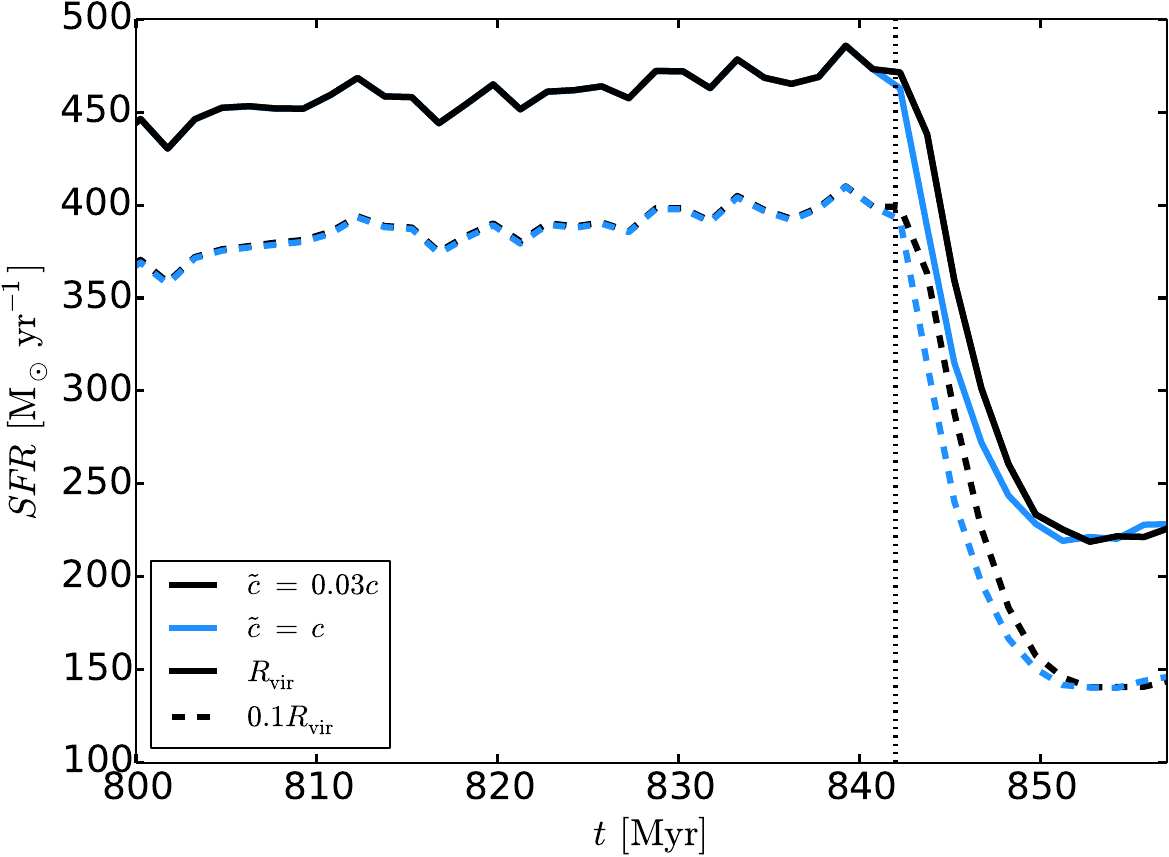}
\includegraphics[scale = 0.425]{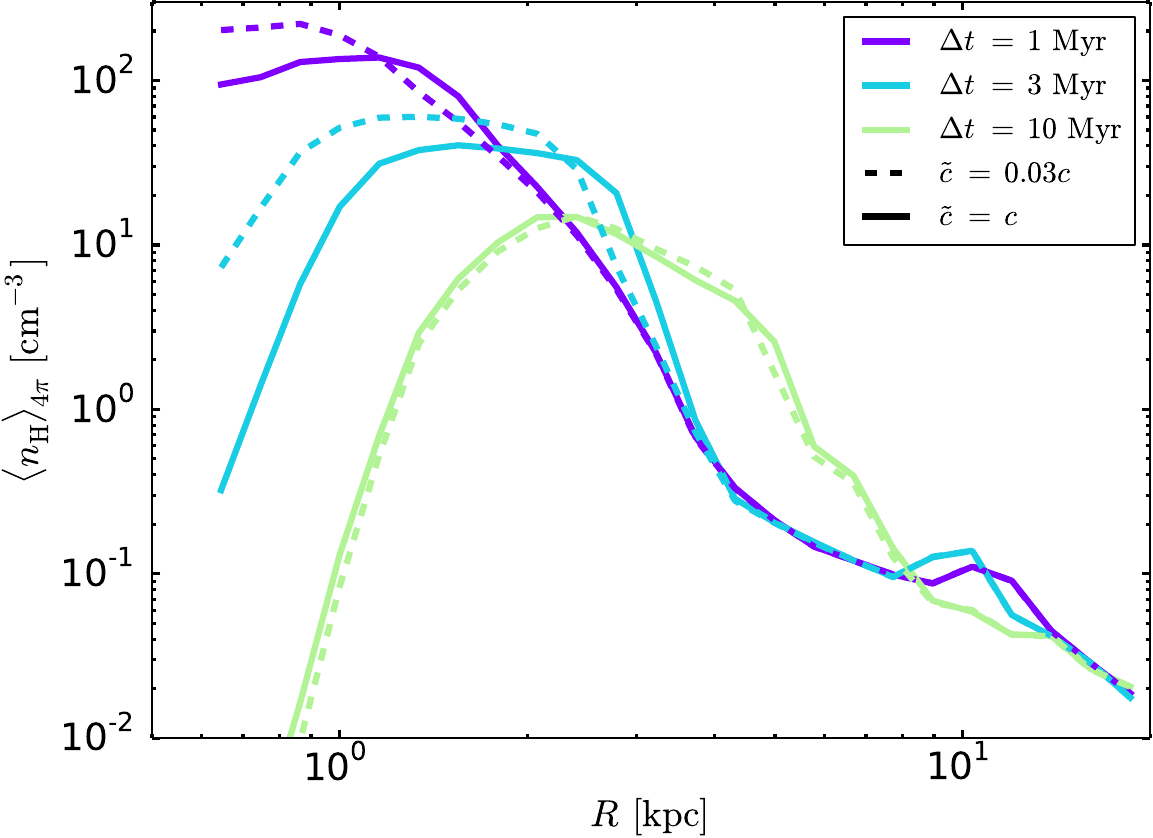}
\caption{\emph{Left:} Star formation rate as a function of time for the UVIR-4e47 performed at a reduced speed of light of $\tilde{c} \, = \, 0.03 c$ (black line) and for the simulation performed at full speed of light (blue line). The star formation rate is evaluated within the virial radius (solid lines) and at $10\%$ the virial radius (dashed lines). The star formation is suppressed slightly more efficiently in the simulation performed at full speed of light. \emph{Right:} Density profiles in the first few Myr after radiation is injected in our simulations. The solid lines correspond to the simulation performed at full speed of light, while the dashed lines give the results for the simulation with a reduced speed of light.}
\label{fig_sfr_c_conv}
\end{figure*}

\begin{figure}
\centering 
\includegraphics[scale = 0.425]{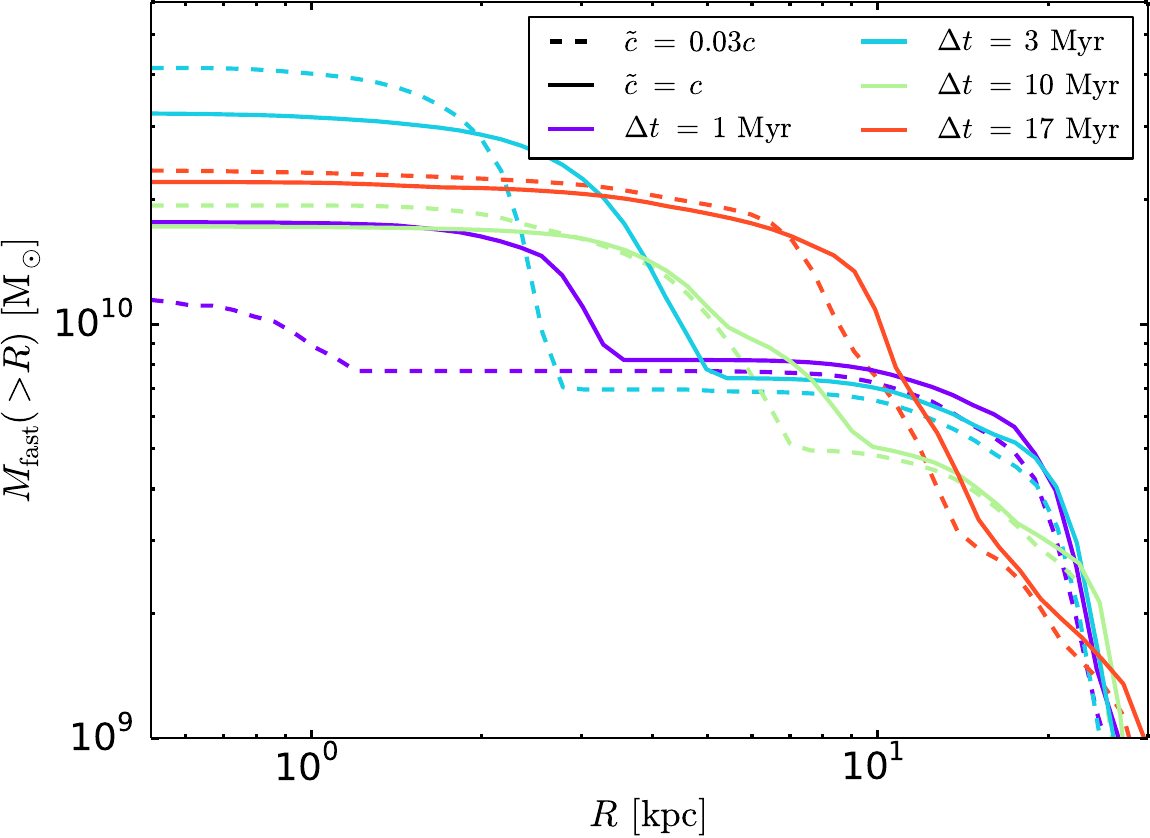}
\caption{Cumulative mass distribution of outflow with $v_{\rm r} > 300 \, \rm km \, s^{-1}$ at various times in our fiducial simulation (dashed curves) and in our simulation performed at full speed of light (solid curves). At $t \, = \, 1 \, \rm Myr$, there is more outflowing gas in the simulation performed at full speed of light, though the outflowing masses converge after a few Myr.}
\label{fig_sfr_c_conv2}
\end{figure}

Since the hydrodynamical time-step is limited by the speed of light in {\sc Ramses-RT}, the present radiation-hydrodynamic simulations rely on the `reduced speed of light' approximation, in which the speed of light is artificially reduced \citep[see][for details]{Rosdahl:13}.
We have adopted a reduced speed of light of $\tilde{c} \, = \, 0.03 c$.
If IR multi-scattering is dynamically important, there is a risk the radiation force could be artificially damped because of the reduced speed of light approximation \citep[see e.g.][]{Bieri:17, Costa:17} and it is therefore crucial to examine how our results are affected by our choice of reduced speed of light parameter.

We perform one additional simulation, identical in setup to UVIR-4e47, though performed at full speed of light, i.e. $\tilde{c} \, = \, c$.
Since the typical time-step becomes shorter by a factor of almost $100$ than in the other simulations presented in this paper, we perform it only for a duration of $20 \, \rm Myr$.
This simulation, however, allows us to probe the convergence properties during the early stages of outflow development, when the IR optical depths are highest and our results should be most sensitive to the reduced speed of light approximation.

One of our main conclusions concerns the suppressive effect of trapped IR radiation on star formation.
We thus begin by comparing the star formation history between this new simulation and UVIR-4e47.
This is shown on the left-hand panel of Fig.~\ref{fig_sfr_c_conv}, where we show star formation histories within $R_{\rm vir}$ (solid lines) and $10 \% R_{\rm vir}$ (dashed lines) in our standard simulation (black lines) and the simulation performed at full speed of light (blue curves).
We find that both star formation histories are well converged, but star formation is suppressed slightly more rapidly in the simulation performed at full speed of light.
The lower diffusion times of trapped radiation in this simulation ensure that the optically thick material feels a higher radiation force \citep[see][]{Costa:17} such that the central density drops more efficiently.
The more rapid drop in central density is also shown on the right-hand panel of Fig.~\ref{fig_sfr_c_conv}, where we plot density profiles at various times in both simulations.
Thus, concerning the ability to suppress star formation and reduce gas density, the results presented in the paper are robust and are, in fact, strengthened somewhat by adopting the full speed of light.

Finally, we also verify that the outflow masses are well converged with respect to the reduced speed of light parameter. 
In Fig.~\ref{fig_sfr_c_conv2}, we plot cumulative mass distributions for outflowing gas with $v_{\rm r} > 300 \, \rm km \, s^{-1}$ at different times.
Generally, we find outflows driven in simulations performed at full speed of light to be slightly more spatially extended (by about $\sim 1 \, \rm kpc$) at any given time.
In the first Myr (purple lines), we find $\lesssim 2$ times higher outflow masses in the simulation with the full speed of light, as a result of the somewhat higher radiation force.
At later times, however, the cumulative velocity distributions of outflowing gas in both simulations converge (green and red lines), indicating that our main results should not be very sensitive to the reduced speed of light approximation.
We have also verified that the outflow rates are well converged after a few Myr from radiation injection, with the main difference occurring in the earliest stages, where there is a time delay on the order of $\sim 1 \, \rm Myr$ between the development of an outflow in UVIR-4e47 with respect to that seen in the simulation performed at full speed of light.

In summary, we find our main results to be robust with respect to variations in the reduced speed of light parameter $\tilde{c}$ in the sense that they are well converged for most of the simulation. Unsurprisingly, the main differences arise in the first few Myr from radiation injection, where the outflow driven in the simulation performed with $\tilde{c} \, = \, c$ develops somewhat more rapidly than in that with $\tilde{c} \, = \, 0.03 c$. Our choice of $\tilde{c}$ therefore gives a lower limit to the efficiency of outflows driven by trapped IR radiation.

\section{Numerical resolution}
\label{sec_convergence}

\begin{figure*}
\centering 
\includegraphics[scale = 0.425]{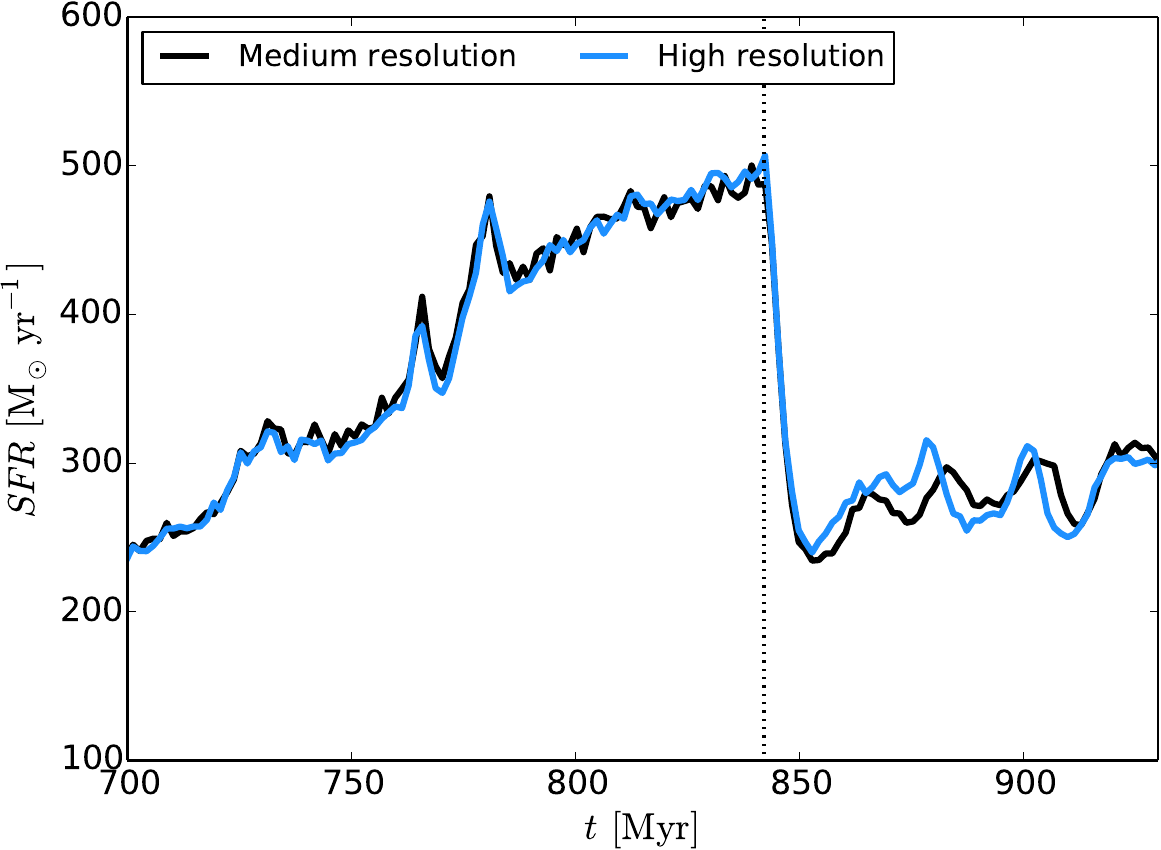}
\includegraphics[scale = 0.425]{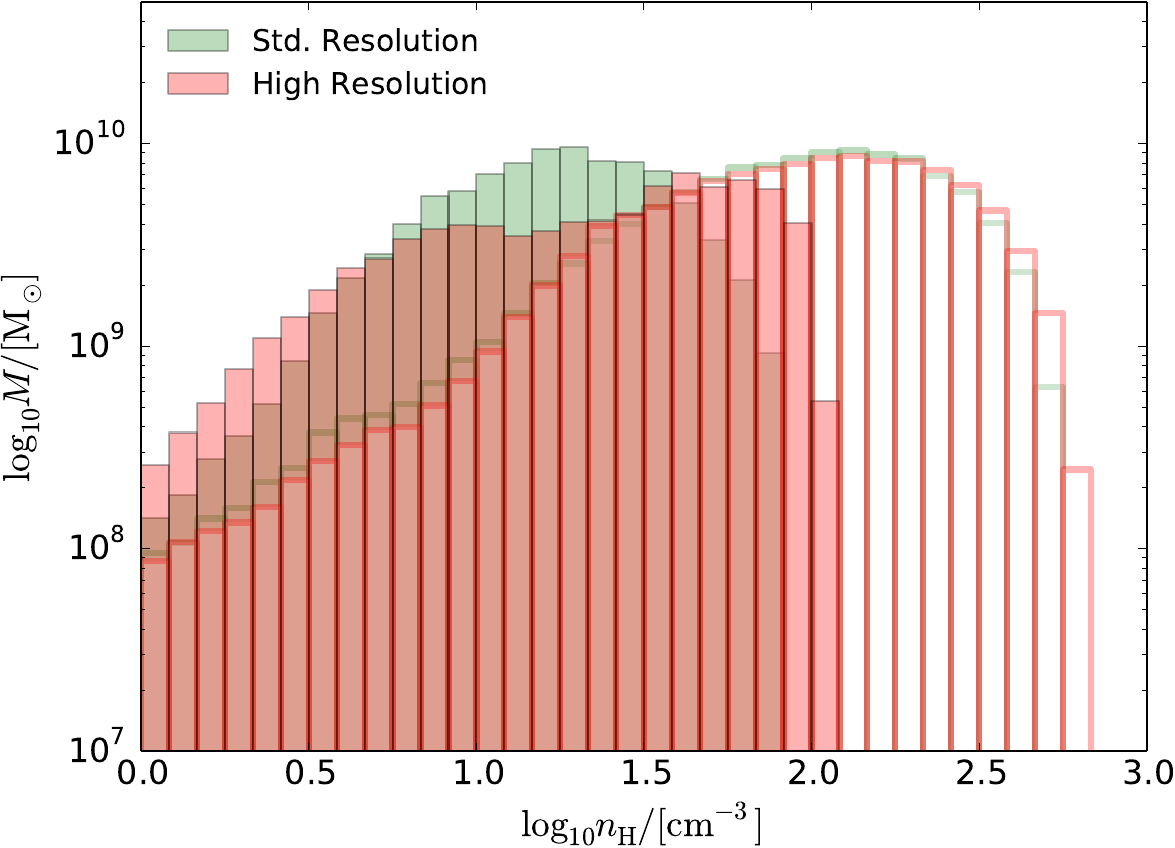}
\caption{\emph{Left:} Star formation rate within the virial radius as a function of time for the UVIR-4e47 performed at standard resolution (black line) and for the high resolution simulation (blue line). The evolution prior to radiation injection (marked with a vertical dashed line) is virtually identical in both simulations. After the quasar starts injecting radiation, the qualitative evolution of the star formation rate is also comparable. \emph{Right:} Density distribution within 5 kpc of the quasar for low velocity gas with $|v_{\rm r}| < 300 \, \rm km \, s^{-1}$ in our standard resolution simulations (green histograms) and the high resolution simulations (red histograms). Open histograms give results at $z \, = \, 6.5$ and filled histograms give results at $z \, = \, 6.4$.}
\label{fig_sfr_res}
\end{figure*}

\begin{figure}
\centering 
\includegraphics[scale = 0.48]{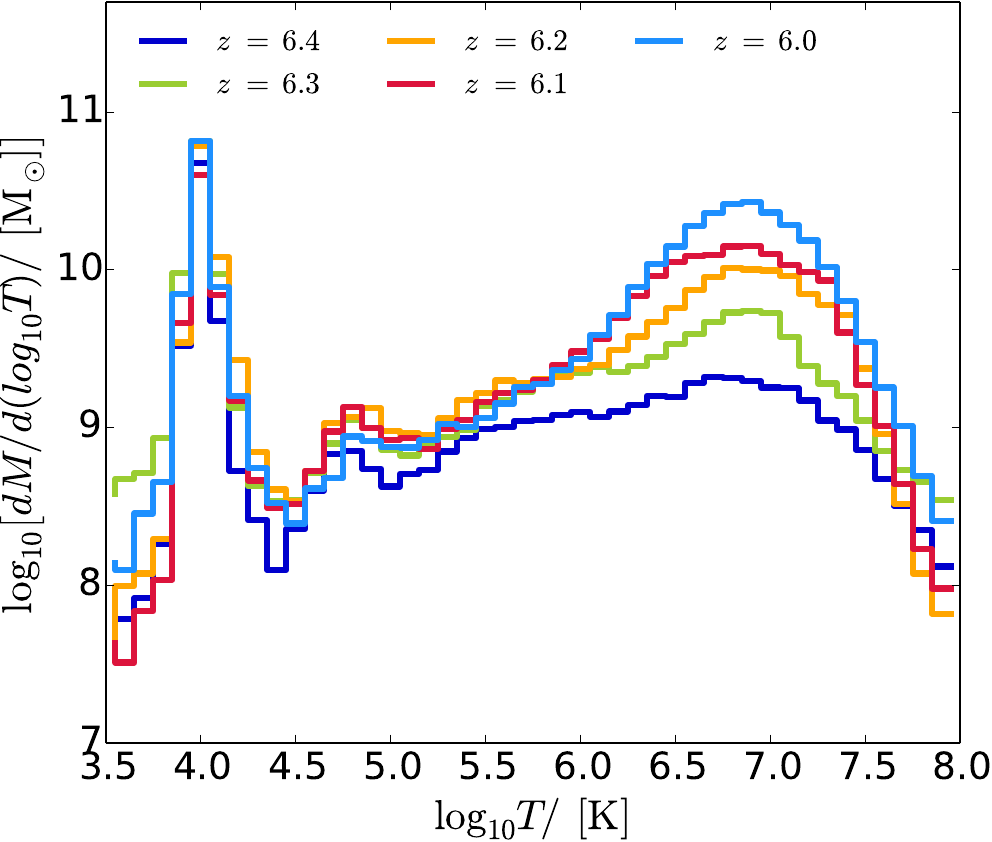}
\caption{The mass distribution of outflowing gas shown as a function of temperature in our high resolution simulation. Only gas with radial velocity $> 500 \, \rm km \, s^{-1}$ is considered. The bulk of the outflowing gas is still in the `cool phase' with temperatures $\lesssim 3 \times 10^4 \, \rm K$. Note that in UVIR-4e47-hires, there is no cut-off in the dust opacity, which explains why the hot component is more prominent than in Fig.~\ref{fig_outflowprops}.}
\label{fig_dmdt_res}
\end{figure}

We performed cosmological `zoom-in' simulations in order to follow the evolution of a massive halo at $z \, = \, 6$. 
The advantage of this approach is the self-consistent treatment of the cosmological environment of the quasar host galaxy, which, as we have seen, plays a major role in determining its structure as well as the impact of feedback on its gas content and the propagation of outflows.
There is, however, inevitable lack of resolution and the nuclear regions, where a large-scale outflow is ultimately powered, are only poorly resolved.

We here verify how/if we expect our main results to change with increasing resolution.
We probe the effects of changing resolution by re-running simulation UVIR-4e47, which includes single- and multi-scattering radiation pressure, and has been shown to launch galactic outflows in this study, at higher resolution.
We increase the maximum refinement level by a factor $2$, keeping all other parameters fixed, such that the minimum cell size drops to $\Delta x \, = \, 62.5 h^{-1} \, \rm pc$.
In particular, note the region into which radiation is injected is kept fixed and is still on the order of $100 \, \rm pc$.

We compare the star formation histories (computed within the virial radius) of UVIR-4e47 to UVIR-4e47-hires in the left-hand panel of Fig.~\ref{fig_sfr_res}.
The evolution prior to the time at which radiation injection begins is virtually identical in both simulations. After the quasar starts injecting radiation, the qualitative evolution of the star formation rate is also comparable, i.e. there is a sharp decrease down to $SFR \, \approx \, 220 \, \rm M_\odot \, yr^{-1}$ followed by a series of (an equal number) of oscillations around $SFR \, \approx \, 250 \, \rm M_\odot \, yr^{-1}$. The main difference is the timescale over which the SFR oscillates, which is somewhat shorter in the high resolution simulation.

The density distribution of low velocity gas ($|v_{\rm r}| < 300 \, \rm km \, s^{-1}$) within 5 kpc of the quasar is shown in the right-hand panel of Fig.~\ref{fig_sfr_res} at $z \, = \, 6.5$ (open histograms) and $z \, = \, 6.4$ (filled histograms). The initial density distribution in UVIR-4e47 (green histograms) and UVIR-4e47-hires (red histograms) is similar, though slightly enhanced (by $\approx 0.1 \, \rm dex$) in the high resolution simulation. Density suppression is efficient in both simulations at $z \, = \, 6.4$, as shown by the development of a bump at low density in both simulations. 
The density distribution is very flat in simulation UVIR-4e47-hires, but extends over a somewhat higher density range, populating the low- and high-density ends more than in UVIR-4e47.
This analysis suggests that as resolution increases, the impact of IR radiation pressure may become more localised and may be expected to predominantly affect the nuclear regions of the galaxy.

Finally, we verify how the temperature probability distribution function of outflowing material with $v_{\rm r} > 500 \, \rm km \, s^{-1}$ contained within the virial radius changes with resolution. This is plotted in Fig.~\ref{fig_dmdt_res} at various redshifts.
We find that the cool phase with $T \, \approx \, 10^4 \, \rm K$ remains the most prominent of all outflowing components.
The main difference with respect to Fig.~\ref{fig_outflowprops} (which refers to UVIR-4e47-dust) is the more pronounced hot component seen in UVIR-4e47-hires. However, note that there is no temperature cut-off in the dust opacity here, which is why hot gas accelerates more efficiently in this case, shocking to higher temperatures.

\section{Varying the `Jeans' pressure floor}
\label{sec_jeansp}

\begin{figure}
\centering 
\includegraphics[scale = 0.42]{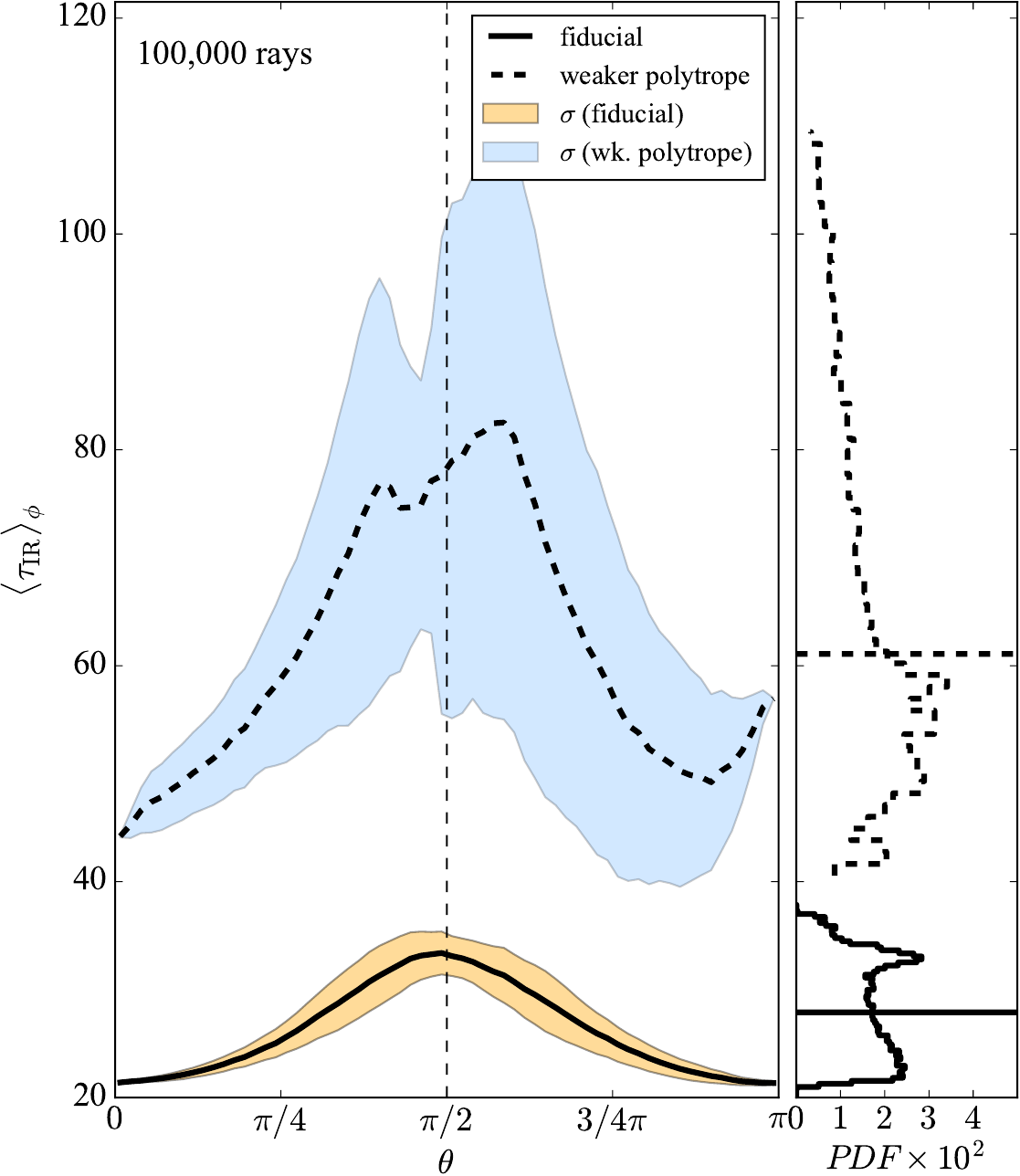}
\caption{IR optical depth distribution at $z \, = \, 6.5$ (at the time at which radiation begins to be injected) as a function of inclination angle $\theta$ (measured from the polar axis in the positive z-direction) obtained by integrating along $10^5$ rays cast in random directions. We show mean optical depths (lines) as well as the $1\sigma$ fluctuation amplitude (shades). The dashed line and the blue shade show results for the simulation with the weak polytrope, while the solid line and the yellow shade give results for our fiducial simulations, with horizontal lines giving the median value in both cases.}
\label{fig_tautheta_poly}
\end{figure}

Here we outline results from a series of simulation tests in which we probe the effect of adopting a weaker polytropic equation of state than employed in the paper.
We perform multiple simulations at different quasar luminosities, in which we take $T_{\rm J} \, = \, 10^3 \, \rm K$ in Eq.~\ref{eq_efftemp}, such that the Jeans length is always resolved with $\approx 8$ cells.
We adopt the same naming convention for these simulations as before but use the names `UVIR-poly' or `UV-poly' in order to emphasise we are employing a different polytropic equation of state. 

As expected, gas collapses down to even smaller scales in these simulations, forming even more compact galaxies.
We show the azimuthally averaged IR optical depths in Fig.~\ref{fig_tautheta_poly}, where we show the distribution obtained in the new simulation with a dashed line (and the blue shade for the standard deviation) at $z \, = \, 6.5$ (at the time at which radiation begins to be injected).
The IR optical depths of the central galaxy now reach $\tau_{\rm IR} \gtrsim 100$ (to be compared with $\tau_{\rm IR} \, \approx \, 40$ in our fiducial simulations) and the azimuthally averaged mean is systematically a factor $2$ greater than in the simulations discussed in the paper.

\begin{figure}
\centering 
\includegraphics[scale = 0.42]{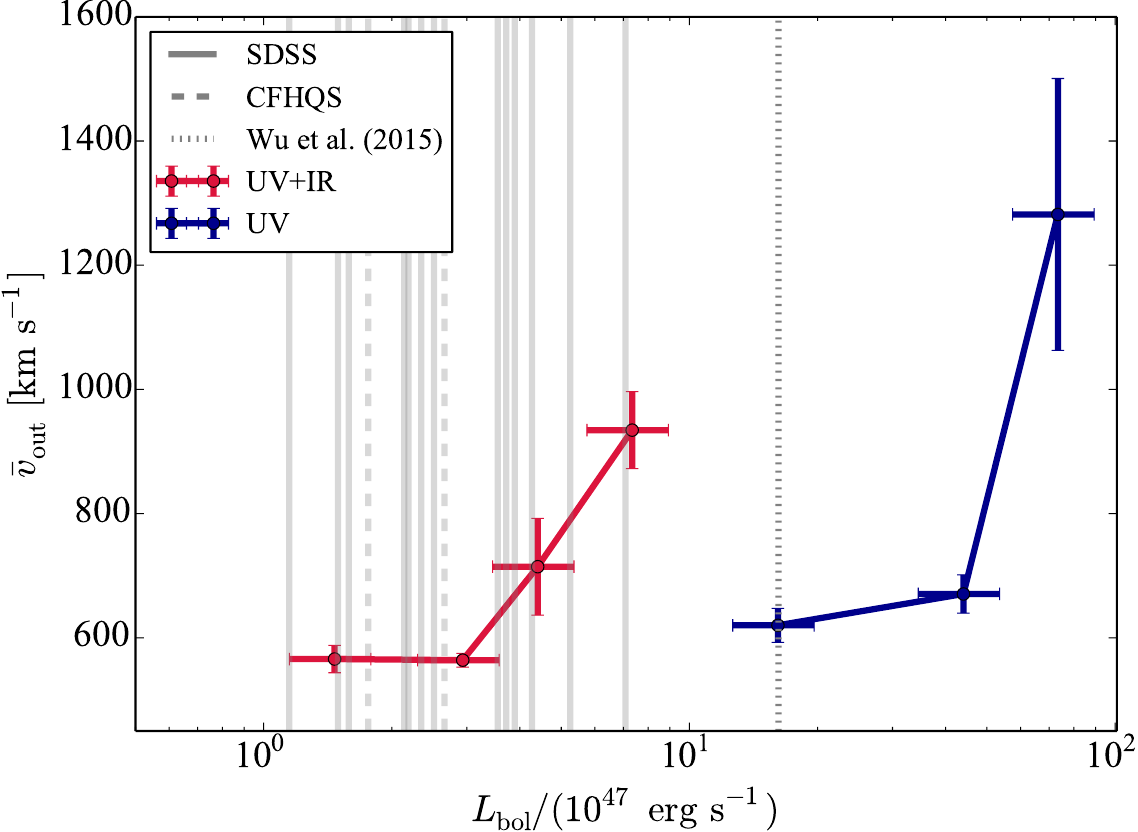}
\includegraphics[scale = 0.41]{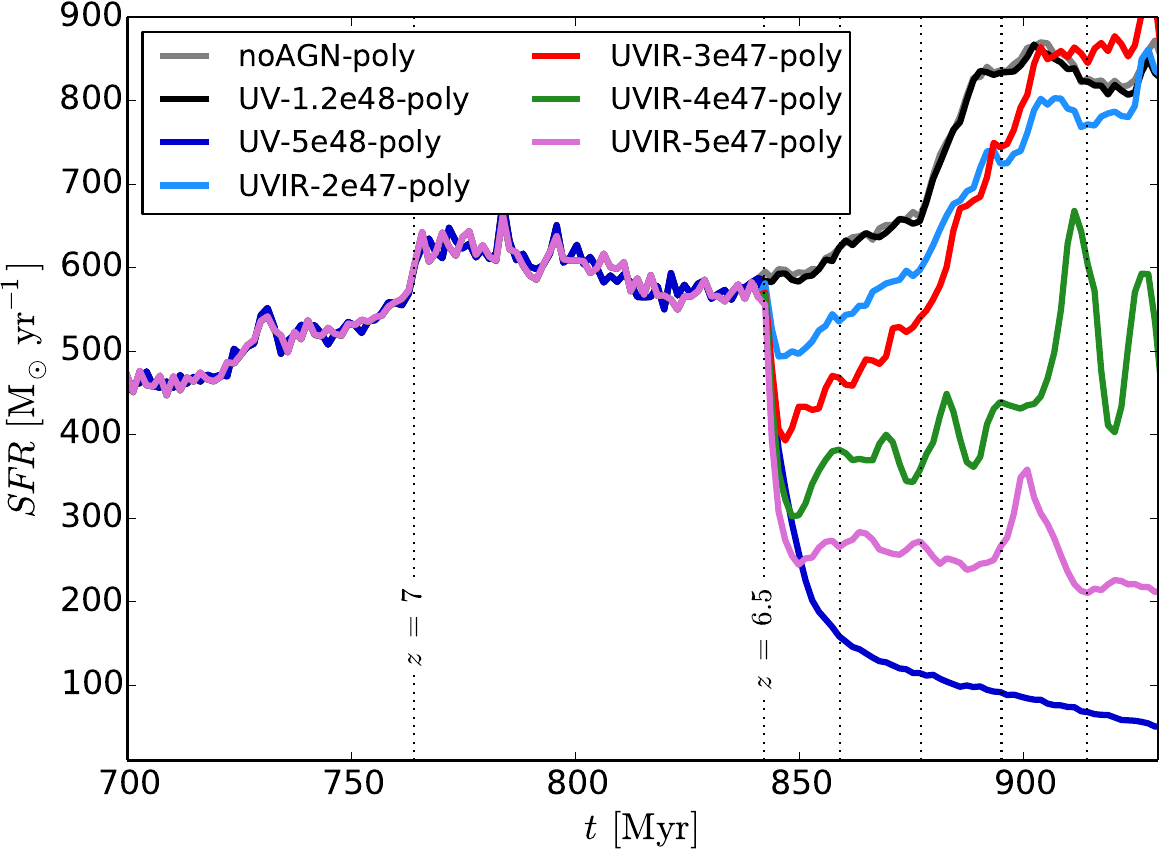}
\includegraphics[scale = 0.42]{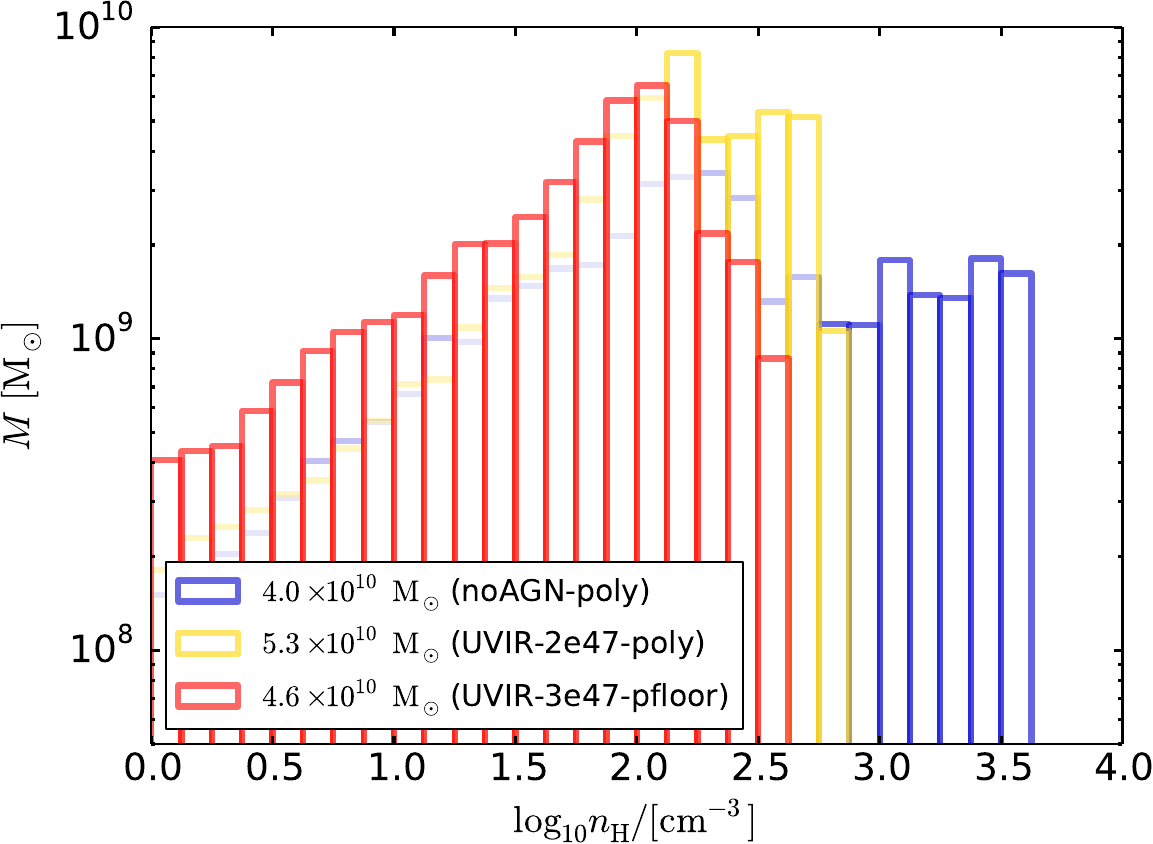}
\caption{\emph{Top:} Mass-weighted mean outflow velocity as a function of quasar bolometric luminosity in our UV-poly simulations (blue line) and in our UVIR-poly simulations (red line). Multi-scattering efficiently launches outflows at reasonable quasar bolometric luminosities $L_{\rm bol} \gtrsim 3 \times10^{47} \, \rm erg \, s^{-1}$, but, due to the deeper potential well, even higher luminosities $L_{\rm bol} \gtrsim 5 \times10^{48} \, \rm erg \, s^{-1}$ are required to power outflows with single-scattering radiation pressure. \emph{Middle:} Star formation history within the virial radius in our various simulations with the weaker polytropic equation of state. Star formation suppression with IR radiation pressure is slightly more efficient than in our fiducial simulations.  \emph{Bottom:} Density distribution function for low velocity gas ($|v_{\rm r}| < 300 \, \rm km \, s^{-1}$) located within 5 kpc of the quasar in noAGN-poly (blue histograms), UVIR-2e47-poly (yellow histograms) and UVIR-3e47-poly (red histograms). Even though the mass in low velocity gas is comparable in all these simulations, the density distribution is shifted to lower values in the presence of IR multi-scattering.}
\label{fig_poly}
\end{figure}

On the one hand, the more efficient gas collapse means that the potential well becomes deeper, such that it becomes even harder to drive outflows with single-scattering radiation pressure.
We find enormous values for the peak (total) circular velocities, which can reach $\approx 850 \, \rm km \, s^{-1}$, most likely as a consequence of unrealistic overcooling in the absence of baryonic feedback.
We estimate an optical/UV critical luminosity (see Section~\ref{sec_critlum}) for outflows driven by single-scattering of $L_{\rm crit} \, = \, 5 \times 10^{48} \, \rm erg \, s^{-1}$, i.e.  $L_{\rm bol, crit} \, = \, (6 \--9) \times 10^{48} \, \rm erg \, s^{-1}$.
On the other hand, the higher optical depths mean that IR radiation is even more efficiently trapped, compensating for the deeper potential well.
Consequently, we find that our conclusion that IR multi-scattering is required to launch powerful outflows becomes \emph{even stronger} when we perform simulations with a weaker polytrope.

While the critical luminosity to launch large-scale outflows dropped by a factor $\approx 4$ with the inclusion of IR multi-scattering in our fiducial simulations, here we find that it drops by a factor $\approx 16$.
In the top panel of Fig.~\ref{fig_poly}, we show the mass-weighted median outflow velocity (like in Fig.~\ref{fig_critlum}) as a function of quasar bolometric luminosity. 
We see that powerful outflows are driven at quasar bolometric luminosities $(3 \-- 8) \times 10^{47} \, \rm erg \, s^{-1}$ if IR multi-scattering is enabled, as in our fiducial simulations.
If multi-scattering is disabled, however, we find that bolometric luminosities on the order of $\gtrsim 5 \times 10^{48} \, \rm erg \, s^{-1}$ are required to launch outflows, in good agreement with our analytic estimate for $L_{\rm bol, crit}$.
For a weaker polytrope, IR multi-scattering therefore becomes even more important.

We show the star formation history (evaluated within the virial radius) in our various simulations using the weaker polytrope in the central panel of Fig.~\ref{fig_poly}.
The star formation rates are, on the whole, higher than in the simulations performed with the stronger polytrope by a factor $\lesssim 2$ at $z \, = \, 7$ and by a factor $\approx 1.2$ at $z \, = \, 6.5$.
The halo-wide star formation rate seen in our `UVIR-poly' simulations with $L_{\rm opt,UV} \,=\, 3 \times 10^{47} \, \rm erg \, s^{-1}$ is about a factor $\approx 2$ lower than in UV-1.2e48-poly, a suppression which is comparable to that seen in the simulations at similar quasar luminosities performed with the stronger polytropic equation of state (see Fig.~\ref{fig_sfr}).
The degree by which the star formation rate is reduced increases in simulation UVIR-5e47-poly (lilac line in the central panel of Fig.~\ref{fig_poly}), where it is $\approx 4$ times lower than in noAGN-poly at $z \, = \, 6$. Suppression of star formation in our simulations employing a weaker polytropic equation of state is therefore somewhat more efficient than in our fiducial simulations. 

Star formation suppression is again due to a combination of gas ejection and a shift in the density distribution towards lower values.
In the bottom panel of Fig.~\ref{fig_poly}, we show the density distribution for low velocity gas ($|v_{\rm r}| < 300 \, \rm km \, s^{-1}$) within the innermost 5 kpc at $z \, = \, 6.4$ in a sub-sample of our simulations. We show results for noAGN-poly (blue), UVIR-2e47-poly (yellow) and UVIR-3e47-poly (red).
Even though the mass of the low velocity gas component is comparable in these simulations, it is clear that the peak of the density distribution shifts towards lower values by more than an order of magnitude if IR multi-scattering is present.

The key conclusions of this paper are robust to changes in the polytropic equation of state, i.e. trapped IR radiation pressure leads to a sharp decrease in the star formation rate of the massive galaxy at reasonable quasar optical/UV luminosities $\sim 3 \times 10^{47} \, \rm erg \, s^{-1}$ in both our fiducial simulations (in which we employ a strong polytropic equation of state) as well as in the less conservative scenario considered here. Given the deliberate exclusion of other baryonic feedback from our simulations, however, selecting a strong polytropic equation of state could be argued to be a good compromise, preventing excessive gas collapse and the formation of unrealistic galaxies. We therefore regard our fiducial simulations as the more realistic scenario.

\section{Trapping infrared radiation}
\label{sec_irtrapeff}

All our results depend on the ability of IR radiation to be trapped within dense galactic gas.
In the main body of the paper, we have seen that in the absence of multi-scattered radiation pressure, we require about $4 \-- 6$ times higher AGN power in order to produce a comparable effect on the star formation rate of the host galaxy and launch large-scale outflows.
We must conclude that trapping of IR radiation is reasonably efficient within the gaseous bulge that forms at the centre of the massive halo.
In this section we measure the IR trapping efficiency directly and calculate the factor by which the radiation force is amplified with respect to $L_{\rm opt, UV}/c$.

We divide our simulation domain into spherical shells centred on the halo centre and with radial spacing of about $1 \, \rm \, kpc$. For each shell, we compute $\langle \rho F_{\rm IR} \rangle_{\rm 4 \pi}$, where $F_{\rm IR}$ is the infrared flux and the average is taken over all {\sc Ramses} cells lying within the shell.
The spherically averaged radiation force is then given by
\begin{equation}
\langle \mathcal{F}_{\rm rad} \rangle_{\rm 4 \pi} \, = \,  4 \pi \frac{\kappa_{\rm IR}}{c} \int_{R_{\rm in}}^{R_{\rm out}}{\langle \rho F_{\rm IR} \rangle_{\rm 4 \pi} R^2 dR} \, ,
\label{eq_radforce}
\end{equation}
while the IR optical depth is
\begin{equation}
\tau_{\rm IR} \, = \, \kappa_{\rm IR} \int_{R_{\rm in}}^{R_{\rm out}}{\rho dR} \, ,
\label{eq_tau}
\end{equation}
where the opacity $\kappa_{\rm IR}$ is assumed to be constant, in line with our simulations.
For a spherical IR radiation front with flux $F_{\rm IR} \, = \, L_{\rm IR} / (4 \pi R^2)$, Eqs.~\ref{eq_radforce} and~\ref{eq_tau} give the expected radiation force
\begin{equation}
\langle \mathcal{F}_{\rm rad, sph} \rangle_{\rm 4 \pi} \, = \,  4 \pi \frac{\kappa_{\rm IR} L_{\rm IR}}{c} \int_{R_{\rm in}}^{R_{\rm out}}{\rho F_{\rm IR} R^2 dR} \, = \, \tau_{\rm IR} \frac{L_{\rm IR}}{c} \, .
\label{eq_taulc}
\end{equation}

Note that since virtually all optical/UV radiation is absorbed at early times in our simulations, we have that $L_{\rm IR} \, \approx \, L_{\rm opt,UV}$.
We begin by testing how the radiation force $\mathcal{F}_{\rm rad}$ actually exerted by the trapped IR component in our simulations compares with the na\"ive expectation that $\mathcal{F}_{\rm rad} \, = \, \tau_{\rm IR} L_{\rm IR}/c$ (Eq.~\ref{eq_taulc}).
We follow \citet{Roth:12} and define an effective optical depth $\bar{\tau}_{\rm IR}$
\begin{equation}
\bar{\tau}_{\rm IR} \, = \, \frac{\langle \mathcal{F}_{\rm rad} \rangle_{\rm 4 \pi}}{L_{\rm UV, opt}/c} \, .
\label{eq_efftau}
\end{equation}

\begin{figure}
\centering 
\includegraphics[scale = 0.44]{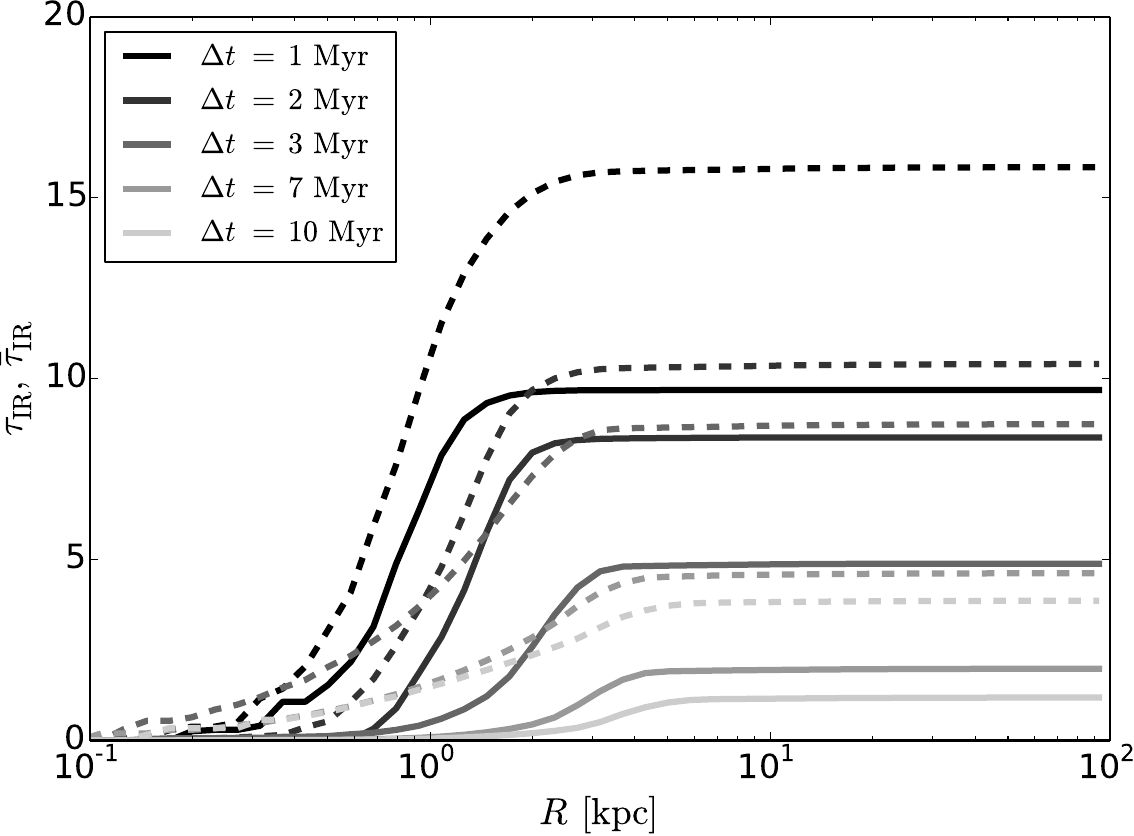}
\caption{The spherically averaged effective optical depth (solid lines) and the actual IR optical depths (dashed lines) for different redshifts in simulation UVIR-4e47. Clearly $\mathcal{F}_{\rm rad} < \tau_{\rm IR} L/c$ throughout the entire galactic halo and at all times. The typical trapping efficiency can, however, be high and here on the order of $60 \-- 90 \%$. The radiation forces are modest, lying in the range of $(1 \-- 10) L_{\rm opt, UV}/c$.}
\label{fig_tau_eff}
\end{figure}

In Fig.~\ref{fig_tau_eff}, we plot the actual, spherically averaged IR optical depth together with the effective optical depth $\bar{\tau}_{\rm IR}$ (Eq.~\ref{eq_efftau}) for simulation UVIR-4e47, at different times.
We see that while $\langle F_{\rm rad} \rangle_{4 \pi}$ can be as high as $\approx 10 L_{\rm opt, UV}/c$ at early times, it is always lower than the optical depth, i.e. in general, $\langle \mathcal{F}_{\rm rad} \rangle < \tau_{\rm IR} L_{\rm UV, opt}/c$.

The optical depth drops very rapidly once gas is expelled, such that the IR radiation force declines to just $(1 \-- 2) L_{\rm opt, UV}/c$ within a few Myr.
At this point, gas becomes optically thin to IR radiation and multi-scattering becomes inefficient.
The `momentum boost' is thus only significant for gas at radial scales smaller than a few kpc.
We can see from Fig.~\ref{fig_tau_eff} that the spatial scale at which the optical depths converge to unity is, in fact, on the order of a few ($1 \-- 4$) kpc, comparable to the radial scales within which star formation is suppressed (see Fig.~\ref{fig_sfr}).

We can make a simple estimate for the maximum scale within which star formation regulation can take place, if we assume that IR radiation pressure is the sole or dominant feedback mechanism.
We consider a galaxy with gas mass $M_{\rm gas}$. If a fraction of $f_{\rm ej}$ of this mass is ejected and swept-up into a shell, the IR optical depth of the outflow is
\begin{equation}
\tau_{\rm IR} \, = \, f_{\rm ej} \kappa_{\rm IR} \frac{M_{\rm gas}}{4 \pi R^2} \, .
\end{equation}
The radius $R_{\rm \tau}$ at which the gas becomes optically thin is obtained by setting $\tau_{\rm IR} \, = \, 1$ in the previous equation, giving 
\begin{equation}
R_{\rm \tau} \, = \, \sqrt{ f_{\rm ej} \kappa_{\rm IR} \frac{M_{\rm gas}}{4 \pi} } \, .
\label{eq_rtau}
\end{equation}
At $z \, = \, 6.5$, we find that $M_{\rm gas} \, = \, 1.2 \times 10^{11} \, \rm M_\odot$. Using $\kappa_{\rm IR} \, = \, 10 \, \rm cm^2 \, g^{-1}$, Eq.~\ref{eq_rtau} thus gives $R_{\rm \tau} \approx 3 (f_{\rm ej} / 0.5)^{1/2} \, \rm kpc$, in reasonable agreement with the spatial scale within which star formation is regulated in our simulations.
Note that even in the hypothetical case that $f_{\rm ej} \, = \, 1$, Eq.~\ref{eq_rtau} still gives $R_{\rm \tau} \lesssim 5 \, \rm kpc$.

It is also interesting to note that the momentum fluxes evaluated in Section~\ref{sec_outflowrate} are considerably lower than the initial IR optical depth measured in the simulations.
The likely reason why $\dot{P}$ appears to be lower than $\mathcal{F}_{\rm IR}$ is that momentum is also used up in slowing down infalling gas and in overcoming gravity and the confining pressure of the ambient medium.

\end{document}